\documentclass[twocolumn]{aastex63}
\usepackage[caption=false]{subfig}
\usepackage{amsmath}
\usepackage{CJK}
\usepackage{multirow}

\newcommand{\Lya}{Ly$\alpha$}
\newcommand{\Ha}{H$\alpha$}
\newcommand{\Hb}{H$\beta$}
\newcommand{\ubar}{$\mu$bar}
\newcommand{\kms}{\rm ~km~s^{-1}}

\newcommand{\ergccs}{\rm ~erg~cm^{-3}~s^{-1}}

\newcommand{\AArm}{\rm~\AA}
\newcommand{\WA}{WASP-121b}

\received{}
\revised{}
\accepted{}

\graphicspath{{./}{figures/}}

\begin{document}
\begin{CJK*}{UTF8}{gkai}

\correspondingauthor{Chenliang Huang and Tommi Koskinen}
\email{huangcl@shao.ac.cn; tommik@arizona.edu}
\title{A hydrodynamic study of the escape of metal species and excited hydrogen from the atmosphere of the hot Jupiter WASP-121b}

\author[0000-0001-9446-6853]{Chenliang Huang (黄辰亮)}
\affil{Lunar and Planetary Laboratory, University of Arizona, 1629 East University Boulevard, Tucson, AZ 85721}
\affil{Shanghai Astronomical Observatory, Chinese Academy of Sciences, Shanghai 200030, China}
\author{Tommi Koskinen}
\affil{Lunar and Planetary Laboratory, University of Arizona, 1629 East University Boulevard, Tucson, AZ 85721}
\author{Panayotis Lavvas}
\affil{Groupe de Spectrom\'etrie Mol\'eculaire et Atmosph\'erique UMR CNRS 6089, Universit\'e Reims Champagne-Ardenne, 51687 Reims, France}
\author{Luca Fossati}
\affil{Space Research Institute, Austrian Academy of Sciences, Schmiedlstrasse 6, 8042 Graz, Austria}

\begin{abstract}
\noindent 
In the near-UV and optical transmission spectrum of the hot Jupiter WASP-121b, recent observations have detected strong absorption features of Mg, Fe, Ca, and \Ha, extending outside of the planet's Roche lobe.  Studying these atomic signatures can directly trace the escaping atmosphere and constrain the energy balance of the upper atmosphere. To understand these features, we introduce a detailed forward model by expanding the capability of a one-dimensional model of the upper atmosphere and hydrodynamic escape to include important processes of atomic metal species.  The hydrodynamic model is coupled to a \Lya\ Monte Carlo radiative transfer calculation to simulate the excited hydrogen population and associated heating/ionization effects.  Using this model, we interpret the detected atomic features in the transmission spectrum of WASP-121b and explore the impact of metals and excited hydrogen on its upper atmosphere.  We demonstrate the use of multiple absorption lines to impose stronger constraints on the properties of the upper atmosphere than the analysis of a single transmission feature can provide.  In addition, the model shows that line broadening due to atmospheric outflow driven by the Roche lobe overflow is necessary to explain the observed line widths and highlights the importance of the high mass-loss rate caused by the Roche lobe overflow that requires careful consideration of the structure of the lower and middle atmosphere.  We also show that metal species and excited state hydrogen can play an important role in the thermal and ionization balance of ultra-hot Jupiter thermospheres.\\
\end{abstract}

\section{Introduction} \label{sec:intro}

\noindent
Heating of the thermospheres of short-period exoplanets by stellar X-ray and extreme UV (XUV) radiation drives mass loss by atmospheric escape that shapes the population of known exoplanets and contributes to features such as the hot Neptune desert and radius gap between Mini-Neptunes with significant H/He envelopes and Super-Earths that have lost their envelopes \citep[e.g.,][]{Owen2017}. An improved understanding of atmospheric escape and how it affects a broad range of planetary atmospheres is therefore necessary for a better understanding of the demographics of planetary systems and their evolution. In this respect, ultra-hot Jupiters such as WASP-121b in extremely close-in orbits around their host stars are particularly interesting because they offer a unique window to rapid hydrodynamic escape that can be observed with transmission spectroscopy. In particular, deep absorption features by atoms and atomic ions directly probe the escaping atmosphere \citep[e.g.,][]{Ehrenreich2015,Ballester2015,Spake2018,Yan2018}. Comparing these observations with models of atmospheric escape sheds light on the key physical processes and the properties of escaping atmospheres in extreme environments \citep[e.g.,][]{Koskinen2013b,Khodachenko2019,GM2019,Shaikhislamov2020,Wang2021}.

Due to its relatively low surface gravity and proximity to an F6V-type host star, the atmosphere of WASP-121b is likely to undergo rapid hydrodynamic escape enhanced by Roche lobe overflow, with a mass-loss rate that is among the highest for known exoplanets \citep{Koskinen2022}. The distance to the system ($\sim$270 pc) and the relatively high effective temperature of the host star do not favor the detection of the escaping atmosphere of WASP-121b at Lyman~$\alpha$ or through the He 1083 nm feature that is often used to probe the upper atmospheres \citep{Oklopcic2019}. However, the high spectral resolution and high signal-to-noise (S/N) detection of other atomic and ion lines in the optical and near-UV (NUV) spectrum of the planet means that WASP-121b offers some of the best observations of an escaping UHJ atmosphere. The interpretation of these observations, however, is complicated by the fact that they include several minor species and excited states that are not typically included in models of exoplanet atmospheric escape. New, more comprehensive models of exoplanet upper atmospheres are therefore required to unlock the full potential of these observations to characterize exoplanet atmospheres.

Using the Space Telescope Imaging Spectrograph (STIS) on board the Hubble Space Telescope (HST), \citet{Sing2019} detected strong absorption features of \ion{Mg}{2} and \ion{Fe}{2} in the NUV spectrum of WASP-121b, with transit depths corresponding to radii larger than the limb of the planetary Roche lobe (RL). Broad-band NUV observations of WASP-121b have also been obtained with the SWIFT telescope \citep{Salz2019}, where a tentative excess absorption signature was also seen. Using three archival transits acquired by the High Accuracy Radial velocity Planet Searcher (HARPS), atomic/ion absorption features of \ion{Fe}{1}, \ion{Mg}{1}, \ion{Ca}{1}, \ion{Na}{1}, \ion{Cr}{1}, \ion{Ni}{1}, \ion{V}{1}, \ion{Fe}{2} were detected using the cross-correlation method \citep{Hoeijmakers2020,Ben-Yami2020,Bourrier2020}, along with the absorption line profiles of H$\alpha$ and Na doublets \citep{Cabot2020}. Observations of two transits with UV-Visual Echelle Spectrograph (UVES) at the Very Large Telescope (VLT) confirmed many of the previous detections and further detected the features of \ion{K}{1}, \ion{Ca}{2}, \ion{Sc}{2} \citep{Gibson2020,Merritt2021}. Finally, \citet{Borsa2021} analyzed two transits observed with the Echelle SPectrograph for Rocky Exoplanets and Stable Spectroscopic Observations ESPRESSO/VLT and observed the absorption line profiles of Na doublets, H$\alpha$, H$\beta$, Ca II H\&K, Mg I, K, Li, and Mn. The signal of the second transit was collected from all four Unit Telescopes (UTs) of the VLT simultaneously (4-UT mode), yielding a 16 m-equivalent telescope.  Using the same ESPRESSO data, \citet{Silva2022} further detected \ion{Ba}{2}, Co, and \ion{Sr}{2}, and also pointed out that \ion{Ca}{2} H\&K absorption features are asymmetric with broader blue wings.

In this work, we present a new multi-species model of the upper atmosphere and escape for WASP-121b. Since escaping upper atmospheres do not exist independently of the lower and middle atmosphere, we use a photochemical model to simulate the composition of the atmosphere from 100 bar pressure to the lower boundary of the escape model at 1 $\mu$bar pressure. In order to compare the results with the observations, we use the combined model density and temperature profiles to simulate transit spectra. As shown by \citet{Koskinen2022}, proper accounting for the extent of the whole atmosphere is critical for calculating the mass-loss rate from planets such as WASP-121b, and is also important for correctly explaining the observations. Here, we focus mostly on the NUV spectrum of the planet where the \ion{Fe}{2} and \ion{Mg}{2} lines clearly probe the escaping plasma. However, we also compare our model results with the absorption lines in the optical observed with the 4-UT mode of the VLT due to the unparalleled S/N of these data and the constraints that they provide on the atomic hydrogen distribution in the atmosphere.

Many of the recent theoretical studies on the escape of hot Jupiter atmospheres simulate the escape of hydrogen and protons, in some cases also including He, and focus on interpreting a single observed feature, such as the H$\alpha$ line or the He 10830\AA~line \citep[e.g.][]{GM2019,Lampon209458,Wang2021a}. For example, \citet{Yan2021} constructed a model of hydrodynamic escape for WASP-121b that only included hydrogen and used the observed H$\alpha$ line alone to constrain the properties of the upper atmosphere. In addition, they did not attempt to include the lower and middle atmosphere in their model. Our model extends the capabilities of similar previous models in two key ways. First, we include multiple minor species and compare our results with observations of several absorption lines simultaneously. Second, we account for the extent of the lower and middle atmosphere that affects the mass-loss rate and the interpretation of the observations. In summary, our focus on metal line absorption with model density profiles that span the whole atmosphere extends our understanding of WASP-121b's upper atmosphere.

The models used in this work are computationally expensive. By necessity at this point, therefore, they are one-dimensional, which can limit their ability to fully interpret the observations. Nevertheless, they provide a valuable tool for identifying the key physics necessary to model UHJ upper atmospheres. They also demonstrate both the capability and limitations of 1D models in explaining observations of planets like WASP-121b. All of this provides guidance for future developments of multi-dimensional models of exoplanet thermospheres that include the relevant minor species with their excited states. This guidance comes in two forms. First, 1D models can include more of the relevant physics and thus the analysis of the results can be used to infer simplified, faster schemes for multi-dimensional models. Second, they point to features in the observations that require more complex models to interpret. In this way, our work provides motivation and focus for the development of 3D models of atmospheric escape that are particularly relevant to planets such as WASP-121b that undergo Roche lobe overflow.

This paper is organized as follows. First, we summarize the setup of the atmospheric model in Section~\ref{sec:model} and discuss a reference atmosphere model for WASP-121b in Section~\ref{sec:hydro}. Next, we explain how we simulate the transmission spectrum that we compare with the observations in Section~\ref{sec:transm-spectr}. Then we discuss our approach to simulating Roche lobe overflow in Section~\ref{sec:tide} and outline several factors that influence the results and present models that can in principle explain the observed features in Section~\ref{sec:compare}. Finally, we discuss the key physical processes and properties of the best-fit model and its implications for the long-term evolution of WASP-121b's atmosphere in Section~\ref{sec:best-fit}.

\section{The atmosphere model} \label{sec:model}
\noindent
Separating the atmosphere of \WA\ into two sections, we use a hydrodynamic model to simulate the upper atmosphere at $P<1~\mu$bar, and a photochemical hydrostatic model to simulate the middle and lower atmosphere at $P>1~\mu$bar.  The transmission spectrum can then be calculated based on the combination of the results from the two models.  In this section, we first discuss the planetary system parameters (Section~\ref{sec:params}) and the stellar spectra (Section~\ref{sec:spectrum}) used in the hydrodynamic model and the photochemical model.  An outline of the hydrodynamic atmospheric model is given in Section~\ref{sec:eqn}.  Then, we explain the new physical mechanisms added to the hydrodynamic model, including the ionization balance of metal species (Section~\ref{sec:ioniz-recomb}) and radiative cooling (Section~\ref{sec:radiative}).  The photochemical hydrostatic model is described in Section~\ref{sec:boundary}.  Finally, we describe the \Lya\ radiation transfer module that provides the excited state H population, as well as the associated heating effect, in Section~\ref{sec:Hn2}.

\subsection{System Parameters}
\label{sec:params}

In this work, we adopted the physical parameters of the planetary system given by \citet{Delrez2016} and summarized in Table~\ref{tab:param}.
It is worth noting that the planetary radius given in the abstract of \citet{Delrez2016} is the planetary mean radius after asphericity correction (see Section~\ref{sec:atm-tide}) instead of the transit radius.
In addition, \citet{Delrez2016} used the mean radius of Jupiter as the planetary radius unit, instead of the equatorial radius as suggested by the XXIXth IAU General Assembly \citep{ResolutionR3_2015}.
Using values provided in their Table 4, we list the system parameters of \WA\ using the units suggested by the IAU in Table~\ref{tab:param}.

\begin{table}
  \caption{Orbital and Physical Parameters of WASP-121 and WASP-121b \citep{Delrez2016}}
  \begin{tabular}{c c}
    \hline
    Item & Value \\
    \hline
    Stellar radius $R_* $ & $ 1.4572\pm 0.03\,R_\odot$ \\
    Stellar mass $M_* $ & $ 1.3521^{+0.080}_{-0.079}\,M_\odot$ \\
    Orbital period $P$ & 1.2749255 day\\
    Planetary mass $M_p $ & $ 1.1824^{+0.064}_{-0.062} \,M_J$ \\
    Planetary radius $R_p$ & $1.766\pm 0.044\,R_J$  \\
    Distance $D$ & $263.2 \pm 0.7$ pc \footnote{\citep{GaiaDR3}} \\
    Orbital separation $a$ & 0.02544$^{+0.00049}_{-0.00050}$ AU \\
    \hline
  \end{tabular}
  \label{tab:param}
\end{table}

\subsection{Stellar Spectrum}
\label{sec:spectrum}

In order to constrain the XUV spectrum of the host star, we use the integrated XUV flux at the planet $F_{\rm XUV}=1.6\times 10^6 ~\rm erg~cm^{-2}~s^{-1}$ from \citet{Salz2019}, which is based on the measured stellar rotation rate.  The spectral energy distribution (SED) at wavelengths shorter than 1700 \AA\ is generated based on the solar SED. The SED is rescaled by the integrated XUV flux and the stellar radius considering that WASP-121 is larger than the Sun.  The shape of the spectrum is consistent with observations of the star Procyon by the Extreme Ultraviolet Explorer (EUVE) but the flux is about 10 times higher than that of Procyon in the same wavelength range, which is expected given that Procyon is likely less active, while WASP-121 is a fast rotator and therefore likely more active \citep{Fossati2018}. The SED at longer wavelengths is generated by LLmodels \citep{Shulyak2004}, which has been specifically designed for simulating the spectra of F, A, and B type stars. The spectra between 1210 \AA\ and 1220 \AA\ have 0.5 \AA\ wavelength bin width, while the rest parts of the spectra have 5 \AA\ bin width.

\begin{figure}
  \begin{center}
    \includegraphics[width=0.5\textwidth]{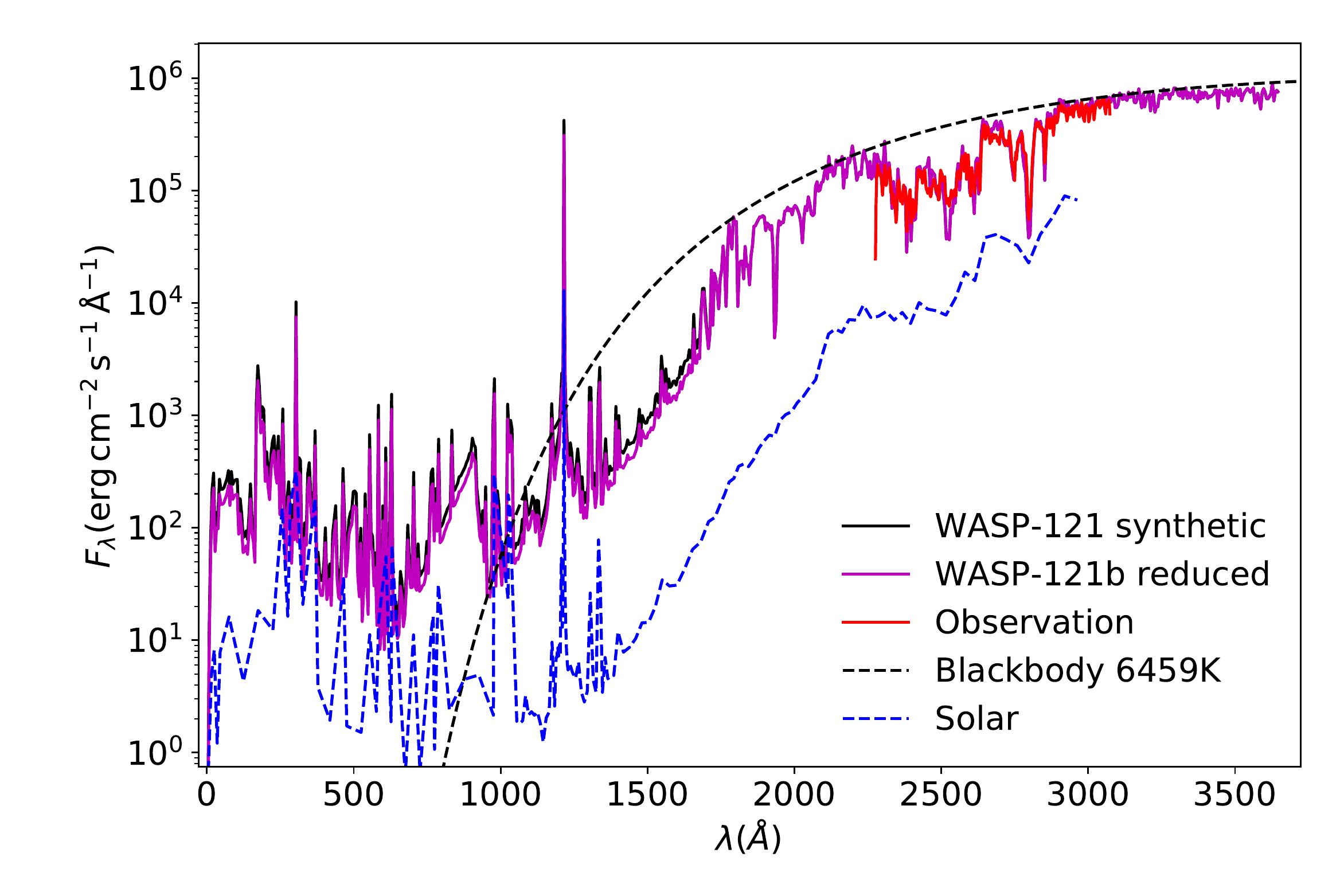}
    \caption{A comparison between the synthetic WASP-121 stellar spectrum at the orbital distance of WASP-121b used in the simulation (black solid line), a 6459 K black body (black dashed line), and the solar spectrum (blue dashed line).  The magenta line shows the spectrum with reduced high-energy intensity used in the best-fit model (see Section~\ref{sec:adjust}).  The observed out-of-transit NUV stellar spectrum is shown by the red line \citep{Sing2019}.  }
    \label{fig:Spec}
  \end{center}
\end{figure}

The synthetic spectrum of WASP-121 is shown by the black line in Figure~\ref{fig:Spec}.  The average solar spectrum from \citet{Koskinen2013a}, a 6459 K blackbody, and the observed out-of-transit NUV spectrum \citep{Sing2019} are shown for comparison.  We note that the Lyman continuum flux of the spectrum is $2.69\times 10^5 ~\rm erg\,cm^{-2}\,s^{-1}$.  We assume full redistribution of energy in the planetary atmosphere model, approximated by using a zenith angle of 60$^\circ$ for incident radiation, with a flux that is divided by a factor of two \citep{Smith1972}.

\begin{figure}
  \begin{center}
    \includegraphics[width=0.5\textwidth]{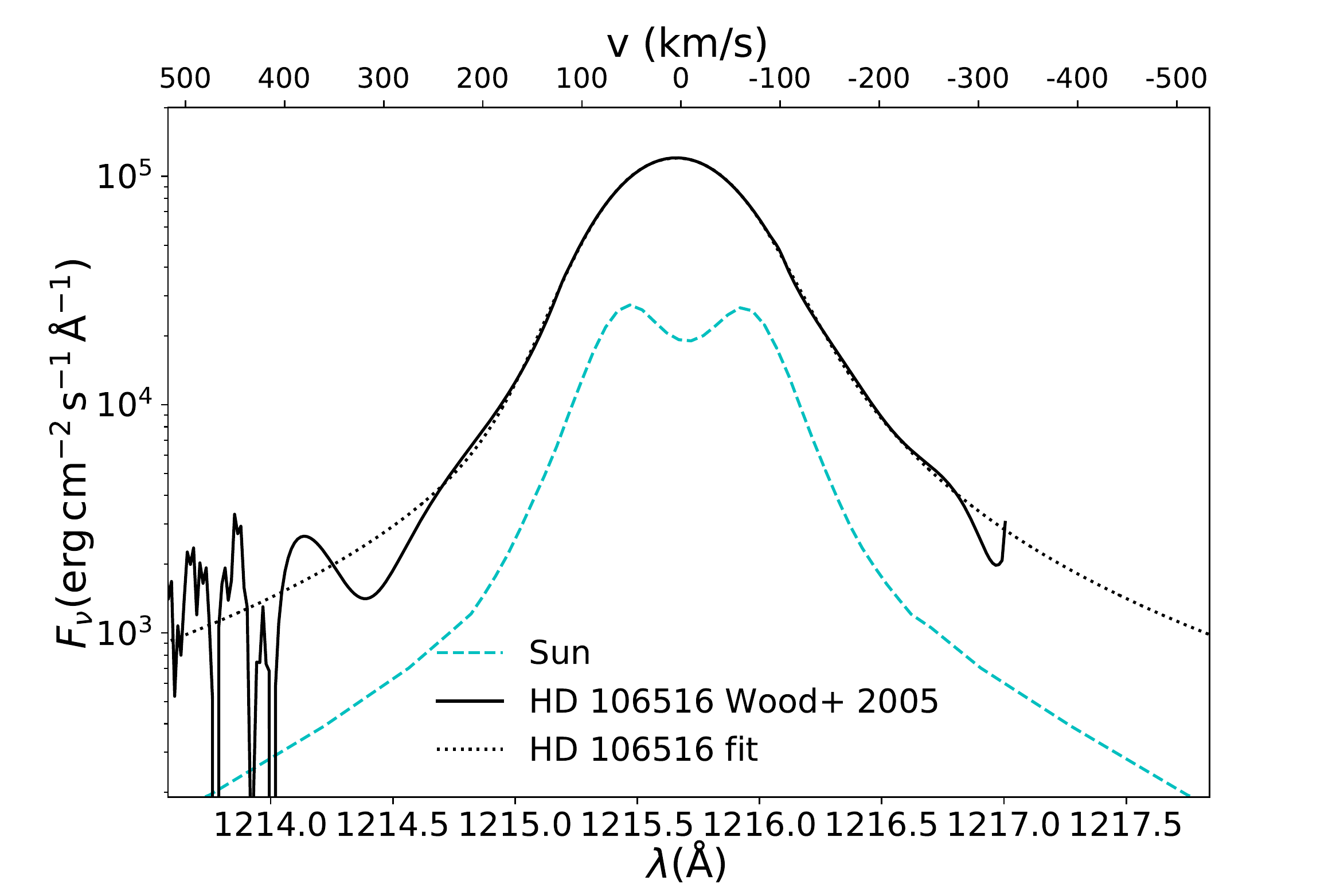}
    \caption{Stellar \Lya\ profile at the planetary orbit. Black solid line: Reconstructed HD 106516 \Lya\ flux \citep{Wood2005} scaled to the stellar radius of WASP-121.  Black dotted line: A double-Voigt profile fit that is used in this calculation.  Cyan dashed line: Moderate solar \Lya\ flux scaled by the radius of WASP-121. }
    \label{fig:lya_input}
  \end{center}
\end{figure}

The stellar \Lya\ emission line has a strong impact on the excited-state hydrogen population in the planet's atmosphere (see Section~\ref{sec:Hn2}).  The \Lya\ emission line of WASP-121 is not observable due to absorption by interstellar hydrogen along the long line of sight to the system from the Earth.  Instead, we construct the \Lya\ line based on the \Lya\ profile of a similar F5V star---HD~106516---provided by \citet{Wood2005}, assuming that they have the same surface \Lya\ intensity.  We fit the line profile with a double-Voigt profile, shown as the dotted line in Figure~\ref{fig:lya_input}.  The \Lya\ flux inferred in this manner at the orbit of \WA\ is $1.0\times 10^5 ~\rm erg\,cm^{-2}\,s^{-1}$.  The \Lya\ emission line for moderately active Sun \citep{Lemaire2005} is also shown for comparison.

To validate the stellar spectrum model described above, we compare the model integrated XUV fluxes in the wavelengths bands of 70 - 80~nm, 80 - 91.2~nm, and 91.2 - 117~nm with the fluxes based on the empirical XUV/\Lya\ flux equations suggested by \citet{Linsky2014}. Based on the empirical flux ratio, we can estimate that the stellar fluxes in the bands of 70 - 80~nm, 80 - 91.2~nm, and 91.2 - 117~nm at the distance of the planet should be $1.2\times 10^4$, $4.0\times 10^4$, and $1.3\times 10^4 ~\rm erg\,cm^{-2}\,s^{-1}$ respectively. For comparison, the fluxes of the synthetic spectrum used in the calculation at these three bands are $1.4\times 10^4$, $3.6\times 10^4$, and $6.3\times 10^4 ~\rm erg\,cm^{-2}\,s^{-1}$, respectively.  The excess emission in the 91.2 - 117.0~nm band of the synthetic spectrum is dominated by emission in the Ly$\beta$, \ion{C}{3} 97.7~nm, and \ion{O}{4} 103.2 and 103.8 nm lines.

\subsection{Hydrodynamic Atmospheric Escape Model}\label{sec:eqn}

We use a grid-based time-dependent one-dimensional multi-species hydrodynamic atmosphere model CETIMB \citep{Koskinen2013a,Koskinen2013b,Koskinen2022} to simulate the coupled thermosphere and ionosphere of WASP-121b, including hydrodynamic escape.  
The model includes photoionization, thermal ionization, recombination, charge exchange, advection and thermal escape of ions and neutrals, diffusion, viscous drag, heating by photoionization, adiabatic heating and cooling, conduction, viscous dissipation, recombination cooling, radiative cooling, and free-free cooling.  
The general details of the model are described in Appendix B by Koskinen et al. (2022). Here, we include a description of photoionization, thermal ionization, and related heating (Section~\ref{sec:photoionization}), charge exchange (Section~\ref{sec:ch-ex}), recombination and recombination cooling (Section~\ref{sec:recomb}), and radiative and free-free cooling (Section~\ref{sec:radiative}) that apply specifically to the simulations of WASP-121b. 

The model has 580 stretched grid cells in the radial direction, with a spacing $\Delta r_0 = 10$ km at the lower boundary and a stretch factor of 1.014, spanning from the radius at 1~\ubar\ up to 31.7 R$_J$ above the altitude of the bottom level.  The location of the upper boundary is always outside the sonic point and the atmosphere undergoes hydrodynamic escape. Therefore, the model uses the so-called outflow upper boundary conditions by extrapolating outflow velocity, species densities, and temperature from the model domain to the upper boundary point at a constant slope \citep{Tian2005}. Given that the upper boundary is above the sonic point, the solution at lower altitudes is not affected.

\subsection{Ionization, Recombination and Charge Exchange}
\label{sec:ioniz-recomb}

A total of 12 elements that are relatively abundant in a solar composition atmosphere or have spectral features that have been detected on WASP-121b are included in the model, including H, He, Mg, Fe, Si, O, C, N, S, Ca, Na, and K.  Solar abundances \citep{Asplund2009} are applied to these species by default. Second ionization states of Mg, Fe, Si, and Ca have been taken into account in the ionization/recombination balance.  For the other elements, only neutral atoms and first ionization states are included.  In the ionization level calculation, we consider photoionization, collisional ionization, and radiative and dielectronic recombination of all species listed, as well as charge exchange processes between these species whose rates are available.  In the following, the reaction rates of these processes used in the model are described in detail.

\subsubsection{Photoionization and Collisional Ionization}
\label{sec:photoionization}

The photoionization cross-section tables of atoms and ions are built upon the sum of the fitting formulae for their outer shell (ground state) electron photoionization cross-sections \citep{Verner1996}, and their inner shell electron cross-sections \citep{Verner1995}.  In addition, high-resolution tabulated photoionization cross-sections of \ion{H}{1}, \ion{He}{1}, \ion{C}{1}, \ion{N}{1}, \ion{O}{1}, \ion{Na}{1}, \ion{Mg}{1}, \ion{Mg}{2}, \ion{Si}{1}, \ion{Si}{2}, \ion{S}{1}, \ion{Ca}{1}, \ion{H}{1}(2s), and \ion{H}{1}(2p) from the Opacity Project \citep[TOP,][]{Cunto1993} are used to replace the fitting functions near the ionization threshold.  In some cases, the ionization threshold listed by TOP is inconsistent with the value given by \citet{Verner1995}. In these cases, we scale the photon energy from TOP to match the correct ionization threshold in Verner and Yakovlev (1995).
The cross-sections in TOP are set to zero for photon energies that are lower than the photoionization threshold in TOP.  For \ion{Fe}{1}, we replace the ground state photoionization cross-section near the ionization threshold with the more accurate calculation by \citet{Zatsarinny2019}.
Tabulated cross-sections only cover the photon energy near the ionization threshold and often provide a simple power-law trend at the high-energy end.
In these cases, cross-sections from fitting formulae are applied for energy higher than transitional energies that are chosen so that the two results can be smoothly connected.  Although the cross-section of \ion{Ca}{2} is also available in TOP, it is not adopted because we doubt the reliability of the power-law trend given by TOP which is also significantly inconsistent with the fitting formula.

\begin{figure}
\begin{center}
\includegraphics[width=0.5\textwidth]{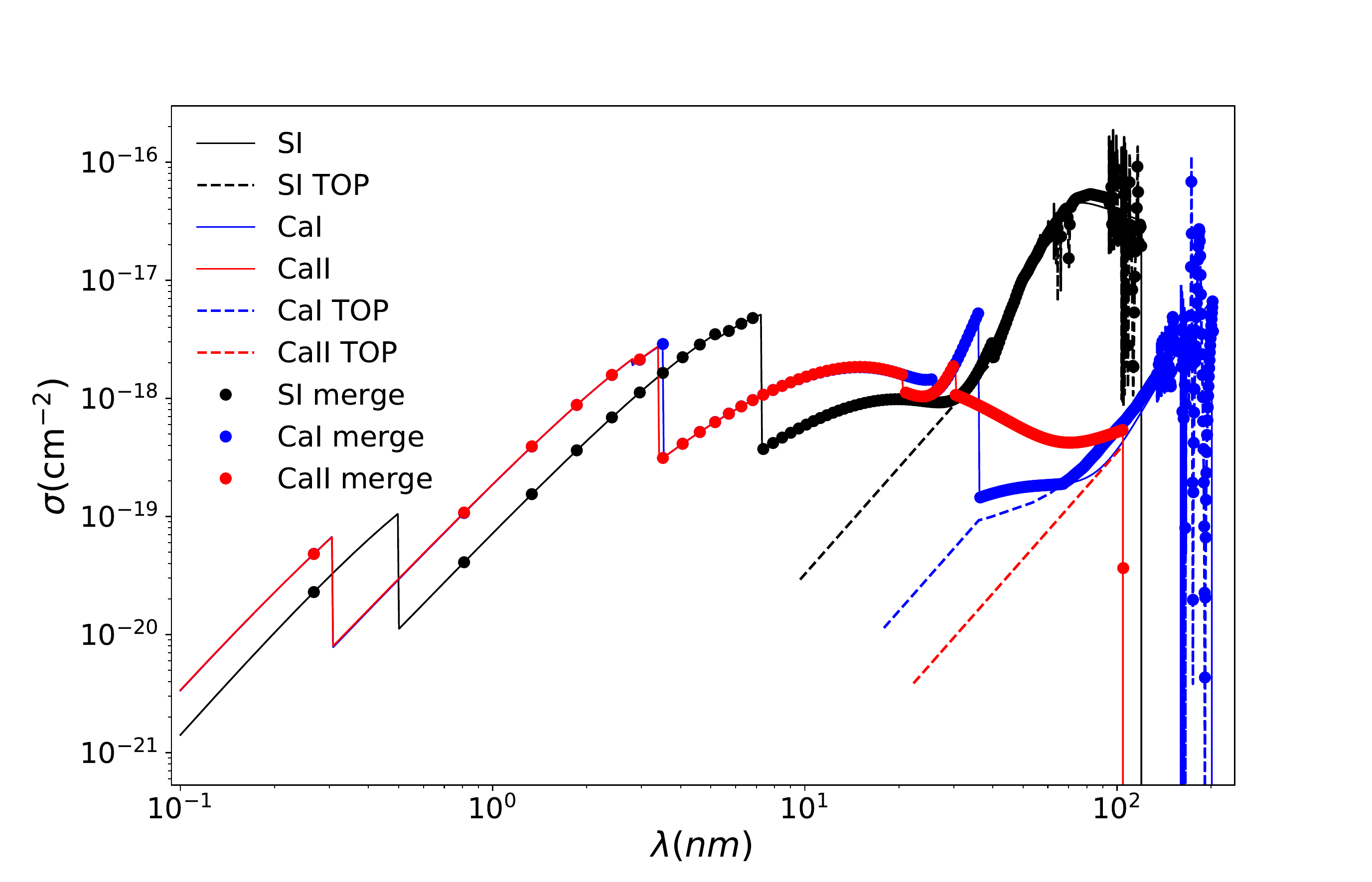}
\caption{Photoionization cross-section of \ion{S}{1}, \ion{Ca}{1}, and \ion{Ca}{2}.  Solid lines show the cross-sections as given by \citet{Verner1995} and \citet{Verner1996}. Dashed lines show the tabulated cross-sections given by TOP.  The cross-section of \ion{Ca}{2} from TOP is not adopted because it is described by a power law that does not agree with the formula from \citet{Verner1995}.  Averaged to the same wavelength bins as the input stellar spectrum, the dots show the cross-section value used by the model.}
\label{fig:pi}
\end{center}
\end{figure}

The high-resolution tabulated cross-sections are averaged to the same wavelength bins as the input stellar spectrum.  As an example, the merged and averaged cross-sections of \ion{S}{1}, \ion{Ca}{1}, and \ion{Ca}{2} used in our model are shown in Figure~\ref{fig:pi}.

If photoelectrons produced by high-energy photons have sufficient energy, they can ionize and excite atoms and ions by collisions.  The fraction of photoelectron energy that is converted to heat through collisions is the photoelectron heating efficiency.  In this work, we assume a constant photoelectron heating efficiency of 93\%, which is suggested for 50 eV photons at an electron mixing ratio of 0.1 \citep{Cecchi2009}.  A relatively high heating efficiency overall is expected in hot, strongly ionized atmospheres in which the mean free path for Coulomb collisions is relatively short for all but the highest photoelectron energies.

Electron collisional ionization rates from \citet{Voronov1997} are applied to all species that are included in the calculation.  The energy consumed by the collisional ionization of hydrogen is accounted for in the calculation of the cooling rate in the atmosphere.  The cooling rates resulting from collisional ionization of other, more minor atoms are assumed to be negligible.

\subsubsection{Charge Exchange}
\label{sec:ch-ex}

A total of 65 charge exchange reactions between different atoms and ions are included in the model. The relevant rate expressions and references are listed in Table~\ref{tab:charge_ex}.
Among these reactions, there are several processes for which reaction rates are only available in the exothermic direction.  For instance, \citet{Kingdon1996} only lists the rate of the forward reaction Mg + H$^+$, but not the rate of the reverse reaction Mg$^+$ + H.  The rates of the reverse reaction are likely to be small because they require a large kinetic energy input from the reactants, but for completeness we nevertheless estimate the reverse rates assuming microscopic balance.

According to microscopic balance, the ratio of the forward reaction to the reverse reaction rate is
\begin{equation}
  \label{eq:1}
  K(T) = \frac{k_f}{k_r} = \exp\left(-\frac{\Delta G}{RT}\right),
\end{equation}
where
\begin{equation}
  \label{eq:2}
  \Delta G = G_{\rm products} - G_{\rm reactants} = \Delta (H-ST)
\end{equation}
is the change in Gibbs free energy through the forward reaction, R is the gas constant, H is the enthalpy, and S is the entropy.  The enthalpy and entropy of atoms and ions are calculated using fits to the results obtained by using empirical equations given by \citet{gordon1999}. Here, the fitting function for $\ln(K) = \ln(a) + b\ln(T) - c/T$, where $a, b$, and $c$ are the fitted parameters.

\subsubsection{Recombination}
\label{sec:recomb}

We apply the case B recombination rate for H \citep[e.g,][]{MC2009}.  Because the electron number density varies slowly in the upper atmosphere (see Figure~\ref{fig:np_noT}), we ignore the weak dependence of the recombination rate coefficient on the electron density.  Fitting the values given by \citet{Storey1995} at $n_e=10^8~{\rm cm}^{-3}$, we have
\begin{equation}
  \label{eq:recomb_H}
\alpha_{B}=1.00\times 10^{-11}(T/300)^{-1.02}~{\rm cm^{3}\,s^{-1}}.
\end{equation}
For each hydrogen recombination, the average contribution to atmospheric cooling is \citep{Draine2011}
\begin{equation}
  \label{eq:8}
  \langle E_{rr}\rangle=[0.684-0.0416\ln(T_4)]k_{\rm B}T,
\end{equation}
where $T_4=T/10^4\ \rm K$.

For He, we use the recombination rate from \citet{Yelle2004}. For Mg, Si, S, and Ca, we use the recombination rates from \citet{Shull1992}. The fitting formulae from UMIST \citep{UMIST} are used for O, C, and N.  The recombination rates for Na and K are from \citet{Verner1996b} and \citet{Landini1991} respectively.

We fit functions to the tabulated radiative and dielectric recombination rates of \ion{Fe}{1} in \citet{Nahar1997b} and rates of \ion{Fe}{2} in \citet{Nahar1997a}, and obtain
\begin{equation}
  \label{eq:recomb_FeI}
  \begin{split}
    \alpha_{FeI}&=2.833\times 10^{-8} T^{-1.5} \exp\left(-\frac{5.731\times 10^4 \mathrm{K}}{T}\right) \\
    &\times \left(1  + 1.383\times 10^4\exp\left(-\frac{120.4\mathrm{K}}{T}\right) \right)  \\
  &+ 1.248\times 10^{-12} (T/10^4\mathrm{K})^{-0.485} \mathrm{~cm^3~s^{-1}},
  \end{split}
\end{equation}
and
\begin{equation}
  \label{eq:recomb_FeII}
  \begin{split}
    \alpha_{FeII}&=1.094\times 10^{-5} T^{-1.5} \exp\left(-\frac{1.490\times 10^4 \mathrm{K}}{T}\right) \\
    &\times \left(1  + 36.74\exp\left(-\frac{1.153\times 10^5\mathrm{K}}{T}\right) \right)  \\
    &+ 1.728\times 10^{-12} (T/10^4\mathrm{K})^{-0.618} \mathrm{~cm^3~s^{-1}}
  \end{split}
\end{equation}
for the recombination rates of \ion{Fe}{1} and \ion{Fe}{2}.

\subsection{Radiative Cooling of Mg, Na, Ca, Fe, and H}
\label{sec:radiative}

When an atom or ion de-excites radiatively from an excited state that is collisionally excited by, say, an electron or hydrogen atom, the kinetic energy of the colliding particle is converted to radiation, and this effectively cools the atmosphere. To estimate the radiative cooling rates, we consider an approximate two-level system for each atom or ion, consisting of the two energy levels associated with the relevant transitions.

Under the two-level approximation, the lower and upper level populations satisfy,
\begin{equation}
  \label{eq:7}
  \begin{split}
  n_l (&n_eC_{lu}^{(e)}+n_HC_{lu}^{(H)}+B_{lu}\bar{J}_{lu}) \\
  &= n_u(A_{ul}+n_eC_{ul}^{(e)}+n_HC_{ul}^{(H)}),
  \end{split}
\end{equation}
where $\bar{J}_{lu}$ is the line profile averaged mean intensity of the radiation field, $C_{ul}$ and $C_{lu}$ are collisional excitation and de-excitation rates, respectively, and the superscript indicates the type of collider particle, whether it is an electron or a hydrogen atom.
Then, the radiative cooling rate of the transition can be written as
\begin{equation}
  \label{eq:3}
  \begin{split}
    \Lambda(T,n_e,n_H) n_e n_l &= \Delta E (C_{lu}^{(e)} n_en_l+C_{lu}^{(H)} n_Hn_l\\
    &-C_{ul}^{(e)} n_en_u-C_{ul}^{(H)} n_Hn_u)\\
    &= n_e n_l\Delta E \frac{C_{ul}^{(e)} + (n_H/n_e)C_{ul}^{(H)}}{A_{ul}+n_eC_{ul}^{(e)}+n_HC_{ul}^{(H)}} \\
    &\times \left(\frac{g_u}{g_l}\exp\left(-\frac{\Delta E}{k_BT}\right) A_{ul} -B_{lu}\bar{J}_{lu}\right),    
  \end{split}
\end{equation}
where $\Delta E$ is the energy difference between two levels.

Adopting the same method, \citet{Huang2017} estimated the radiative cooling rates of Mg, Na, K, O, C, Si, and S in the upper atmosphere of the hot Jupiter HD~189733b.  The radiative excitation rates $B_{lu}\bar{J}_{lu}$ are ignored because permitted transitions in the optical and near-UV bands, which dominate the radiative cooling, are always associated with strong absorption features in the stellar spectrum.  The results show that Mg is the major coolant throughout the upper atmosphere, while cooling by Na can dominate at P$>$0.5 $\mu$bar where Na is less strongly ionized. Cooling by Na may have limited relevance, however, because on many Hot Jupiters, strong molecular coolants are likely to dominate in the middle atmosphere.

\begin{figure}
\begin{center}
\includegraphics[width=0.5\textwidth]{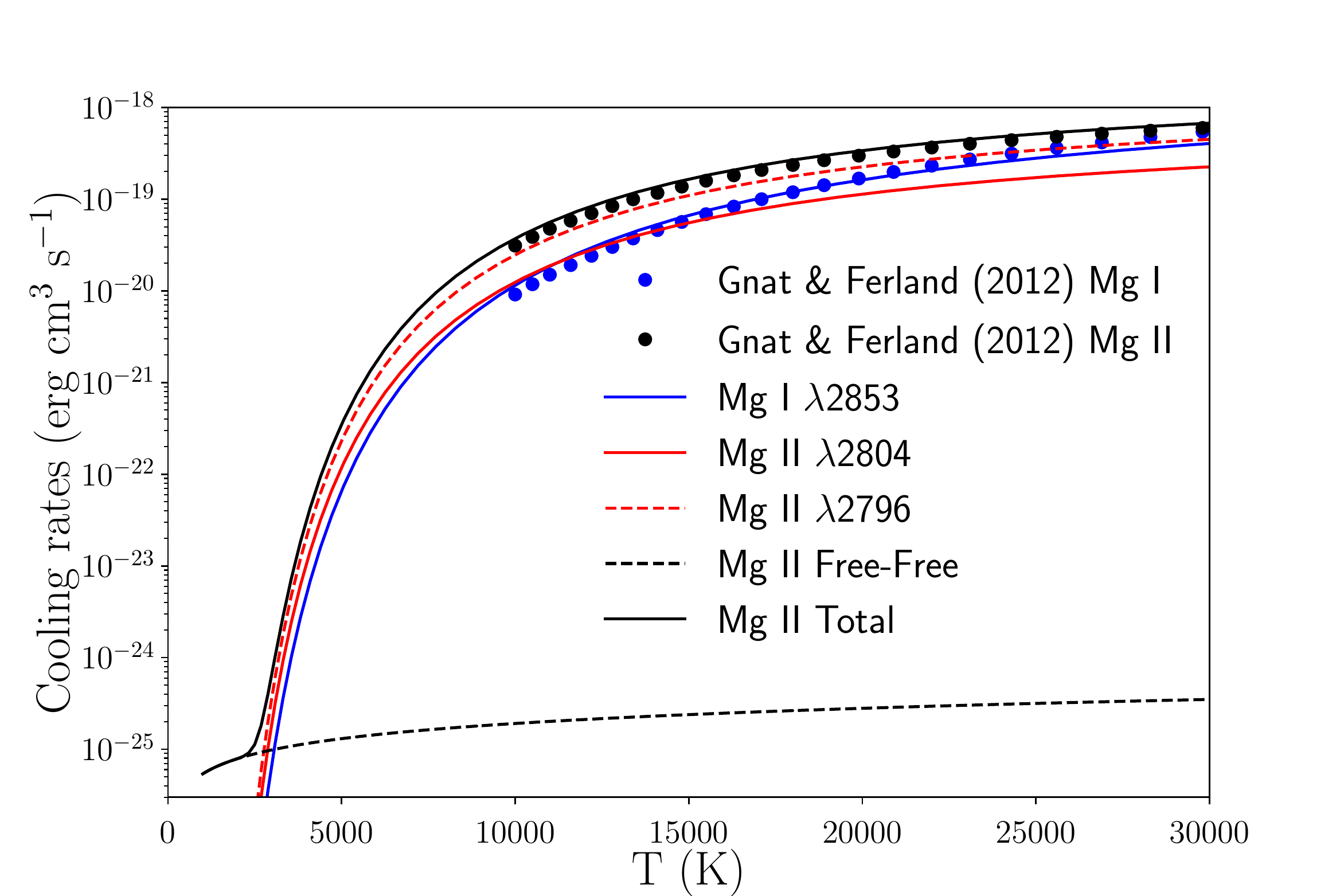}
\caption{Comparison of \ion{Mg}{1} and \ion{Mg}{2} cooling rates used in the hydrodynamic atmosphere model (blue and black solid line) with the tabulated value in \citet{Gnat2012}. }
\label{fig:Mg_cool}
\end{center}
\end{figure}

In this work, we estimate the \ion{Mg}{1}, \ion{Mg}{2}, and \ion{Na}{1} cooling rates in the same way as \citet{Huang2017}.  The line cooling of \ion{Mg}{1} $\lambda$2853 transition is included for \ion{Mg}{1}, and \ion{Mg}{2} $\lambda$2796 and $\lambda$2804 transitions are included for \ion{Mg}{2}.  For ion species, we also consider the free-free cooling rate using the formula \citep{Osterbrock2006,Draine2011}
\begin{equation}
  \label{eq:9}
  \Lambda_{ff}=1.9095\times10^{-25} Z^2T_4^{0.55}(\ergccs),
\end{equation}
where $Z$ is the charge number of the ion.
To validate the calculated cooling rate, we compare our \ion{Mg}{1} and \ion{Mg}{2} cooling rates with the rates estimated by \citet{Gnat2012} using Cloudy~\citep[version 10.00]{Ferland1996}, shown in Figure~\ref{fig:Mg_cool}.  \citet{Gnat2012}'s calculation, which also neglects radiative excitation, applies to a low-density gas ($n<1 ~\rm cm^{-3}$) with a temperature above 10,000K.  This shows that our simplified approach provides a good approximation of the most important metal line cooling rate in the upper atmosphere, which is due to Mg II.

Although radiative cooling by Ca and Fe is not considered by \citet{Huang2017} because of their low abundance in the upper atmosphere of HD~189733b, the detection of Ca and Fe absorption lines in the upper atmosphere of \WA\ indicates that their radiative cooling should be evaluated in more detail.  To estimate the cooling rate of \ion{Ca}{2}, we use the electron collisional rates of $\lambda$3934 and $\lambda$3968 transitions in the CHIANTI database \citep{Dere1997,Zanna2021} as well as the free-free cooling.

We also use the electron collisional rates from the CHIANTI database to estimate the \ion{Fe}{2} radiative cooling rate.  Of the 4339 \ion{Fe}{2} transitions whose collisional rates are provided in the database, 1105 transitions that also have Einstein A coefficient or non-zero oscillator strength are included in the model.  

Since the CHIANTI database does not include the collisional excitation rates of \ion{Fe}{1}, we calculate the electron collisional excitation rates of 1025 permitted transitions with a minimum Einstein A of $10^6$~s$^{-1}$ by using the van Regemorter formula \citep{Regemorter1962,Jefferies1968}.  The formula,
\begin{equation}
  \label{eq:4}
  C_{lu}^{(e)} = 2.16\alpha^{-1.68}e^{-\alpha}T^{-3/2}f ~\rm cm^{3}~s^{-1},  
\end{equation}
where $\alpha=\Delta E/kT$, expresses the collisional excitation rate in terms of the oscillator strength $f$, which are acquired from the NIST atomic spectra database \citep{NIST_ASD}. Although the van Regemorter formula does not apply to the collisional rates for dipole-forbidden transitions, their Einstein A are small so that the contribution of these transitions to cooling is generally negligible.  In addition, the electron and neutral hydrogen collisional excitation rates of fine-structure and forbidden lines at 1.44, 1.36, 14.2, 22.3, 24.0, and 34.2 $\mu$m given by \citet{Hollenbach1989} are included.

Besides the collisional excitation rates and spontaneous decay rates of each transition, the radiative cooling rate of a species also depends on the population of the lower levels that may of course differ from the ground state.  For \ion{Fe}{1} and \ion{Fe}{2}, we assume that the energy levels that do not connect to any lower states with electric dipole transitions follow the Boltzmann distribution. For example, the lowest energy level that has an electric dipole transition with the ground state of \ion{Fe}{2} is the 64th level ($3d^64p$ $^6\mathrm{D}^o$) at the energy of 4.8 eV.  Thus, we assume that the 64th level of \ion{Fe}{2} and all states above it are depleted, while states below follow the Boltzmann distribution.  Similarly, we assume that the 54th level of \ion{Fe}{1} ($3d^64s4p$ $^5\mathrm{D}^o$) at the energy of 3.2eV and all states above it are depleted.   The same level population treatment is applied to the transmission spectrum calculation described in Section~\ref{sec:Fei-abs}.

Using a more sophisticated level population model in Cloudy, \citet{Verner1999} calculated the departure of each \ion{Fe}{2} level from its LTE population in an iron dominant gas.  In the radiation-dominated case, such as the situation discussed in this work, they show that the 64th level and levels above it are depleted, while the lowest 63 levels are only slightly under-populated compared to LTE ($<$ 30\% for the 45th level).

\begin{figure}
\begin{center}
\includegraphics[width=0.5\textwidth]{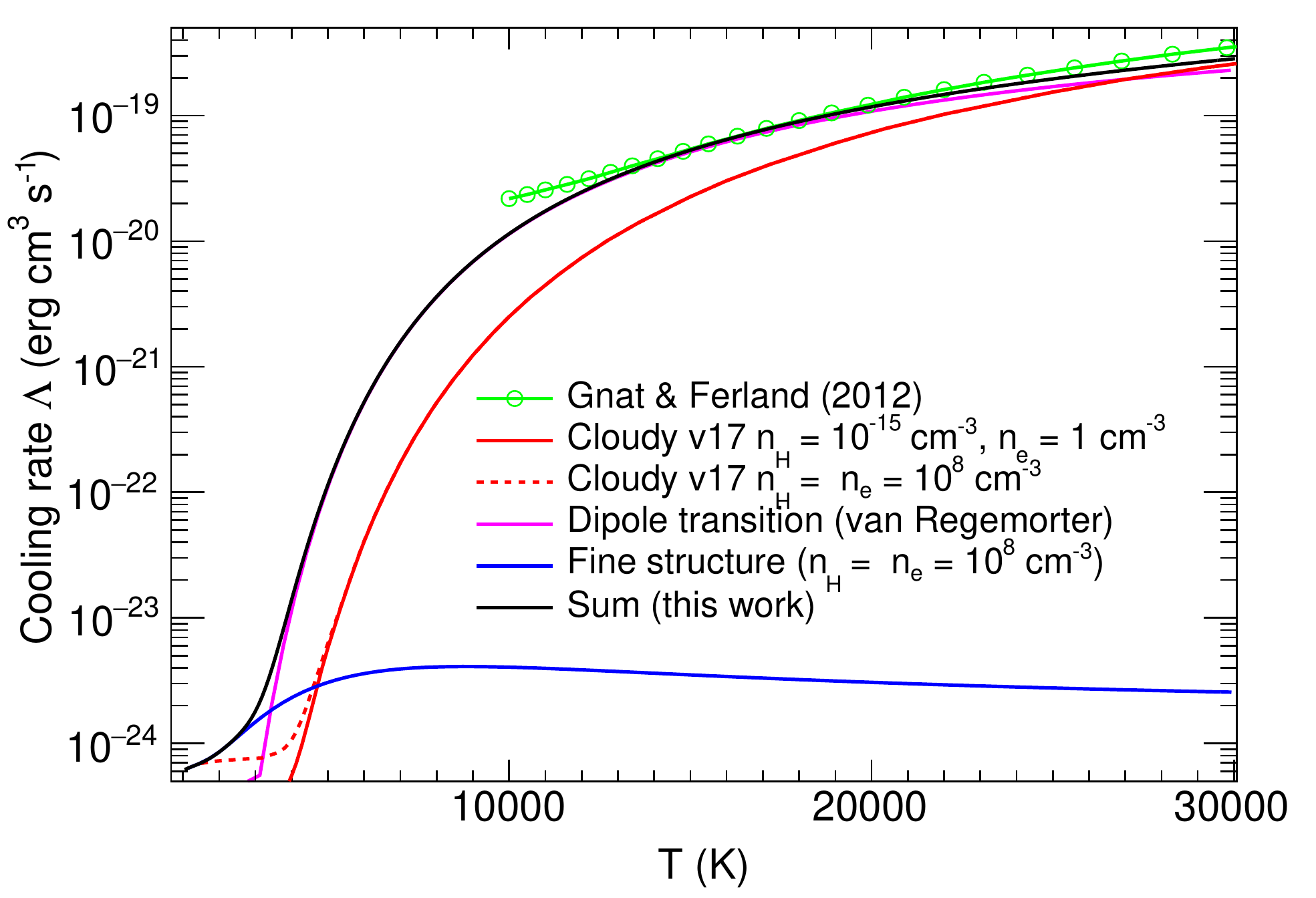}
\caption{Comparison of \ion{Fe}{1} cooling rate used in the hydrodynamic atmosphere model (black solid line) with the tabulated value in \citet{Gnat2012} (green line with circles), reproduced value using a newer version of Cloudy at the original $n_{\rm H}$ and $n_{\rm e}$ number densities (red solid line), and higher number densities (red dashed line). The contributions to cooling by dipole transitions (magenta) and fine-structure/forbidden lines (blue) are also shown separately.}
\label{fig:FeI}
\end{center}
\end{figure}

\begin{figure}
\begin{center}
\includegraphics[width=0.5\textwidth]{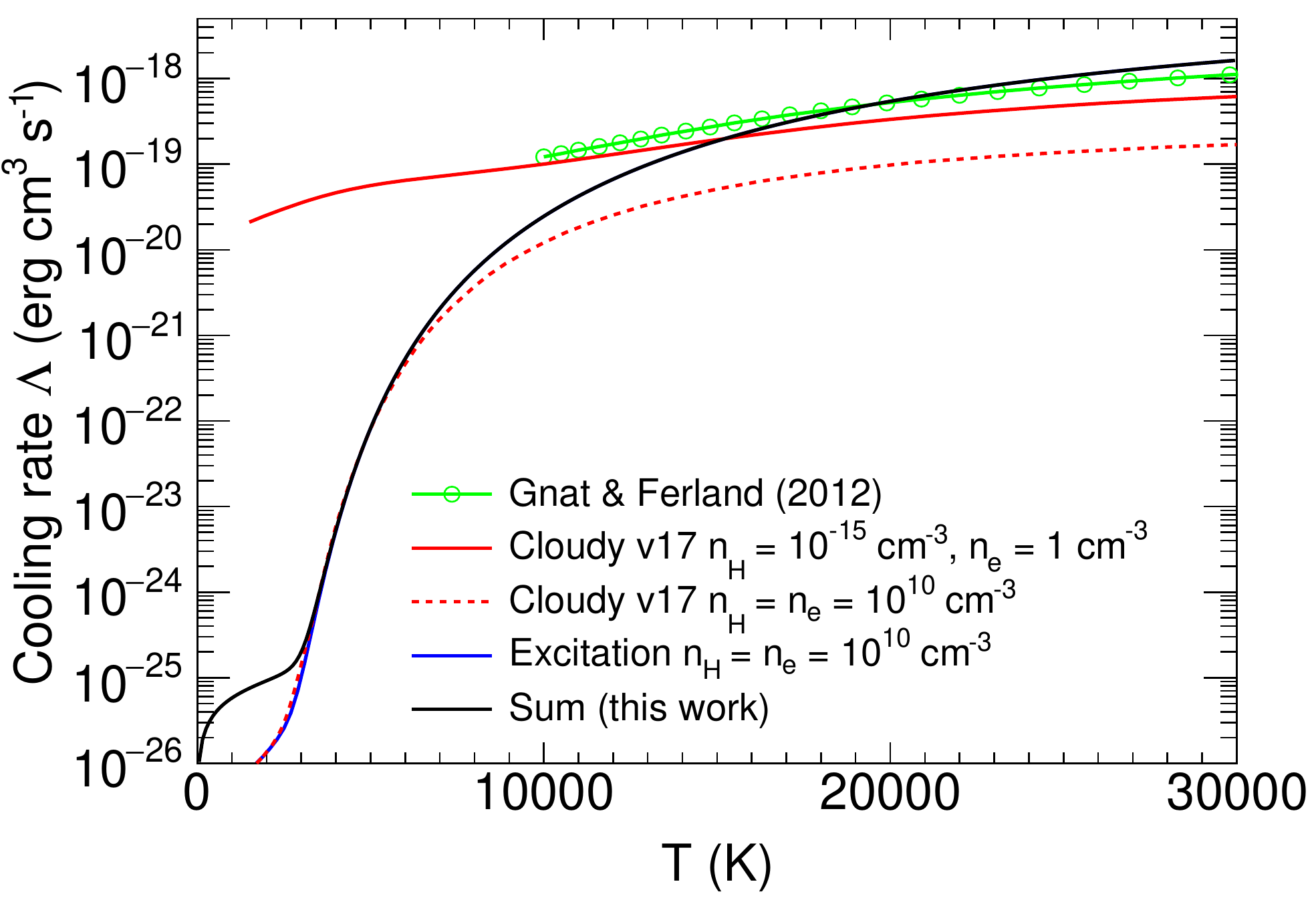}
\caption{Comparison of \ion{Fe}{2} cooling rate used in the hydrodynamic atmosphere model (black solid line) with the tabulated value in \citet{Gnat2012} (green line with circles), reproduced value using a newer version of Cloudy at original number densities (red solid line), and higher number densities (red dashed line). The contributions to cooling by dipole transitions (blue) are shown separately. The free-free cooling is important at $T<3000$ K.}
\label{fig:FeII}
\end{center}
\end{figure}

We likewise compare the cooling rates of \ion{Fe}{1} and \ion{Fe}{2} thus calculated with the results given by \citet{Gnat2012}, shown in Figure~\ref{fig:FeI} and \ref{fig:FeII}.  In the low number density environment ($n_e=n_{\rm Fe}=1~\rm cm^{-3}$, $n_{\rm H}=10^{-15}~\rm cm^{-3}$) that they discuss, the spontaneous decay time is much shorter than the collisional excitation-time and thus the gas is in the coronal limit, meaning that only the ground state is significantly populated.

To compare with our Fe cooling rate at a broader temperature range, we repeat \citet{Gnat2012}'s calculation using Cloudy (version 17.01) and extend the temperature sampling down to 1500 K.  Besides the low number densities used in \citet{Gnat2012}, we also calculate the cooling rates at higher number density environments that are closer to the planetary upper atmosphere for comparison.  Due to the restriction of the output information of Cloudy, to estimate the cooling rate, the number density of the studied ion species has to be much larger than the density of hydrogen.  Therefore, we apply $n_e=n_{\rm H}=10^8~\rm cm^{-3}$ and $n_{\rm Fe I}=10^{15}~\rm cm^{-3}$ for \ion{Fe}{1}, $n_e=n_{\rm H}=10^{10}~\rm cm^{-3}$ and $n_{\rm Fe II}=10^{15}~\rm cm^{-3}$ for \ion{Fe}{2}.

\begin{figure}
\begin{center}
\includegraphics[width=0.5\textwidth]{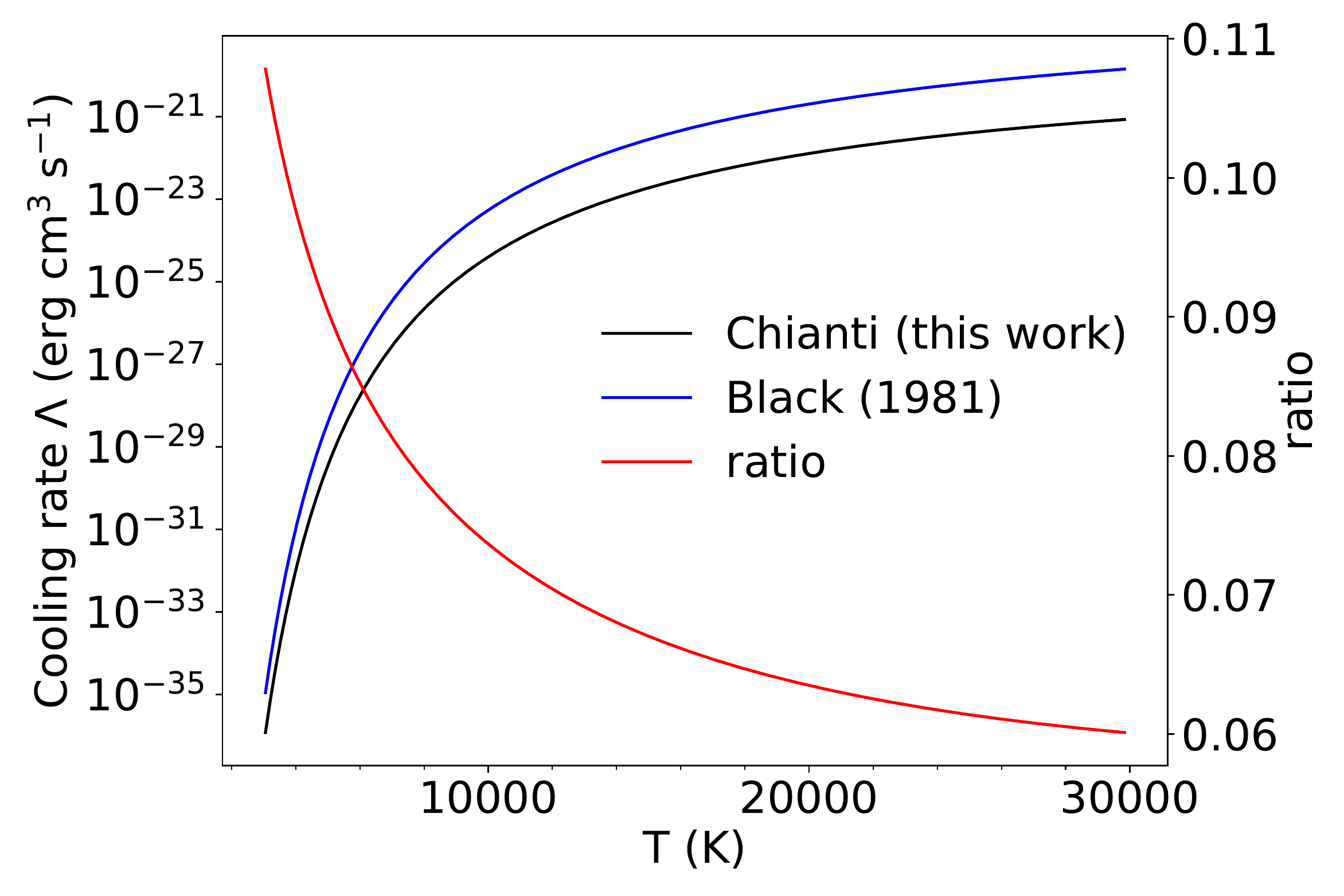}
\caption{Radiative cooling rate due to the emission of \Lya\ photons. The rate provided by \citet{Black1981} (blue) is compared to the cooling rate derived from the collisional excitation rate in the CHIANTI database (black).}
\label{fig:Hcool}
\end{center}
\end{figure}

Previous hydrodynamic atmosphere models have shown that radiative cooling due to \Lya\ emission from the planetary atmosphere is important \citep[e.g.][]{MC2009}.  
In contrast to the cooling rate given by \citet{Black1981} that has been used in many previous models, the \Lya\ cooling rate in this work uses the collisional excitation rates from the CHIANTI database, which are based on updated calculations by \citet{Anderson2002}.  
Figures~\ref{fig:Hcool} compares the cooling rates obtained by the two methods, where \citet{Black1981} gives a rate that is about 10 times higher compared to the new results here.

\subsection{Lower and Middle Atmosphere}
\label{sec:boundary}

To reduce complexity, our hydrodynamic model does not include molecules such as CH$_4$, CO, NH$_3$, or H$_2$O, and associated photochemistry.  Therefore, we set the bottom boundary of the hydrodynamic model at a pressure of 1~$\mu$bar, above which these molecules are completely dissociated.  A hydrostatic photochemical/thermochemical model, which includes diffusion, condensation, more than 100 species, and more than 1500 reactions \citep{Lavvas2014}, is used to model the atmosphere at pressures of $10^{-6}<P<100$ bar to provide the radius and mixing ratios of the atoms at the lower boundary of the hydrodynamic escape model. 

\begin{figure}
    \begin{center}
      \includegraphics[width=0.5\textwidth]{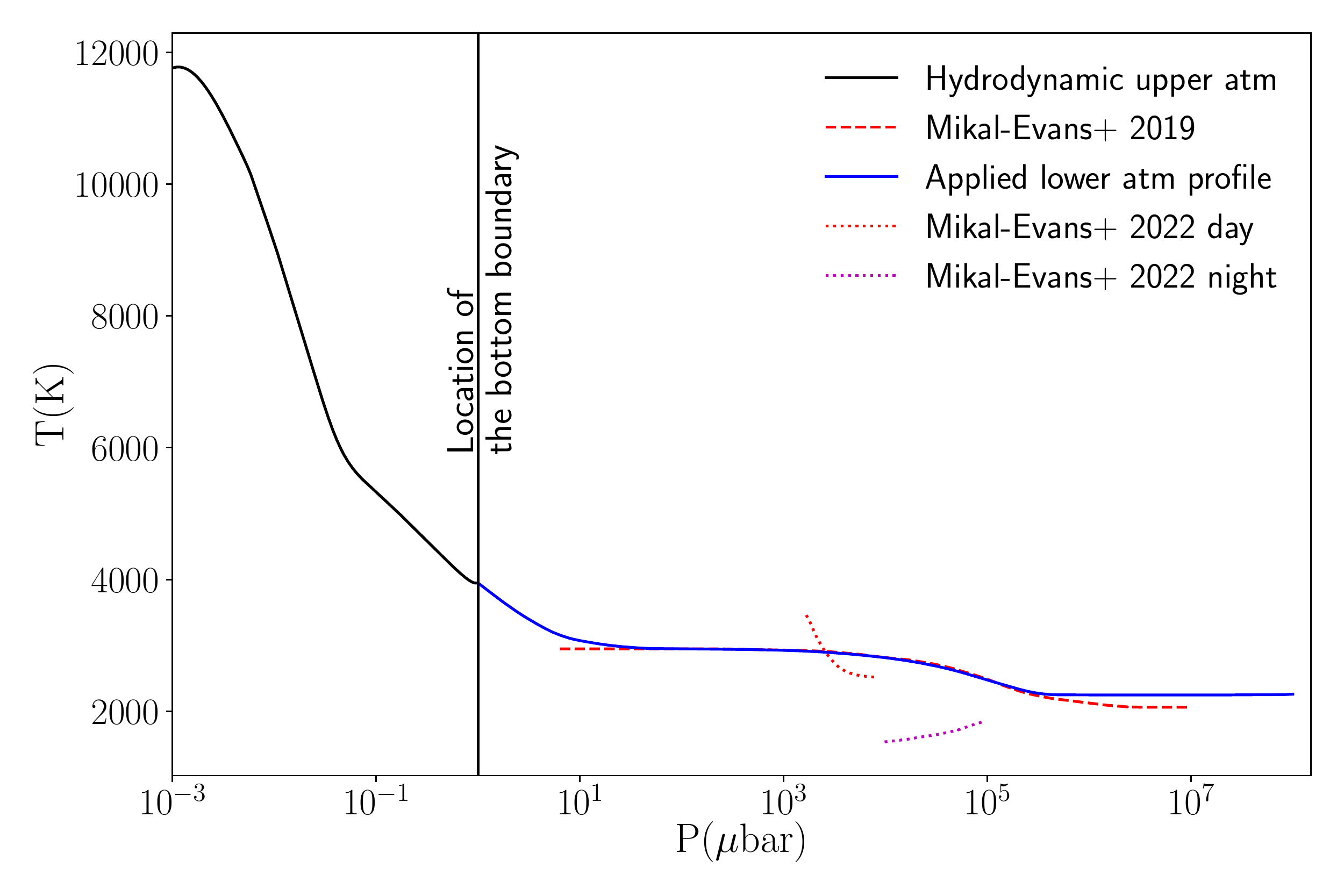}
      \caption{Temperature profiles applied in the lower and middle atmosphere photochemical model. The lower atmospheric temperature profile presented in \citet{Evans2019} as well as the dayside and nightside atmospheric temperature profiles presented in \citet{Mikal-Evans2022} are also shown for comparison.}
      \label{fig:Tcomb_notide}
    \end{center}
  \end{figure}

Shown by the solid blue line in Figure~\ref{fig:Tcomb_notide}, the temperature profile of the photochemical model is assigned as follows.  In the pressure range $10^{-4}<P<10$ bar, we adopt the Markov chain Monte Carlo (MCMC) median temperature profile of the secondary eclipse observation retrieval analysis given by \citet{Evans2019}, shown as the magenta dashed line in Figure~\ref{fig:Tcomb_notide}.  On the high pressure end, this retrieved temperature is slightly below the Fe condensation temperature, making the model yield a very low abundance of Fe that would not be detectable in the upper atmosphere.  To avoid Fe condensation, we increase the temperature at 10 bar by $\sim$200 K, which is comparable to the 1-$\sigma$ uncertainty in \citet{Evans2019}. A constant temperature is extrapolated from $P>$10~bar up to $P=100$~bar.  Between $10^{-6}<P<10^{-4}$ bar, the temperature is smoothly connected to the calculated upper atmosphere temperature profile at 1~$\mu$bar.

The applied temperature profile is in a similar temperature range as the dayside temperature profile presented in \citet{Mikal-Evans2022} derived from observations in longer wavelengths, which are also shown in Figure~\ref{fig:Tcomb_notide}.

We take the pressure at the visual broadband transit radius $R_p$ to be 4~mbar.  The transit radius in the optical continuum shown in Section~\ref{sec:continuum} supports this choice.  The radius, temperature, and elemental abundances in the photochemical model at the bottom boundary of the hydrodynamic upper atmosphere model are consistent with each other. The lower and middle atmosphere simulations also verify that all molecules are depleted at 1~$\mu$bar of this UHJ atmosphere as assumed. 
  
\subsection{\texorpdfstring{\Lya}{Lya} Radiative Transfer, Excited State H and Associated Heating and Ionization}
\label{sec:Hn2}

The strong \Ha\ absorption feature observed on WASP-121b calls for careful consideration of the distribution of excited state H.  As have been shown in previous works \citep{Huang2017,GM2019,Yan2021,Miroshnichenko2021,Yan2022}, the excitation of the H($n$=2) state is dominated by radiative excitation and the population of H(2$p$) is mostly determined by \Lya\ intensity through
\begin{equation}
  \label{eq:10}
  n_{2p}=n_{1s}\frac{g_{2p}}{g_{1s}}\frac{c^2}{2h\nu^3}\bar{J}_{Ly\alpha},
\end{equation}
where $\bar{J}_{Ly\alpha}=\int_0^\infty J_{Ly\alpha}(\nu) \phi(\nu) d\nu$ is the Voigt profile averaged mean intensity.

In this work, we use the \Lya\ Monte Carlo radiative transfer calculation developed in \citet{Huang2017} to calculate the \Lya\ mean intensity inside a plane-parallel atmosphere.
This calculation is computationally intensive and cannot realistically be coupled to our atmosphere model.  Instead, we calculate the level populations separately based on the model output and iterate until the results are self-consistent.

Unlike the complete frequency redistribution function used by \citet{GM2019} to approximate the probability distribution of scattered photon frequency and direction for a given incident photon, we apply the more accurate Hummer-IIB partial redistribution function \citep{Hummer1962}.  It accounts for the dipole angular dependence of resonant scattering when the initial and final states are the H(1$s$) state, and the intermediate H (2$p$) state has a finite lifetime and its fine-structure splitting is negligible.  The Stokes matrix for Rayleigh scattering is used to track the polarization of propagating photons, which further determines the angular distribution of outgoing photons at scattering. Recoil is included in computing the new frequency of the photon after scattering.

Unpolarized \Lya\ photons are ejected into the scattering calculation through three physical channels.  Stellar \Lya\ photons based on the line profile and flux given in Section~\ref{sec:spectrum} are incident vertically on the top boundary of the medium.  The stellar \Lya\ intensity is divided by a factor of 2, which accounts for the uniform redistribution of stellar \Lya\ photons across the day-side of the atmosphere.  
Two internal \Lya\ photon generation channels are also included, electron-impact excitation and recombination cascade.  We assume that every recombination of H (case B) produces one \Lya\ photon.

Besides exciting an H atom in the ground state, \Lya\ photons may photoionize Si, Mg, Fe, Ca, Na, and K, or photodissociate H$_2$, and thus end the scattering cycle. Although the absorption cross-section of H$_2$ is complex, \Lya\ spectrum within the atmosphere has a broad plateau near the line center within the velocity of $\sim \pm 70\kms$\citep{Huang2017}.  Averaging the cross-section of H$_2$ between 1215.4~\AA\ and 1216~\AA\ at 2500~K, we have $\sigma_{\rm H_2}=4.0\times 10^{-19} \rm cm^{2}$ \citep{Koskinen2021}.
Although \Lya\ can also photoionize excited-state H, we do not include this process in the calculation because of the following two reasons.  First, the number density of excited-state H is much lower compared to the metal species listed above, thus it is not an effective sink for \Lya\ photon.  In addition, because the \Lya\ intensity is negligible compared to the Balmer continuum, the effect of this process on the photoionization of excited hydrogen is also negligible.

Instead of resonant scattering, \Lya\ photons may also undergo complete redistribution if the excited H(2$p$) transits to H(2$s$) through collisional $\ell$-mixing, which is followed by the reverse process back to H(2$p$).
We fit the $\ell$-mixing rate provided by \citet{Seaton1955} for temperatures between 100 and 100,000 K using an analytical function and thus, the electron and proton collisional mixing rates are
\begin{equation}
  \label{eq:e2s2p}
  \begin{split}
    & C_{2s\to 2p}^{(e)} = 1.394\times 10^{-3}T^{-0.4055}\exp(-13.35/T) \\
    &+ 8.723\times 10^{-4}T^{-0.3803}\exp(-23.97/T) ~\rm cm^{3}~s^{-1} 
  \end{split}
\end{equation}
and
\begin{equation}
  \label{eq:p2s2p}
  \begin{split}
    & C_{2s\to 2p}^{(p)} = 3.706\times 10^{-3}T^{-0.2509}\exp(-188.6/T) \\
    &+ 3.654\times 10^{-4}T^{-0.07835}\exp(-1080/T) ~\rm cm^{3}~s^{-1},
  \end{split}
\end{equation}
respectively.
The scattering \Lya\ photons are considered destroyed if H(2$p$) is photoionized by Balmer continuum photons or de-excited by electron collisions instead of radiative decay.  If the collisional $\ell$-mixing to H(2$s$) is followed by photoionization, collisional de-excitation, or two-photon decay, the \Lya\ photons are also considered destroyed.

To reduce the computational cost of the Monte Carlo simulation, we smooth the $\bar{J}_{Ly\alpha}$ as a function of altitude within the atmosphere with B-spline.  The procedure removes the high-frequency statistical noise while preserving the overall trend of the simulated function.  For comparison, we note that \citet{Yan2022} undertook a 3-D Monte Carlo \Lya\ radiation transfer calculation with a spherically symmetric planetary atmosphere model.  In their study, the \Lya\ radiation transfer calculation was conducted with the LaRT code \citep{LaRT}, which applies the Hummer-IIB partial redistribution function.  The LaRT model does not account for the absorption of photons by metals or any other processes of H(2$p$) besides the emission of \Lya\ photons.  Their results suggest, in agreement with our assumptions, that the distribution of excited state H is approximately spherically symmetric.
Additionally, even when only considering the stellar \Lya\ source, where 3-D effects are most pronounced, the results from our plane-parallel simulation are consistent with the results of the 3-D radiative transfer model of \citet{Yan2022} in the substellar direction.
With the same stellar \Lya\ flux, the peak intensity produced by our plane-parallel model differs by only $\sim$10\% from the peak intensity along the substellar direction presented in Fig. 13 of their work.

Based on the Hummer-IIB partial redistribution function, if the incident photon is a line-wing photon whose frequency is far from the line center, the scattered photon is most likely to be a line-wing photon whose frequency is within one Doppler width of the incident photon.  In combination with the small scattering cross-section to line-wing photons, this allows the photons to travel large distances with a few scattering events \citep[see][chap. 15.7]{Draine2011}.  In comparison, the frequency of photons based on the complete redistribution function is always in the vicinity of the line core, making it much less likely that the photons escape.  As a result, the \Lya\ intensity calculated using the complete frequency redistribution function for the resonant scattering process is typically higher than that obtained using the Hummer-IIB partial redistribution function.

Solving the rate equilibrium, we obtain the H(2$s$) number density
\begin{equation}
\footnotesize
  \label{eq:11}
  \begin{split}
    &n_{2s} =\\
    &\frac{n_{2p}(n_eC_{2p\to 2s}^{(e)}+n_pC_{2p\to 2s}^{(p)})+n_{\rm H}n_eC_{1s\to 2s}+n_pn_e\alpha_{2s}}{A_{2s}+\Gamma_{2s}+n_eC_{2s\to 2p}^{(e)}+n_pC_{2s\to 2p}^{(p)}+n_eC_{2s\to 1s}},
  \end{split}
\end{equation}
where $C_{1s\to 2s}$ and $C_{2s\to 1s}$ are collisional excitation and de-excitation rates given by CHIANTI, $A_{2s}=8.26~\rm s^{-1}$ is the two-photon decay rate, $\Gamma_{2s}= 576 ~\rm s^{-1}$ is the photoionization rate of H(2$s$), and $\alpha_{2s}$ is the recombination rate to H(2$s$) as the function of $n_e$ and $T$.
Based on the tabulated rates in \citet{Storey2015} with the largest fitting parameter $\kappa$, which applies to a Maxwell-Boltzmann electron distribution, we use the following formula in the calculation:
\begin{equation}
  \label{eq:6}
  \begin{split}
     \log_{10} \left(\frac{\alpha_{2s}}{\rm cm^{3}\,s^{-1}}\right) &= \frac{0.697(\log_{10} n_e)^{2.675}}{T}\\
     & - 0.269 (\log_{10} T)^{1.428} -11.136.
  \end{split}
\end{equation}

Having calculated $n_{2s}$, and $n_{2p}$, we update their contribution to hydrogen ionization as well as heating.  Photoelectric heating from the H($n$=2) state by the Balmer continuum as well as collisional de-excitation are included in the heating rates.  As stated above, we iterate between the Monte Carlo model and the hydrodynamic model to reach self-consistency.

\section{The Reference Atmosphere Model}
\label{sec:hydro}

\begin{figure}
    \begin{center}
      \includegraphics[width=0.5\textwidth]{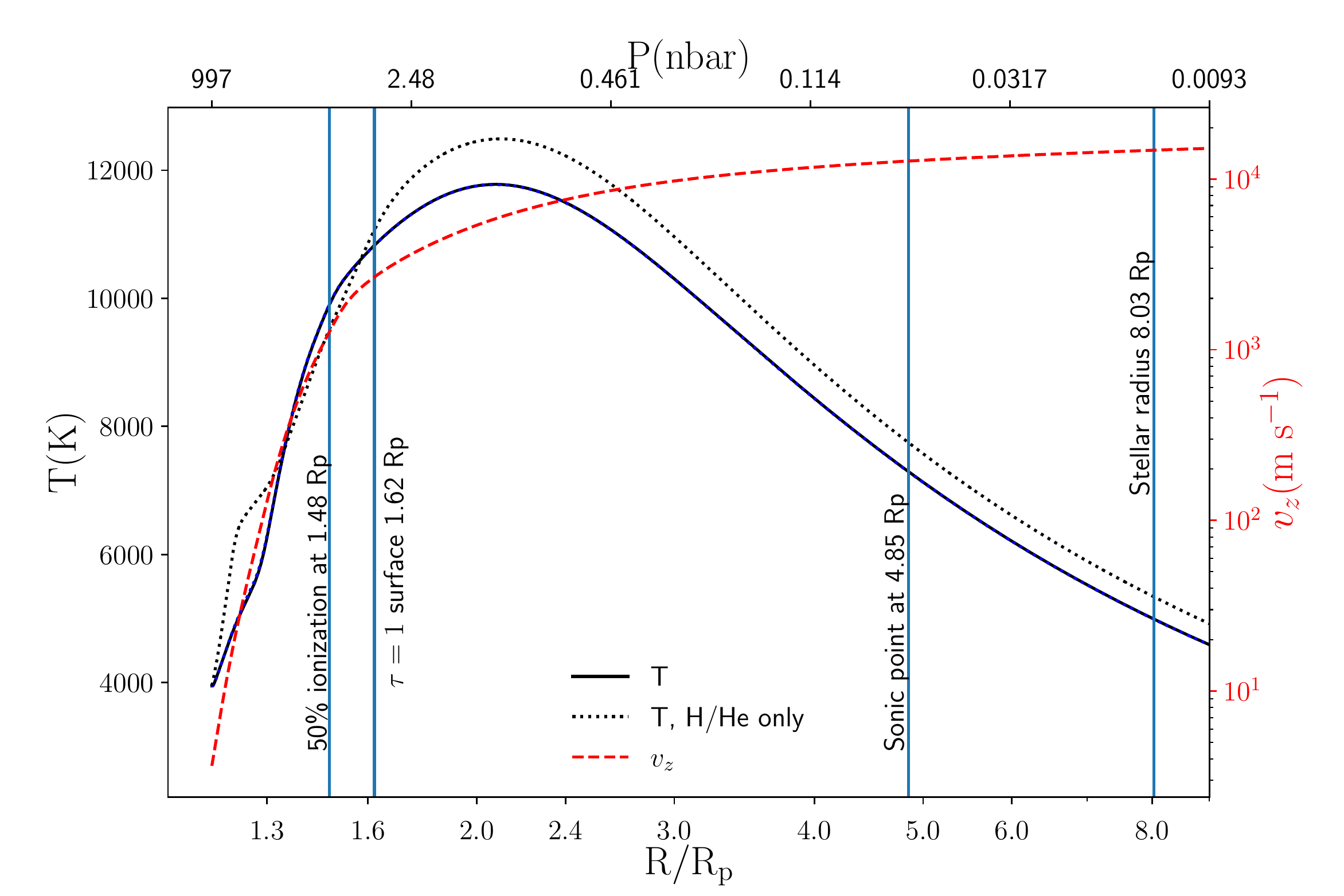}
      \caption{Temperature and radial velocity simulated by the hydrodynamic model for Case A after three iterations.  The blue dotted line that overlaps exactly with the black solid line shows the temperature after two iterations. The black dotted line shows the temperature of the case A' model that simulates an atmosphere with H and He only, excluding metals.}
      \label{fig:Tvr_noT}
    \end{center}
  \end{figure}

We first run the model under spherical symmetry with the observed system parameters of WASP-121b (see Table~\ref{tab:param}).  Hereafter, we will refer to this model as \textbf{Case A} and it represents the globally averaged atmosphere of WASP-121b under the strong irradiation experienced by the planet at its current orbital distance, without Roche lobe effects.
Figure~\ref{fig:Tvr_noT} shows the temperature and radial velocity profile predicted by the model.  The consistency between the temperature profile after two and three iterations demonstrates the efficiency of the radiative transfer and hydrodynamic simulations in achieving rapid convergence.  The locations where H atoms are 50\% ionized, where $\tau=1$ for stellar 13.6 eV radiation, where the outflow speed equals the local adiabatic sound speed, and the size of the star are marked with vertical lines.  In this case, the predicted mass-loss rate is $3.72\times 10^{12}~\rm g~s^{-1}$ or 0.052 $M_p$/Gyr.

\begin{figure}
  \begin{center}
    \includegraphics[width=0.5\textwidth]{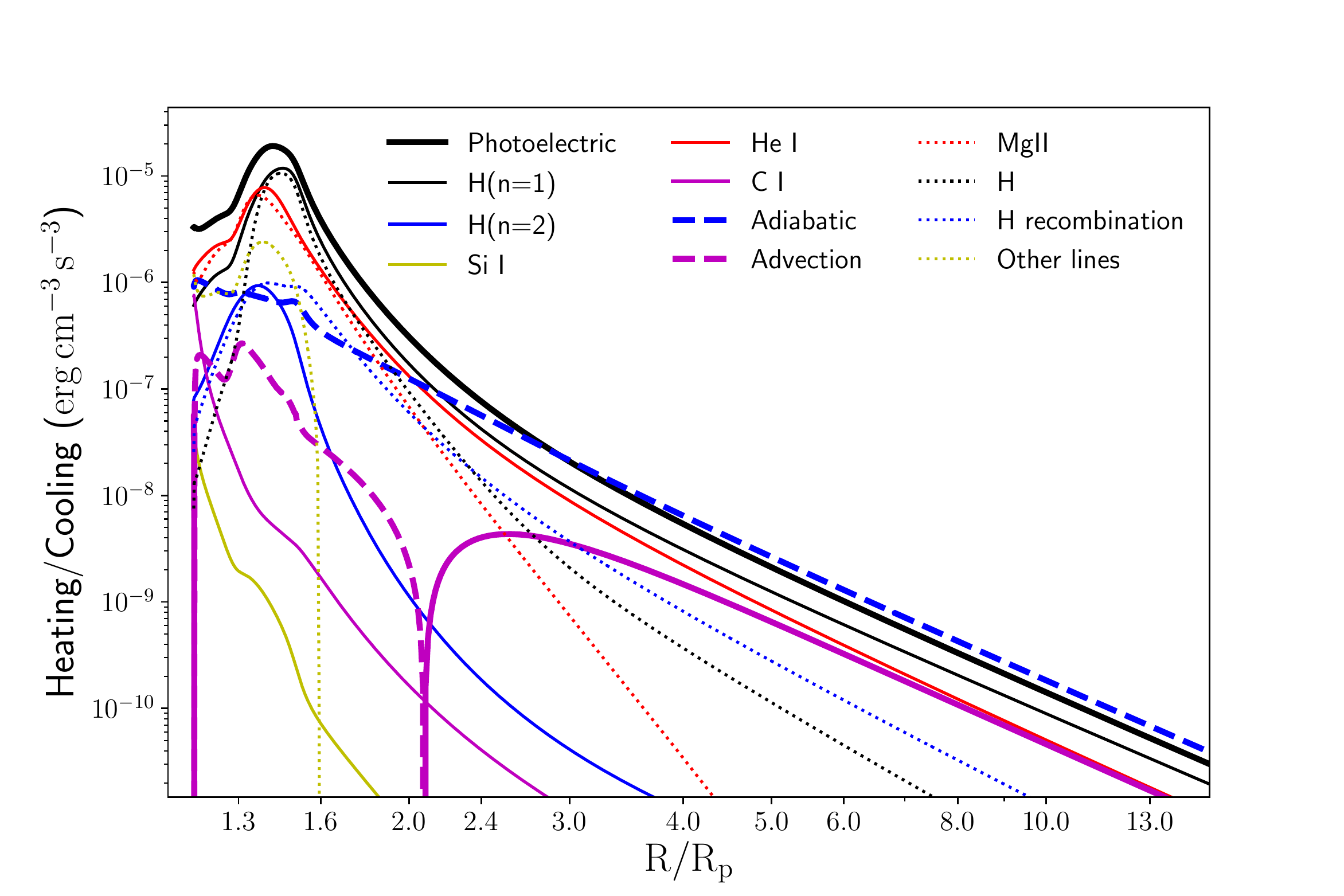}
    \caption{Rate of heating (solid line) and cooling (dashed or dotted line) process for case A. The thin solid lines in the figure represent the contributions of several major species to the total photoionization heating (thick black line).}
    \label{fig:Qr_noT}
  \end{center}
\end{figure}

Figure~\ref{fig:Qr_noT} shows the key heating and cooling rates.  The thick black line shows the total photoionization heating rate while the contributions to this heating rate by ground state H, excited H, He, Si, and C are shown by the thin solid lines.
Photoionization of ground state H and He is the primary source of heating.
Adiabatic cooling is the dominant cooling mechanism at high altitudes because of the relatively high mass-loss rate.
Radiative cooling dominates at radii below about 2.2~$R_p$, around the heating peak and below.

\begin{figure}
  \begin{center}
    \includegraphics[width=0.5\textwidth]{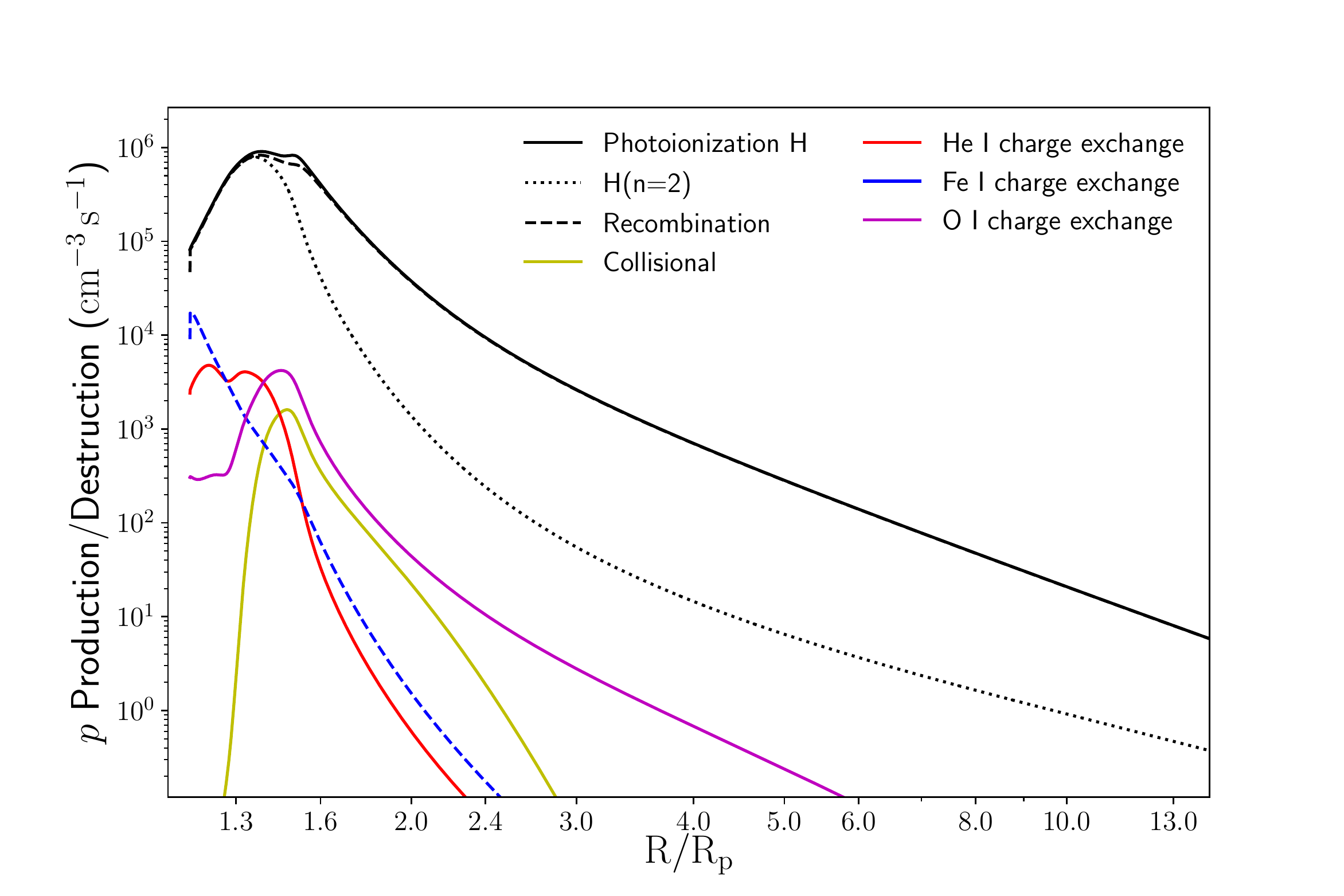}
    \caption{Rates of major proton production (solid) and destruction (dashed) for case A. As a component of H photoionization, photoionization of H($n=2$) is shown with the black dotted line. }
    \label{fig:ion_noT}
  \end{center}
\end{figure}

Figure~\ref{fig:ion_noT} shows the major proton production rates and destruction rates.  For a pair of charge exchange processes of H with other species (see Section~\ref{sec:ch-ex}), if the net effect is ionizing H, the net rates are shown with solid lines.  Otherwise, the net rates are shown with dashed lines.  In case A, the photoionization is well balanced by local recombination.  This is in contrast to the ionosphere of HD~209458b where advection from below is more efficient in replenishing H than recombination \citep{Koskinen2013a}.  The difference is mainly due to the much higher ionization fraction of \WA's atmosphere caused by significantly higher stellar XUV flux incident on \WA.

  \begin{figure}
    \begin{center}
      \includegraphics[width=0.5\textwidth]{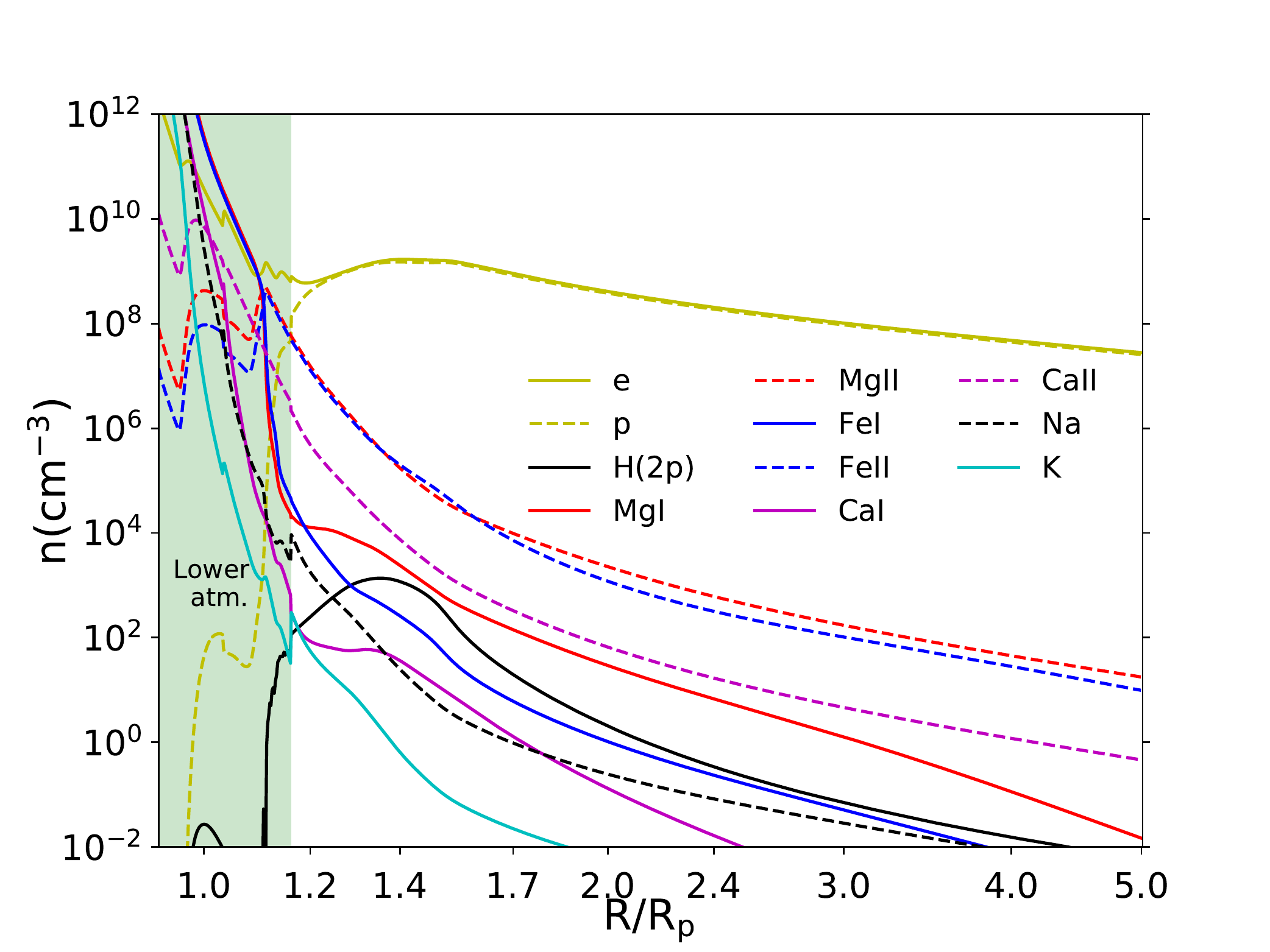}
      \caption{Number density distributions of major species in both the upper atmosphere and lower atmosphere of case A.  The region of the lower atmosphere calculated with the photochemical model is marked with green background.}
      \label{fig:np_noT}
    \end{center}
  \end{figure}

Figure~\ref{fig:np_noT} shows the number density distributions of major species that contribute to absorption features calculated by the hydrodynamic model for the upper atmosphere and the hydrostatic atmosphere model for the lower atmosphere.
Although the ionization balance is calculated independently in the two models, the number densities of most atoms and ions from the two models connect smoothly.  Metal species are mostly ionized throughout the upper atmosphere.

To analyze the impact of metal species on the model, we construct another H/He-only hydrodynamic upper atmospheric model based on the same lower/middle atmosphere setup and iteration procedure as in case A, which will be referred to as \textbf{case A'}.  
The temperature of case A' is shown as the black dotted line in Figure~\ref{fig:Tvr_noT}. 

Metal species can impact the model through the following four mechanisms: (1) The absence of radiative cooling by metals, mostly by \ion{Mg}{2} and \ion{Fe}{2}, that dominate cooling between 1.15 and 1.4 $R_p$ (see Figure~\ref{fig:Qr_noT}), should make the temperatures higher near the base of the model, (2) \ion{Si}{1}, \ion{Mg}{1}, \ion{Fe}{1}, and \ion{Ca}{1} can absorb \Lya\ photons, leading to a reduction in the \Lya\ mean intensity and ionization fraction, (3) heating by photoionization of \ion{Si}{1} and \ion{C}{1} can heat the atmosphere near the transition from the molecule-dominated to the atom-dominated layer, and (4) charge exchange between \ion{Fe}{1} and H$^+$ may reduce the H ionization fraction. On WASP-121b, the case A' temperatures are warmer than the reference model temperatures near the base of the model and at higher altitudes. However, the presence of metals does not lead to an appreciable reduction in the Ly$\alpha$ intensity because the metals are mostly ionized throughout the model. For the same reason, heating by metal photoionization is not important either. Also, the charge exchange rate between Fe I and H$^+$ is lower than the H photoionization and recombination rates. We expect that mechanisms (2)-(4) can be more important on planets orbiting stars with lower XUV flux and mass-loss rates but they do not play a significant role on WASP-121b.

\section{Transmission Spectrum}
\label{sec:transm-spectr}

To calculate the transmission spectrum, we connect the upper atmosphere structure simulated by the hydrodynamic model and the lower atmosphere structure simulated with the photochemical model as shown in Figure~\ref{fig:np_noT}. 
The outflow velocity of the lower and middle atmosphere photochemical model is set to conserve the overall mass-loss rate, given by the hydrodynamic model.
Thus, we have radial distributions of temperature, outflow velocity, and number densities of the atmosphere at a total of 781 grid points spanning from 100~bar, which radius is referred to as $R_{core}$ in the following, up to $\sim10^{-6}\mu$bar at $\sim$19 $R_p$.  We place this simulated planetary atmosphere at the center of the stellar disk to calculate absorption by the atmosphere and the transit depths.

\subsection{Transit Continuum}
\label{sec:continuum}
\begin{figure}
    \begin{center}
      \includegraphics[width=0.5\textwidth]{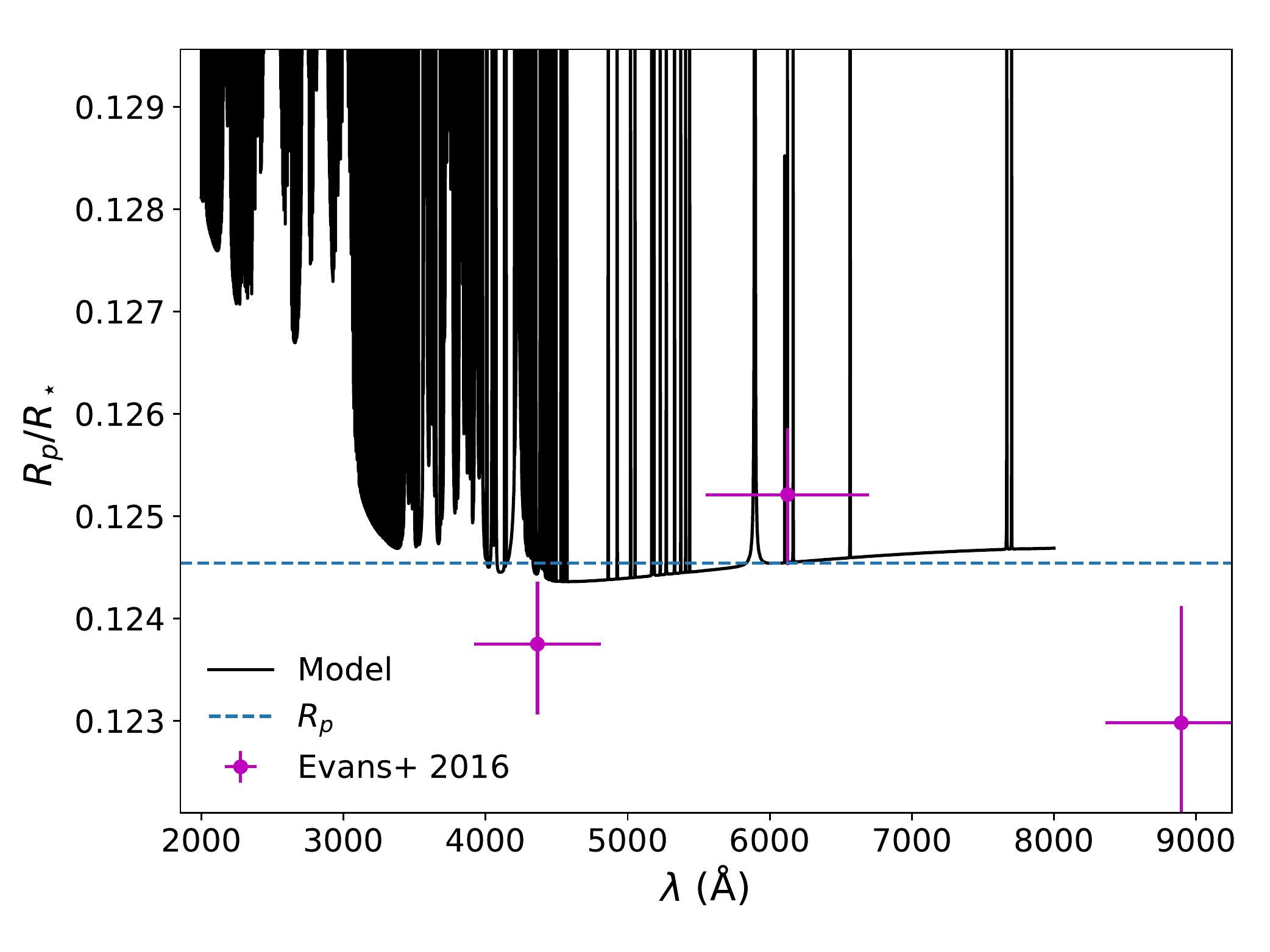}
      \caption{Continuum of the simulated transmission spectrum. The horizontal dashed line represents the value of $R_p/R_*$ reported by \citet{Delrez2016}.  The broadband radius ratios derived from the same observational data by \citet{Evans2016} are also depicted.}
      \label{fig:continuum}
    \end{center}
  \end{figure}

For the transit continuum that underlies the individual absorption lines, we include extinction by H$^-$ \citep{John1988} and Rayleigh scattering by H \citep{Lee2004}, He \citep{Fisak2017}, and H$_2$ \citep{Dalgarno1965}.  Figure~\ref{fig:continuum} shows a section of the simulated continuum profile, which is in agreement with the measured planet radius marked by the horizontal dashed line, or broad-band optical transit depths indicated by the markers within the error bar.  This indicates that our choice of $P=4$ mbar at $R_p$ is appropriate. Rayleigh scattering and extinction by H$^-$ are responsible for the continuum slope of the blue end and the red end, respectively. We note that our model does not include possible molecular absorbers, as our primary focus is on the interpretation of the upper atmosphere observations (see below).

\subsection{Absorption Lines}
\label{sec:lines}

For the line profile of each transition in the reference frame comoving with the bulk outflow of the atmosphere, we apply a Voigt profile with thermal broadening based on the local temperature and natural broadening based on each transition. We ignore collisional broadening since our focus is in the planetary upper atmosphere where it is negligible. We note that pressure broadening is obviously important for strong metal lines, such as those of Na and K, probing the middle and lower atmosphere.  Since transit depths are not cumulative with altitude, however, we can still accurately represent the profile of the line core even if our middle atmosphere transit depths are not fully accurate as long as the transit is driven by the upper atmosphere.

In calculating the transmission spectrum, we include the H Balmer, \ion{Mg}{1}, \ion{Mg}{2}, \ion{Ca}{1}, \ion{Ca}{2}, \ion{Fe}{1}, \ion{Fe}{2}, \ion{Na}{1}, and \ion{K}{1} lines.  Here, it is simple to estimate the features of hydrogen-like metal atoms/ions because only a few electric dipole transitions from the ground state need to be considered.  The included transitions for these species are \ion{Mg}{2} $\lambda$2796, $\lambda$2804, \ion{Ca}{2} $\lambda$3935, $\lambda$3970, \ion{Na}{1} $\lambda$3303, $\lambda$3304,$\lambda$5892, $\lambda$5898, and \ion{K}{1}$\lambda$4045, $\lambda$4048, $\lambda$7667, $\lambda$7701.  We apply vacuum wavelengths throughout the simulation.  The methods used to calculate the absorption features of other species in the model are discussed below.  

\subsubsection{Balmer Lines}
\label{sec:balmer-abs}

The Balmer line (\Ha, \Hb, and H$\gamma$) absorption profiles are calculated based on the H($2s$) and H($2p$) distributions (see black solid line in Figure~\ref{fig:np_noT}).  The procedure to estimate these distributions in the hydrodynamic atmosphere model is described in Section~\ref{sec:Hn2}.
Due to the high electron number density and low \Lya\ intensity in the lower atmosphere, the electron collisional processes may be more effective in exciting H atoms compared to radiative processes and thus H atom level population may follow Boltzmann distribution.  Therefore, in the lower atmosphere  where P$>1\mu$bar, LTE H($n=2$) number densities are applied if they are higher than the number density based on the \Lya\ intensity, obtained using equation~\ref{eq:10}.
Because the lower atmosphere is relatively thin with a low number density of excited H, the contribution of the lower atmosphere to the absorption by the Balmer lines on \WA\ is negligible.  

\subsubsection{\texorpdfstring{\ion{Mg}{1}}{Mg I}}
\label{sec:mgi-abs}

In UHJ upper atmospheres, the electron collisional rates of almost all transitions are much smaller than radiative rates.
The first excited state of \ion{Mg}{1} at 2.71 eV ($3s3p$ $^3\mathrm{P}^o$) cannot spontaneously decay to the ground state through an electric dipole transition.
We assume that the $^3\mathrm{P}^o$ state and the ground state $^1\mathrm{S}$ can be described as a two-level system, using equation~(\ref{eq:7}) with the radiative and collisional transitional rates between these two levels.
An electron collisional de-excitation rate of $3\times 10^{-8} \mathrm{~cm^3~s^{-1}}$ is applied for the temperature range of the model \citep{Osorio2015}.
The hydrogen atom collisional rate is always small for this transition \citep{Barklem2012}, and is thus ignored.

The second excited state of \ion{Mg}{1} at 4.34 eV ($3s3p$ $^1\mathrm{P}^o$) can decay to the ground state through a permitted transition with a rate of $4.9\times 10^{8}~\rm s^{-1}$, which is significantly larger than any collisional rates.
Therefore, this excited state can only be populated through radiative excitation.
However, because it is in a deep absorption line in the stellar spectrum, the number density of \ion{Mg}{1} in this excited state can be neglected. For a similar reason, we assume all higher energy levels of \ion{Mg}{1} are depleted in our model.

From the NIST atomic spectra database, we retrieve a total of 31 \ion{Mg}{1} transitions with the lower energy level being the ground state or $^3\mathrm{P}^o$ and the radiative decay rate greater than $10^6~\rm s^{-1}$.
  
\subsubsection{\texorpdfstring{\ion{Ca}{1}}{Ca I}}
\label{sec:cai-abs}

Another potentially abundant species that has an intricate radiative network is \ion{Ca}{1}.  Based on the NIST atomic spectra database, we supplement and update the list of radiative transitions of \ion{Ca}{1} according to the recent calculation results of \citet{Yu2018}.  
Similar to \ion{Mg}{1}, we assume that the \ion{Ca}{1} $3s4p$ $^3\mathrm{P}^o$ state at 1.88 eV and the ground state $^1\mathrm{S}$ can be described as a two-level system, with electron collisional de-excitation rate of $8\times 10^{-8} \mathrm{~cm^3~s^{-1}}$ \citep{Osorio2019}, hydrogen collisional de-excitation rate of $3\times 10^{-10} \mathrm{~cm^3~s^{-1}}$ \citep{Barklem2016} and $A=2.74\times 10^3~\rm s^{-1}$.  In the lower atmosphere, the hydrogen atom collisional de-excitation rate of this transition can be larger than the spontaneous radiative transition which is rarely seen in models.  We assume that other \ion{Ca}{1} triplet states are depleted because they can spontaneously decay to a lower level.  Among the singlets, the first excited state $4s3d$ $^1$D$_2$ is a meta-stable state since it cannot decay to the ground state by emitting a single photon.  However, it can decay to the $^3\mathrm{P}^o$ state through a forbidden transition with $A\sim 400~\rm s^{-1}$, faster than the production rate by collisional excitation.

In this work, we include 55 \ion{Ca}{1} transitions with lower levels no higher than $^3\mathrm{P}^o$ and radiative decay rates greater than $5\times10^5~\rm s^{-1}$.  This is in contrast to  large number of absorption lines in the \ion{Ca}{1} cross-correlation template created based on an isothermal LTE atmosphere \citep[e.g.][]{Hoeijmakers2020}.  Although not observed, the model suggests that the \ion{Ca}{1} $\lambda$4226\AArm\ absorption feature should be stronger than \ion{K}{1} $\lambda$7699\AArm\ feature, and may be accompanied by a broad line wing.
\subsubsection{Fe}
\label{sec:Fei-abs}
As described in Section~\ref{sec:radiative}, we assume that the $3d^64s4p$ $^5\mathrm{D}^o$ state and higher energy levels of \ion{Fe}{1} and the $3d^64p$ $^6\mathrm{D}^o$ state and higher of \ion{Fe}{2} are depleted.  Combining this with the criteria of spontaneous rates greater than $10^6~\rm s^{-1}$, we include a total of 492 \ion{Fe}{1} transitions and 829 \ion{Fe}{2} transitions from the NIST atomic spectra database in the calculation.

\subsection{Line Broadening Due to Bulk Outflow Velocity and Planet Rotation}
\label{sec:broad}

Several broadening mechanisms, including natural broadening, collisional broadening, thermal broadening, and velocity broadening, determine the line width of spectral features.
In the low-pressure environment of the planetary upper atmosphere, the effects of natural broadening and collisional broadening of metal lines are negligible compared to the observed FWHM of tens of km s$^{-1}$ of the spectral line probing the upper atmosphere \citep{Borsa2021}.

\begin{figure}
  \begin{center}
    \includegraphics[width=0.48\textwidth]{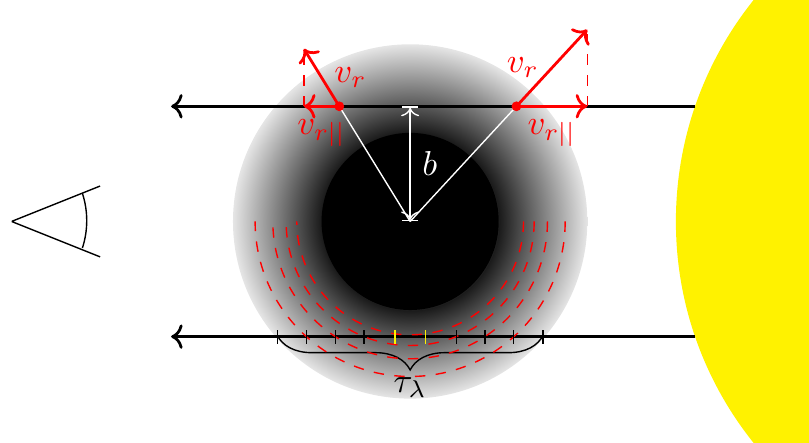}
    \caption{Schematic diagram illustrates the velocity broadening mechanism of spectral lines and the line-of-sight optical depth integral.  A light ray originating from the star on the right passes through the planet's atmosphere with an impact parameter $b$ to reach the observer on the left.  The width of the velocity broadening corresponds to the line-of-sight component of the atmospheric radial outflow velocity $v_{r\|}$.  The optical depth $\tau_\lambda$ along a light trajectory is the sum of the product of the length of the light propagation path within each radial shell of the model, illustrated with red dashed circles, and the opacity of that shell.}
    \label{fig:expand}
  \end{center}
\end{figure}

The width of many observed features cannot be explained by thermal broadening either.
For example, the required atmospheric temperature to explain the observed $\sim80 \rm km~s^{-1}$ FWHM of \ion{Ca}{2} features with thermal broadening is unrealistically high at $2.8\times 10^6$~K.
In addition, because of the higher mass of Na compared to H, the thermal broadening of the Na spectral lines should produce significantly narrower lines than those of hydrogen.
This contradicts the observations showing that the width of Na absorption line profiles is similar to \Ha\ \citep{Borsa2021, Cabot2020}.  To account for this, we introduce the line-broadening effect of radial outflow and planetary rotation.  Both of them can exceed 10 $\rm km~s^{-1}$ and are much higher than the thermal velocity of the metal species. We assume that the direction of the outflow (escape) velocity is radially outward. The line-of-sight component of this outflow velocity leads to line broadening, as illustrated by Figure~\ref{fig:expand}.

\begin{figure}
  \begin{center}
    \includegraphics[width=0.48\textwidth]{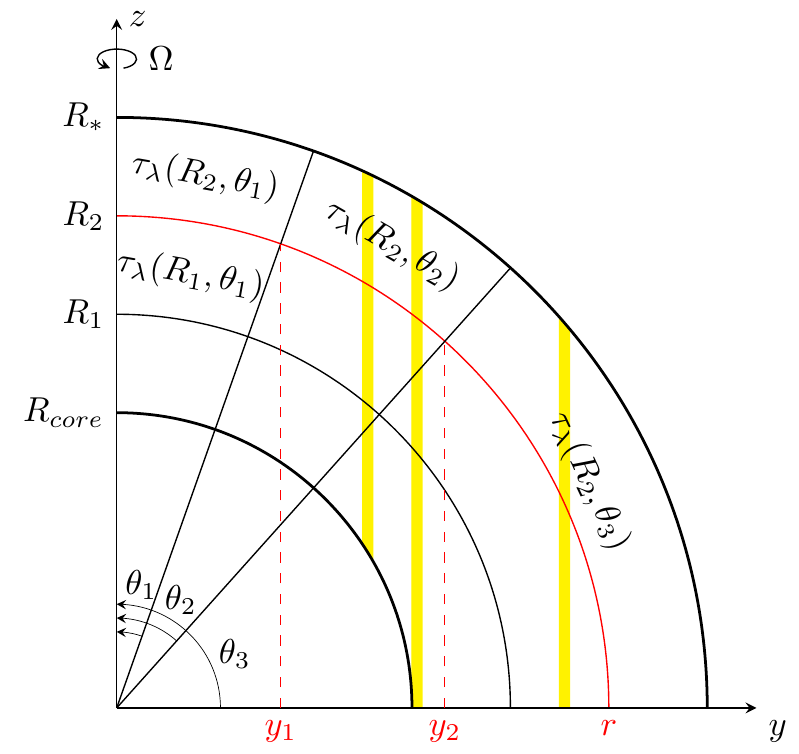}
    \caption{A cross-section view of a quarter of the planetary atmosphere in the $y$-$z$ plane.  The observer is outside the paper.  This sketch illustrates the grids in radial and polar angular directions that are used to calculate the optical depth and transmission spectrum.  The line-of-sight rotational velocity is $\Omega y$.  The three vertical yellow stripes represent three atmospheric slices, each with the same line-of-sight rotational velocity. }
    \label{fig:sector}
  \end{center}
\end{figure}

The rotational velocity of a tidally locked planet can dominate line broadening in the atmosphere if the outflow velocity is small.
In order to include rotational broadening in our calculation, we divide each quarter of the atmospheric annulus in the $y$-$z$ plane into 20 circular sectors to calculate the atmospheric absorption, as illustrated by Figure~\ref{fig:sector} that shows a simplified model with only 3 sectors.
Each sector has the same width in the $y$-direction --- $\Delta y_i=r\Delta\sin(\theta_i)$ is a constant, where $\theta_i$ is the polar angle of the sector in the $y-z$ plane.  The line-of-sight rotational velocity is $v_{\rm rot,i}=\Omega y_i$, where $\Omega$ is the planet's rotational angular frequency.  Thus, when rotational velocity dominates line broadening, the transit depth at a distance of $\Delta \nu/\nu_0$ away from the line center of an absorption feature corresponds to the absorption of stellar light by a planetary atmospheric slice with constant $y=\frac{\Delta \nu c}{\nu_0\Omega}$.

This is the case for the spectral profiles with small transit depths that probe the deeper atmosphere.  They typically show a dip at the line center (e.g. Na, K, and \Hb\ in Figure~\ref{fig:optical_tide}) due to the geometry effect: an atmospheric slice with a small $y$ has less area in the $y$-$z$ plane because the opaque core of the planet that does not contribute to the spectral features fills a large fraction.
The slice that is approximately tangent to the planet's surface, shown as the middle vertical stripe in Figure~\ref{fig:sector}, has the largest area.  Thus, the split of the two peaks roughly corresponds to the rotational velocity at the planet's equator, which is approximately $7.3~\rm km~s^{-1}$. A case study on the impact of these two broadening mechanisms on \ion{Ca}{2} K$\lambda$3934\AArm\ line is presented in Section~\ref{sec:spectral-features}.

\subsection{Algorithm for calculating the transit spectrum}
\label{sec:alg}

In the wavelength domain, we divide the simulated spectral range of 2000 -- 8000~\AA\ into more than $2\times 10^5$ wavelength points.  Wavelength bands near the center of absorption features are divided into finer spacing, with a spectral resolution of up to $R = 2\times 10^6$.
First, a 2-D opacity array in the Lagrangian frame that does not include the bulk velocity is constructed for each shell and wavelength point according to the local number densities and temperature.

Because of the velocity shifts described in Section \ref{sec:broad}, the atmospheric optical depth $\tau_\lambda$ is a function of the impact parameter $b$ of the light trajectory and its polar angle.
To calculate the absorption in transit, we divide the cross-section annulus of the atmosphere into a 2-D grid in the radial and angular directions, as illustrated in Figure~\ref{fig:sector}, and calculate $\tau_\lambda$ in each grid cell.
In the angular direction, a quarter of the atmosphere is divided into 20 sectors as described in Section~\ref{sec:broad}.
In the radial direction, we uniformly sample 2000 grid points between $R_{\rm core}$ and the stellar radius $R_*$.  A layout with three radial grid points is shown in Figure~\ref{fig:sector}.
The opacity $\kappa_\lambda$ in the proper frame of each radial atmospheric shell, which is attached to the bulk outflow gas, is obtained by interpolating the 2-D opacity array to the wavelength points accounting for the blueshift/redshift caused by the bulk outflow and planet rotation.
Multiplying the distance that the trajectory passes through each shell and integrating along the line of sight, we obtain the optical depth at a given wavelength, impact parameter, and polar angle $\tau_\lambda(b,\theta)$. 

Trajectories with large impact parameters may have long segments in the innermost few shells they pass through, which leads to a large dispersion of the line-of-sight component of the atmospheric outflow velocity within the same shell.  Because the single line-of-sight velocity values cannot adequately represent the whole segment, these segments are further divided along the trajectory to improve the characterization of spectral line shapes, illustrated with short yellow vertical bars in Figure~\ref{fig:expand}.

\begin{figure}
  \begin{center}
    \includegraphics[width=0.5\textwidth]{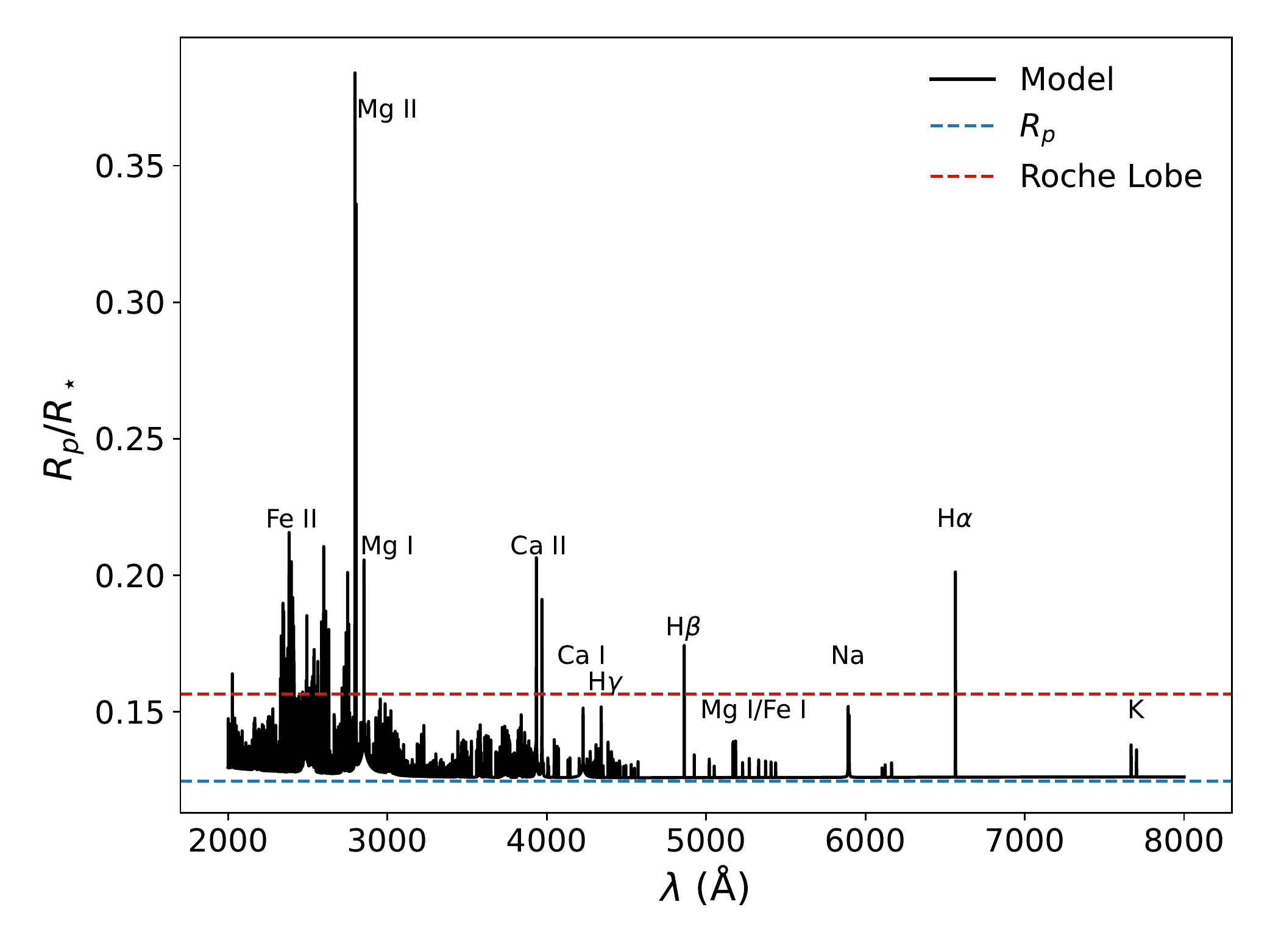}
    \caption{Simulated transmission spectrum of case A.  The red dashed line indicates the radius of the Roche Lobe at the terminator.}
    \label{fig:full_noT}
  \end{center}
\end{figure}

The absorption of light is symmetric between the northern and southern hemispheres.  Combining the optical depth of each trajectory with the cross-sectional area it represents, and summing over the impact parameters $b_j$ as well as the polar angles $\theta_i$ in one hemisphere, we obtain the square of the planetary apparent radius or stellar light attenuation fraction: 
\begin{equation}
\small
\label{eq:5}
\begin{split}
\left[\frac{R_p}{R_*}(\lambda)\right]^2 &= \frac{1}{R_*^2}\left\{\vphantom{\sum_{i,j}b_j}R_{\rm core}^2 \right. \\
&\left.+2\Delta b\sum_{i,j}b_j\Delta \theta_i\{1-\exp[-\tau_\lambda(b_j,\theta_i)]\}\right\}.
\end{split}
\end{equation}
For example, the transmission spectrum calculated based on case A is shown in Figure~\ref{fig:full_noT}. We note that the cores of the Mg II lines here correspond to an obstacle with a radius of about 3.08 $R_p$ ($R_p/R_* =$~0.38) viewed in transit while the cores of the other strong lines correspond to an obstacle of about 1.65 $R_p$ ($R_p/R_* =$~0.21).

\subsection{Comparison with the NUV Observations}
\label{sec:observation}

\begin{figure}
  \begin{center}
    \includegraphics[width=0.5\textwidth]{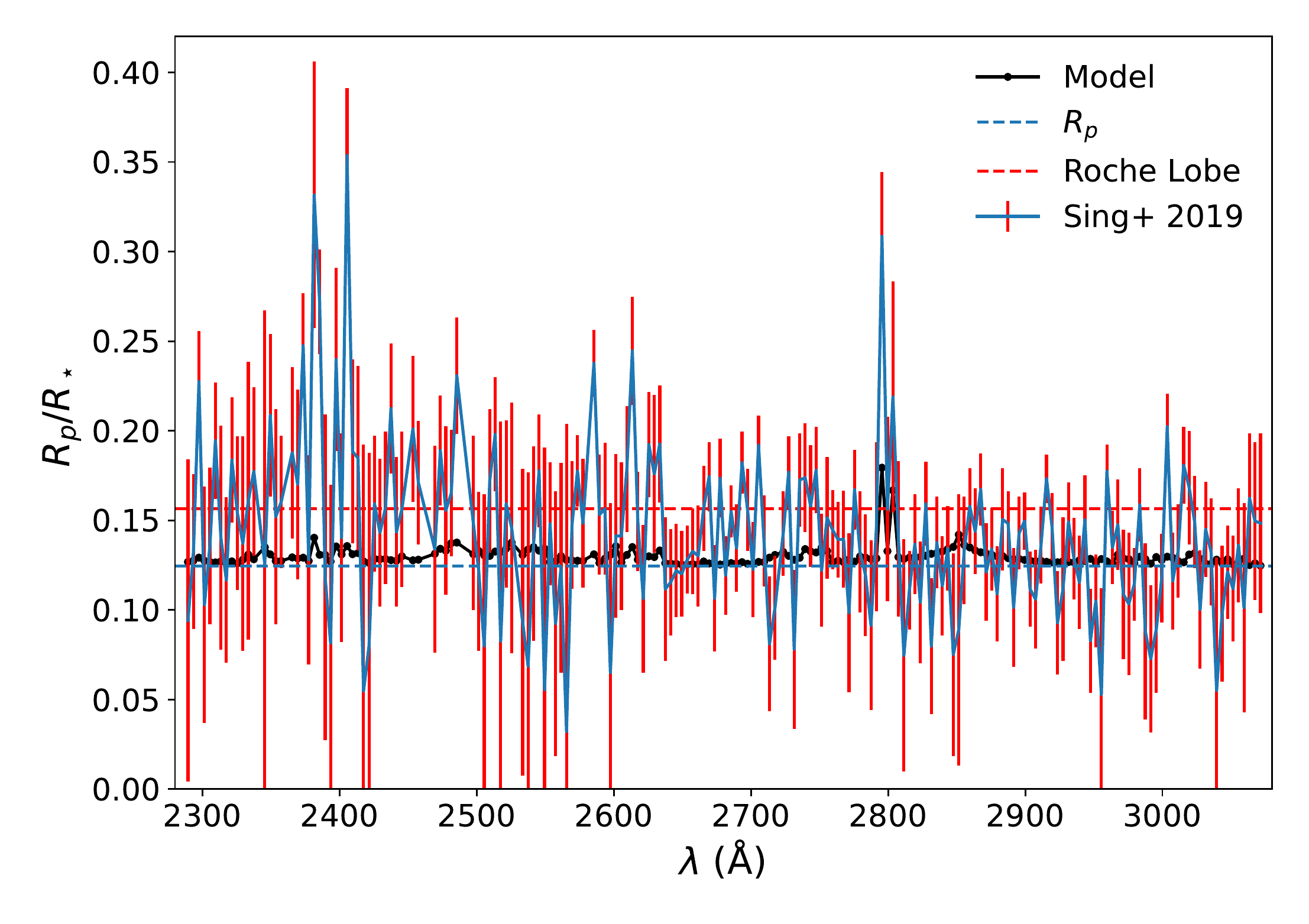}
    \caption{Binned simulated NUV transmission spectrum of case A compared to HST observation \citep{Sing2019}. $\chi^2/N=2.07$.}
    \label{fig:NUV_noT}
  \end{center}
\end{figure}

\begin{figure*}
  \begin{center}
    \includegraphics[width=0.9\textwidth]{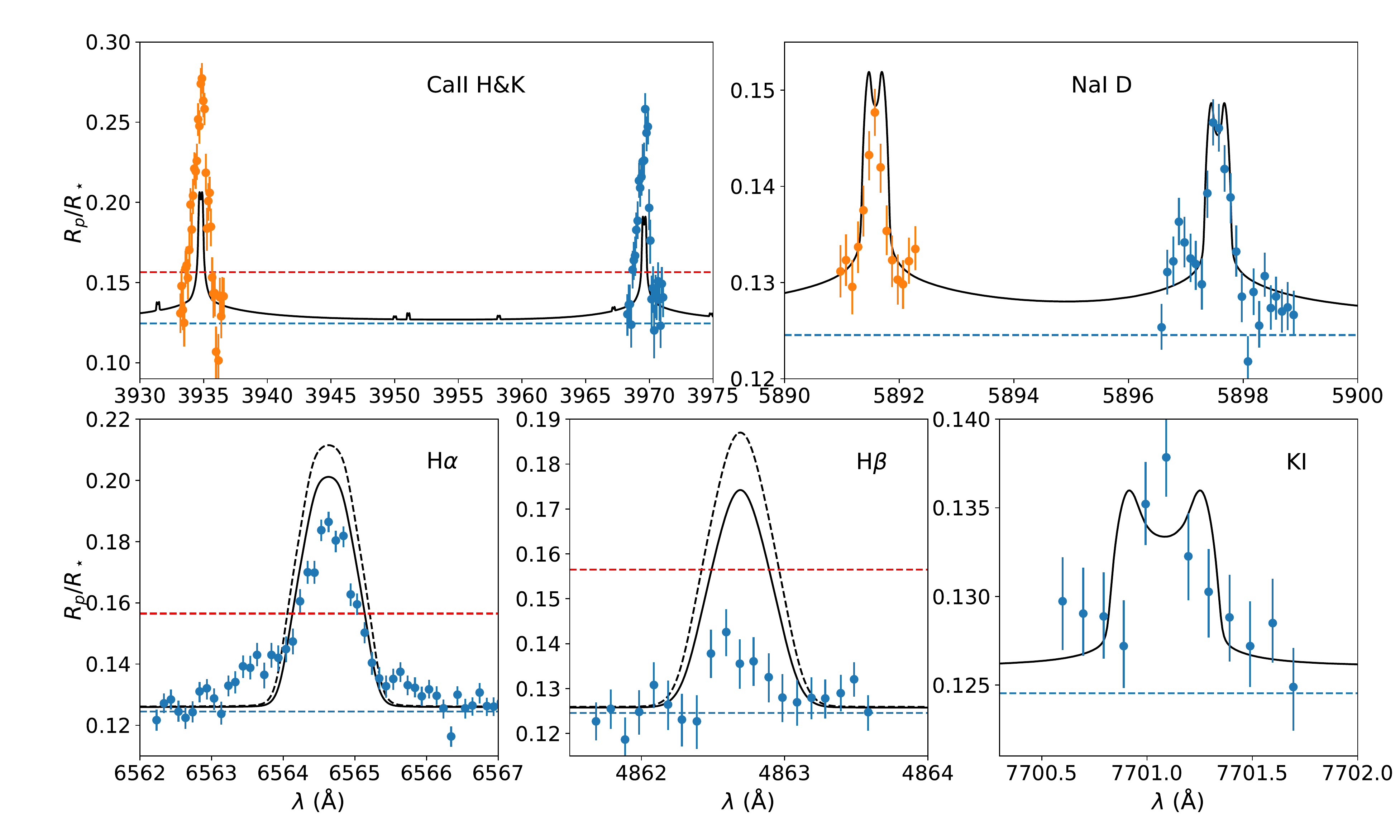}
    \caption{Comparison of optical features observed by \citet{Borsa2021} with simulated profiles of case A.  The horizontal dashed lines indicate $R_p$ and the radius of the Roche Lobe at the terminator.  The black dashed lines in the panels of \Ha\ and \Hb\ show the profile produced with the H/He only A' model.}
    \label{fig:optical_noT}
  \end{center}
\end{figure*}

Since the spectral resolution of HST NUV observations is not high enough to resolve individual spectral features, we processed our simulated spectrum as follows to approximate the treatment of the data by \citet{Sing2019}.  First, we multiply the simulated stellar light attenuation fraction $[R_p/R_*(\lambda)]^2$ by the stellar spectrum to get the expected spectral energy distribution obscured by the planet during transit.  Then, the obscured and stellar spectra are both convolved with a Gaussian smoothing function having an FWHM of 0.09\AA, which mimics the HST spectral resolution of R=30,000.  The ratio of these two spectra reproduces the expected transit radius to be observed.  Finally, following the data reduction process, the convolved NUV spectra are divided into 187 4\AA-wide bins with the same wavelength grid as the processed observed spectrum.  The boxcar averaged transit depths for each bin are shown with black dots connected by a black solid line in Figure~\ref{fig:NUV_noT}.  The red horizontal dashed line marks the equipotential surface that coincides with the first Lagrangian point (L1) i.e., the boundary of the Roche lobe along the terminator that is lower than the distance to the L1 point indicated in Figure 8 of \citet{Sing2019}.

Compared to the observations, the model has a reduced $\chi^2$ of 2.08 with 187 degrees of freedom.  The spectrum is dominated by \ion{Fe}{2} and strong \ion{Mg}{2} lines. Although, for example, the size of the planet in the \ion{Mg}{2} lines at full model resolution exceeds 35\% of the stellar radius, the transit depths of all lines in the binned model spectrum are significantly lower than the observed transit depths.
This is because the line widths based on the Case A model are too narrow to match the data.

\subsection{Comparison with the Optical Observations}
\label{sec:normalization}

The absorption lines of the different species in the optical transmission spectrum provide an additional check on the model and much more robust constraints on the upper atmosphere than the NUV data alone.
From Figures 9--12 in \citet{Borsa2021}, we use \emph{webplotdigitizer} \citep{Rohatgi2020} to extract the observed spectral features.  We extract the profiles of absorption features as well as 5 additional data points in the continuum region on each side of the spectral lines.  When comparing with the model, we normalize the observed features based on the outermost 4 points on both sides to match them with the simulated profile.  
\citet{Borsa2021} noted that all detected absorption features are blueshifted by different degrees from the expected wavelengths of the lines.  Due to the 1D nature of the model, our simulated spectra cannot reproduce these shifts.
Therefore, in addition to shifting the observed wavelengths to vacuum values by applying air refractive index $n=1.0002762$, we also shift the observed features according to the blueshift velocities reported in Table 4 in \citet{Borsa2021}.  This is equivalent to assuming that the blueshift arises in aggregate from horizontal winds in the atmosphere, as suggested by \citet{Borsa2021}, that cannot be directly simulated by our approach.
Because the spectral resolution is high enough to resolve spectral features, we do not convolve or bin the model spectral lines.

As a result, Figure~\ref{fig:optical_noT} shows the comparison of several observed spectral features with the simulated transmission spectra obtained based on the Case A model.
It is worth noting that although we show the Ca and Na doublet features in single panels, the two observed absorption peaks are normalized separately. The model provides a good fit to the Na and K lines, as they do not probe the escaping atmosphere where Na and K are mostly ionized. In line with the NUV spectrum, however, the model underestimates the strength and broadening of the Ca II lines that probe the escaping part of the atmosphere.

In contrast to the lines of metal ions, the model overestimates absorption in the H Balmer lines, while not matching the broad wing of the H$\alpha$ line.  It is the result of the combination of many factors, including the high stellar \Lya\ intensity and the close proximity between the peak of the temperature and the peak of the neutral atomic hydrogen number density.  This problem would be worse for models that do not include metals. The \Ha\ and \Hb\ profiles produced with the metal-free A' model are depicted with black dashed lines in Figure~\ref{fig:optical_noT}. Due to the higher temperatures and the resulting more extended atmosphere, the Balmer lines exhibit deeper transit depths.

The comparison of our model spectrum with the NUV and optical observations indicates that a higher density and velocity of the escaping material are required to explain the observations. This implies a significantly higher mass-loss rate at the terminator of the planet than that predicted by our one-dimensional globally averaged model. Below, we argue that Roche lobe overflow (RLOF) can sufficiently enhance the model transit depths to match the observations and thus, the observations of WASP-121b provide a confirmation that the atmospheric escape from this planet is enhanced by RLOF. We caution the reader that RLOF is an inherently multi-dimensional problem. In the absence of a multi-dimensional model that includes the required photochemistry and radiative transfer, we use results from our one-dimensional model to demonstrate the density and velocity enhancements that are required to match the data and discuss their relationship to what would be expected under RLOF.

\section{Tidal Potential}
\label{sec:tide}

Because of the small surface gravity and the proximity to its host star, the stellar tide can significantly affect WASP-121b, as it does on WASP-12b \citep{Dwivedi2019}. As mentioned above, a larger atmospheric outflow velocity is necessary to match the observed spectral line profiles. A strong tidal force can reduce the average effective gravity, boost the mass-loss rate, and increase the outflow velocity. Therefore, in the following, we add the tidal (Roche) potential to the atmospheric escape model and simulation of the transmission spectrum to investigate its possible impact on the atmospheric properties and spectral line profiles.

\subsection{Atmosphere Model with Tidal Potential}
\label{sec:atm-tide}

In the planetary atmosphere model, we define a Cartesian coordinate $x$ that points towards the host star, $y$ is the direction of the orbital motion, and $z$ points to the north pole. The tidal (Roche) potential of the two-body system in the co-rotating frame, centered at the tidally locked planet, is
\begin{equation}
  \footnotesize
  \label{eq:13}
  \begin{split}
    \phi(x,y,z) &= -\frac{GM_p}{(x^2+y^2+z^2)^{1/2}}-\frac{GM_*}{[(a-x)^2+y^2+z^2]^{1/2}} \\
    &-\frac{1}{2}\Omega^2[(\frac{M_*a}{M_*+M_p}-x)^2+y^2].
  \end{split}
\end{equation}
The planet radius $R_p$ is determined based on the transit depth assuming a circular planetary cross-section.  However, because of the tidal potential, the planet $y$-axis size $R_{py}$ is larger than its equal-potential point in the $z$-axis $R_{pz}$.  Considering that the surface of the planet is a nearly elliptical equipotential surface, we solve $R_{py}R_{pz}=R_p^2$ to obtain $R_{py}$ and $R_{pz}$, and then calculate the radius $R_{px}$ of the equipotential surface along the $x$-axis. The triaxial radii $R_{px}$, $R_{py}$, and $R_{pz}$ of the planet obtained after considering the tidal potential factor are listed in Table~\ref{tab:derived}.  These values agree with the structural parameters corrected for asphericity listed in \citet{Delrez2016}, except for the different values of Jupiter radius units adopted, as mentioned in Section~\ref{sec:params}.

  \begin{figure}
    \begin{center}
      \subfloat[x-y plane]{\includegraphics[width=0.2\textwidth]{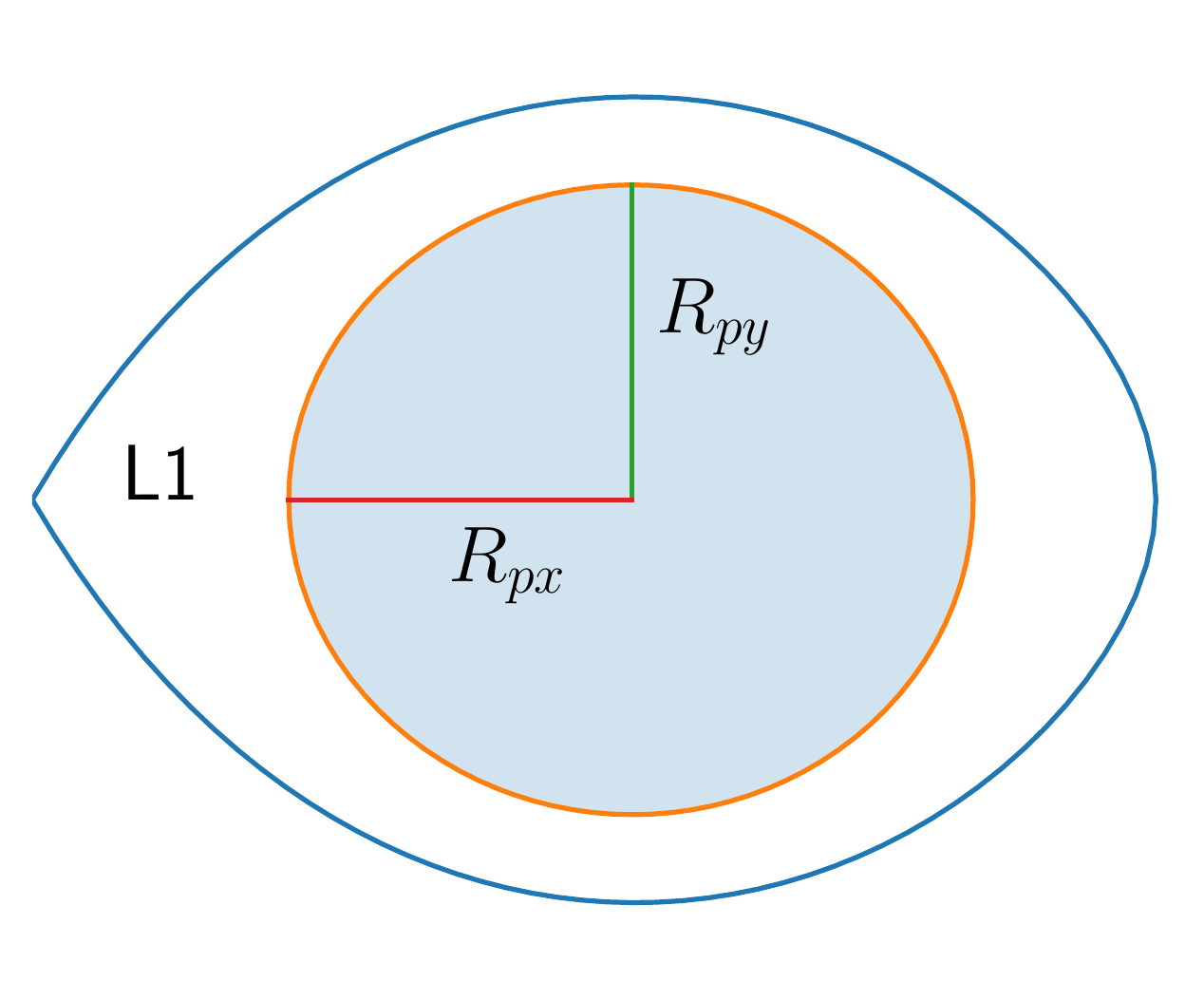}}
      \subfloat[y-z plane]{\includegraphics[width=0.3\textwidth]{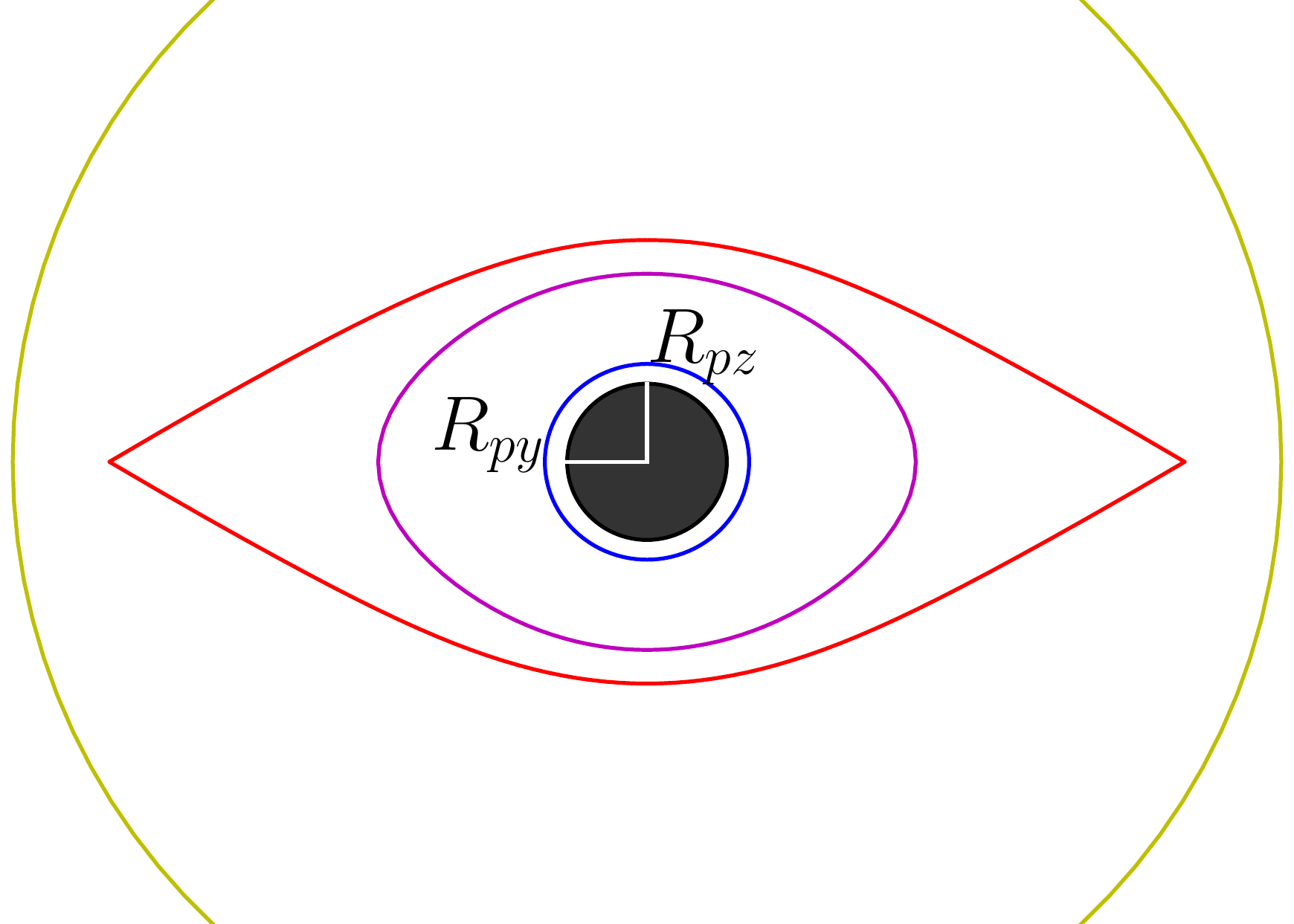}}
      \caption{Geometry of WASP-121b system and its Roche lobe (RL). {\bf (a)} The planet viewed from the pole. The host star is towards the left.  The disk shows the planet and the blue curve shows the location of the RL. $L1$ indicates the location of the first Lagrangian point. {\bf (b)} The planet as viewed by an observer in the middle of transit. The filled black disk shows the planet. The first blue elliptical shape indicates the RL. The outermost yellow circle shows the size of the host star. The two curves in between show the contours of two equipotential surfaces.}
      \label{fig:demo}
    \end{center}
  \end{figure}

  \begin{table}
  \caption{Derived Values and Best-fit Values for the Orbital and Physical Parameters of WASP-121 and WASP-121b}
  \begin{tabular}{c c}
    \hline
    \multicolumn{2}{l}{Derived quantities using the canonical value}\\
    \hline
    $R_{px} $ & $ 1.945\,R_J$\\
    $R_{py} $ & $ 1.785\,R_J = 0.1259\,R_*$\\
    $R_{pz} $ & $ 1.747\,R_J = 0.1232\,R_*$\\
    $R_{L1} $ & $ 1.748\,R_{px}$  \\
    $R_{RL} $ & $ 2.233\,R_J = 0.1575\,R_*$  \\ 
    \hline
    \multicolumn{2}{l}{Parameters that provide a better match}\\
    \hline
    $R_* $ & $ 1.457\,R_\odot$ \\
    $M_* $ & $ 1.352\,M_\odot$ \\
    P & 1.2749255 day\\
    $M_p $ & $ 1.120 \,M_J$ \\
    $R_p$ & $1.766\,R_J$  \\
    a & 0.02545 AU \\
    $R_{px} $ & $ 1.959\,R_J$\\
    $R_{py} $ & $ 1.786\,R_J$\\
    $R_{pz} $ & $ 1.746\,R_J$\\
    $R_{L1} $ & $ 1.705\,R_{px}$  \\
    $R_{RL} $ & $ 2.194\,R_J$  \\ 
    \hline
  \end{tabular}
  \label{tab:derived}
\end{table}

The two schematics in Figure~\ref{fig:demo} visualize, from two angles, the relative size of the planet with respect to the Roche lobe (RL) and the star when using the canonical parameters.
The distance to the L1 point $R_{L1}$, and the radius of the RL in the terminator plane $R_{RL}$ (see Section~\ref{sec:spec-tide}) are also listed in Table~\ref{tab:derived}.  The modified system parameters that can provide a better fit to the observations, as described in Section~\ref{sec:adjust}, are listed in the bottom half of Table~\ref{tab:derived}.  

In the following, we calculate escape through the L1 point along the substellar streamline by using gravitational potential based on the Roche potential:
\begin{equation}
  \label{eq:12}
  \phi(x) = -\frac{GM_p}{x}-\frac{GM_*}{a-x}-\frac{1}{2}\Omega^2(\frac{M_*a}{M_*+M_p}-x)^2.
\end{equation}
Because the photochemical model of the lower atmosphere does not include the tidal force, the pressure-radius relationship at pressures higher than 1 $\mu$bar  is recalibrated to account for it. Using the temperature and mean molecular weight of each pressure level, we integrate the equation of hydrostatic equilibrium along the substellar direction both outward and inward starting from the radius $R_{px}$ where pressure is set to 4~mbar.

In contrast to the case A model, the stellar radiation is incident vertically on the atmosphere instead of at a 60$^\circ$ zenith angle.  Accordingly, to represent globally averaged stellar radiation, the incident stellar flux is divided by a factor of 4 instead of a factor of 2.  The atmosphere model that includes the tidal potential with canonical system parameters will be referred to as case B.

\begin{figure}
    \begin{center}
      \includegraphics[width=0.5\textwidth]{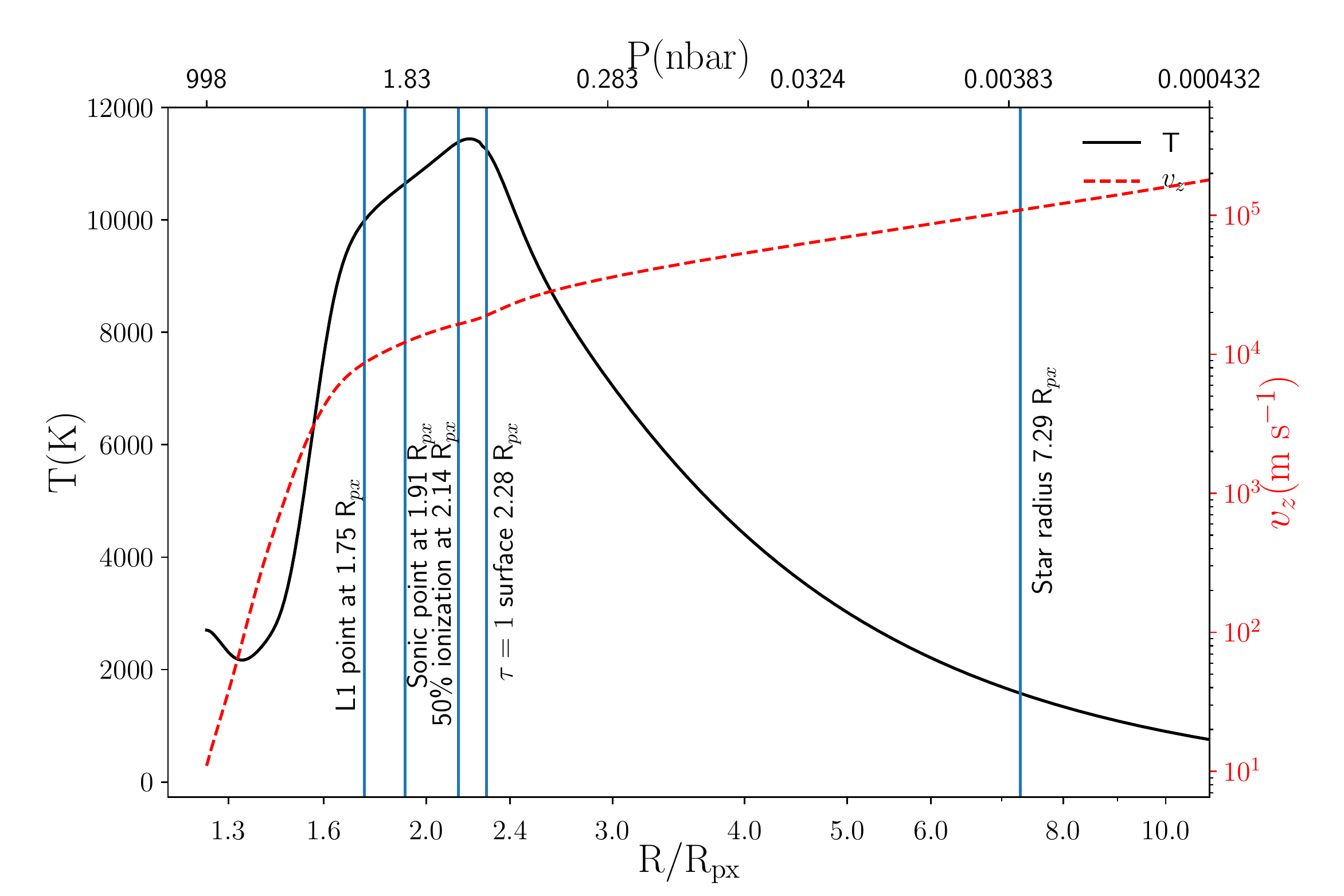}
      \caption{Temperature and radial velocity along the substellar direction simulated by the hydrodynamic model for case B.}
      \label{fig:Tvr_63}
    \end{center}
  \end{figure}

The temperature and radial velocity profile of case B is shown in Figure~\ref{fig:Tvr_63}.
At the radius equal to the stellar radius, the radial outflow velocity is over 7 times higher than in case A. The higher outflow velocity results in more effective adiabatic cooling and thus lower atmospheric temperature. Outside of the temperature peak, the hydrogen ionization fraction becomes lower than in case A due to the larger recombination coefficient caused by the lower temperatures. The radius at which $\tau=1$ for stellar radiation with the energy of 13.6~eV, the ionization threshold of H, is at higher altitudes due to the higher neutral hydrogen column density. To connect smoothly to the lower temperature near the base of the hydrodynamic model when the tidal potential is included, the dotted line in Figure~\ref{fig:Tcomb} is applied as the temperature for the photochemical lower atmosphere model.

\subsection{Model inputs and results}
\label{sec:tabul}

We generate a suite of atmospheric models, the key information of which is listed in Table~\ref{tab:summary}.
The input parameters listed include whether the tidal potential is considered, the masses and radii of the planet and the star, orbital separations, temperature adjustment to the lower and middle atmosphere, planetary gravity at 1 $\mu$bar, multiplier of the stellar XUV flux, \Lya\ flux, and multiplier of the Fe abundance relative to the default solar value.
More details on how these parameters are chosen are given in Section~\ref{sec:adjust}. The purpose of these models is to explore how different parameter choices and model inputs in general can change the predicted properties of the planetary atmosphere and the transit depths.

Correspondingly, we extract representative information from the result of each atmospheric model and the simulated transmission spectrum and also list them in Table~\ref{tab:summary}.  The atmospheric properties included in the table are the simulated global mass-loss rate (see below), mass-loss rates according to energy-limited escape, without and with the RL correction factor (see Section~\ref{sec:mdot}), and the atmospheric pressure at the L1 point.

To represent the features in the simulated transmission spectrum, we select the line center transit depths of \ion{Ca}{2} K ($\lambda$3935 \AA), \Ha, \Hb, and Na D$_2$ ($\lambda$5890 \AA).  As a comparison, the depths observed by \citet{Borsa2021} using 4-UT mode are 0.281$\pm$0.009, 0.186$\pm$0.003, 0.143$\pm$0.005, and 0.147$\pm$0.002, respectively.  In addition, we list the \ion{Mg}{2} $\lambda$2796 line center transit depth binned to a 4 \AA\ passband, as shown in Figure~\ref{fig:NUV_noT}, which can be compared to the observed value of 0.309$\pm$0.036 obtained by \citet{Sing2019} using the same passband at the same wavelength.  The reduced $\chi^2$ that compares binned NUV simulated spectrum with observation is also listed.

We chose these spectral features because they can be directly compared with high-quality observations that trace species inside and escaping from the RL. Together, these features yield more information about the properties of the escaping atmosphere. The Na D$_2$ line is a good tracer of the lower and middle atmosphere below the RL boundary.
In the escaping atmosphere, the most abundant ionization states of the metals, such as Mg$^{2+}$, Na$^+$, Ca$^{2+}$, or Fe$^{2+}$, do not have strong detectable spectral features at the observable wavelengths.
The line center transit depth of the \ion{Ca}{2} K line reflects the extent of the Ca$^+$ population, which is the second most abundant ionization level of the element.
Combining it with the binned \ion{Mg}{2} transit depths can further constrain the width of the spectral lines.
On the other hand, \Ha\ and \Hb\ features together reveal the extent and slope of the H($n=2$) population, which are sensitive to the atmospheric temperature and the \Lya\ intensity in the atmosphere.

\subsection{Mass-Loss Rates}
\label{sec:mdot}

If the mass-loss rate in each direction is assumed to be the same as the rate along the simulated substellar streamline, then the mass-loss rate of the atmosphere is $\dot{M}=4\pi r^2v_r\rho$, where $v_r$ is the bulk radial outflow velocity.
The mass-loss rate given by each model is listed in the first column of the bottom half of Table~\ref{tab:summary}.
This assumption is reasonable for planets with spherical gravitational potential, as long as the globally averaged stellar flux is used in the simulations.
However, for systems with a non-negligible tidal potential, the mass-loss rate estimated using the substellar atmospheric streamline is an upper limit.
For example, \citet{Guo2013} compared results from 1D and 2D versions of his model to explore RL effects on a planet like HD209458b. The simulations indicate that a 1D model with substellar gravity overestimates the mass-loss rate by a factor of $\sim$2 compared to a 2D model with uniform heating around the planet. According to the same study, significant day/night temperature differences in the upper atmosphere could reduce the mass-loss rate by a factor of 7 compared to a 1D model with substellar gravity.  

We compare the mass-loss rates with the energy-limited mass-loss rates of the system, which are often assumed in the literature for atmospheric escape that is driven by stellar XUV flux.
Considering only the gravitational potential energy of the planet itself and assuming the mass-loss efficiency $\epsilon=1$, we have
\begin{equation}
  \label{eq:14}
  \dot{M}_{\rm XUV}=\frac{R_p^3L_{\rm XUV}}{4a^2GM_p},
\end{equation}
where $L_{\rm XUV}$ is the stellar luminosity at energies higher than the ionization threshold of H at 13.6 eV.
To account for the Roche potential, \citet{Erkaev2007} proposed a correction factor for the potential $K(\eta) = 1-\frac{3}{2\eta}+\frac{1}{2\eta^3}$, where $\eta=R_{L1}/R_p$.  The energy-limited $\dot{M}_{\rm XUV}$ and $\dot{M}_{\rm XUV}/K(\eta)$ are also listed in Table~\ref{tab:summary}.

\begin{splitdeluxetable*}{ccccccccccccBcccccccccc}
  \tablecaption{Input parameters of each model and key result features \label{tab:summary}}
  \tablehead{
    \colhead{\multirow{2}{*}{Model}} & \colhead{Tidal} & \colhead{$M_p$} & \colhead{$R_p$} & \colhead{$M_* $} & \colhead{$R_*$} & \colhead{$a$} & \colhead{dT} & \colhead{$\log g$ \tablenotemark{a}} & \multicolumn{3}{c}{Multiplier} & \colhead{$\dot{M}$} & \colhead{$\dot{M}_{\rm XUV}$} & \colhead{$\frac{\dot{M}_{\rm XUV}}{K(\eta)}$} & \colhead{P @ L1} & \colhead{Ca II K \tablenotemark{b}} & \colhead{\Ha \tablenotemark{b}} & \colhead{\Hb \tablenotemark{b}} & \colhead{Na D$_2$ \tablenotemark{b}} & \colhead{Mg II \tablenotemark{c}} & \colhead{\multirow{2}{*}{$\chi^2/N$}}  \\
    \cline{3-21}
  & potential & ($M_J$) & ($R_J$) & ($M_\odot$) & ($R_\odot$) & (AU) & (K) & @ 1$\mu$bar & \colhead{XUV} & \colhead{\Lya} & \colhead{Fe}  & \multicolumn{3}{c}{($M_p$/Gyr)\tablenotemark{d}}  & (nbar) & \multicolumn{5}{c}{($R_p/R_*$)} & 
  }
  \startdata
   Case A & N & 1.1824 & 1.766 & 1.3521 & 1.4572 & 0.02545 & 0 &  2.84 & 1 & 1 & 1 & 0.052 & 0.16 & 0.55 & $1.07\times 10^{-3}$ & 0.199 & 0.201 & 0.174 & 0.152 & 0.182 & 2.08 \\
   Case B & Y & 1.1824 & 1.766 & 1.3521 & 1.4572 & 0.02545 & 0 & 2.70 & 1 & 1 & 1 & 0.32 & 0.16 & 0.68 & 2.67 & 0.200 & 0.166 & 0.133 & 0.141 & 0.202 & 2.03 \\
   Case C & Y & 1.1204 & 1.766 & 1.3521 & 1.4572 & 0.02545 & 350 & 2.62 & 1 & 1 & 4 & 1.20 & 0.18 & 0.80 & 2.52 & 0.270 & 0.221 & 0.145 & 0.147 & 0.303 & 1.18\\
   \textbf{Case D}\tablenotemark{e} & Y & 1.1204 & 1.766 & 1.3521 & 1.4572 & 0.02545 & 350 & 2.62 & 0.74 & 0.35 & 4 & \textbf{1.03} & 0.13 & 0.59 & 2.02 & 0.278 & 0.185 & 0.135 & 0.147 & 0.302 & 1.20 \\
   Case E & Y & 1.1204 & 1.766 & 1.3521 & 1.4572 & 0.02545 & 350 & 2.62 & 1 & 0.25 & 5 & 1.21 & 0.18 & 0.80 & 2.51 & 0.269 & 0.191 & 0.137 & 0.147 & 0.303 & 1.21 \\ 
   Case F & Y & 1.1204 & 1.766 & 1.3521 & 1.4572 & 0.02545 & 350 & 2.62 & 0.47 & 0.6 & 3.5 & 0.80 & 0.084 & 0.38 & 1.47 & 0.268 & 0.180 & 0.133 & 0.143 & 0.292 & 1.24 \\
   Case G & N & 1.1204 & 1.766 & 1.3521 & 1.4572 & 0.02545 & 350 & 2.80 & 1 & 1 & 1 & 0.073 & 0.18 & 0.63 & $1.31\times 10^{-3}$ & 0.217 & 0.209 & 0.181 & 0.155 & 0.192 & 1.90\\
   Case H & N & 0.9964 & 1.8911 & 1.6603 & 1.5604 & 0.02725 & 1000 & 2.63 & 1 & 1 & 4 & 0.188 & 0.27 & 1.25 & $2.69\times 10^{-3}$ & 0.256 & 0.241 & 0.206 & 0.166 & 0.229 & 1.43\\
   Case I & Y & 1.152  & 1.766 & 1.3521 & 1.4572 & 0.02545 & 550 & 2.63 & 0.74 & 0.35 & 4 & 1.08 & 0.124 & 0.54 & 2.07 & 0.280 & 0.185 & 0.135 & 0.148 & 0.307 & 1.22 \\
   Case J & Y & 1.1204 & 1.785 & 1.396 & 1.473 & 0.02572 & 120 & 2.61 & 0.21 & 0.7 & 3.5 & 0.55 & 0.038 & 0.179 & 0.85 & 0.296 & 0.183 & 0.135 & 0.143 & 0.318 & 1.33\\
   Case K & Y & 1.1204 & 1.785 & 1.396 & 1.473 & 0.02572 & 120 & 2.61 & 1 & 0.15 & 4 & 1.3 & 0.183 & 0.86 & 2.51 & 0.265 & 0.183 & 0.135 & 0.144 & 0.308 & 1.2 \\
   \enddata
   \tablenotetext{a}{$g=GM_p/r$ in $cgs$ unit, where $r$ is the radius where P=1 $\mu$bar in the substellar direction.}
   \tablenotetext{b}{The line center transit depths of Ca II K ($\lambda$3935 \AA), \Ha, \Hb, and Na D$_2$ ($\lambda$5890 \AA) obtained by \citet{Borsa2021} using the 4-UT mode are 0.281$\pm$0.009, 0.186$\pm$0.003, 0.143$\pm$0.005, and 0.147$\pm$0.002 respectively.}
   \tablenotetext{c}{Mg II $\lambda$2796 transit depth binned with 4 \AA\ passband as shown in Figure~\ref{fig:NUV_noT}.  The transit depth obtained by \citet{Sing2019} using the same passband is 0.309$\pm$0.036.}
   \tablenotetext{d}{1 $M_p$/Gyr= $7.1\times 10^{13}~\rm g~s^{-1}$.}
   \tablenotemark{e}{Preferred model.}
\end{splitdeluxetable*}

\subsection{Transmission Spectrum with Tidal Potential}
\label{sec:spec-tide}

  \begin{figure}
    \begin{center}
      \includegraphics[width=0.5\textwidth]{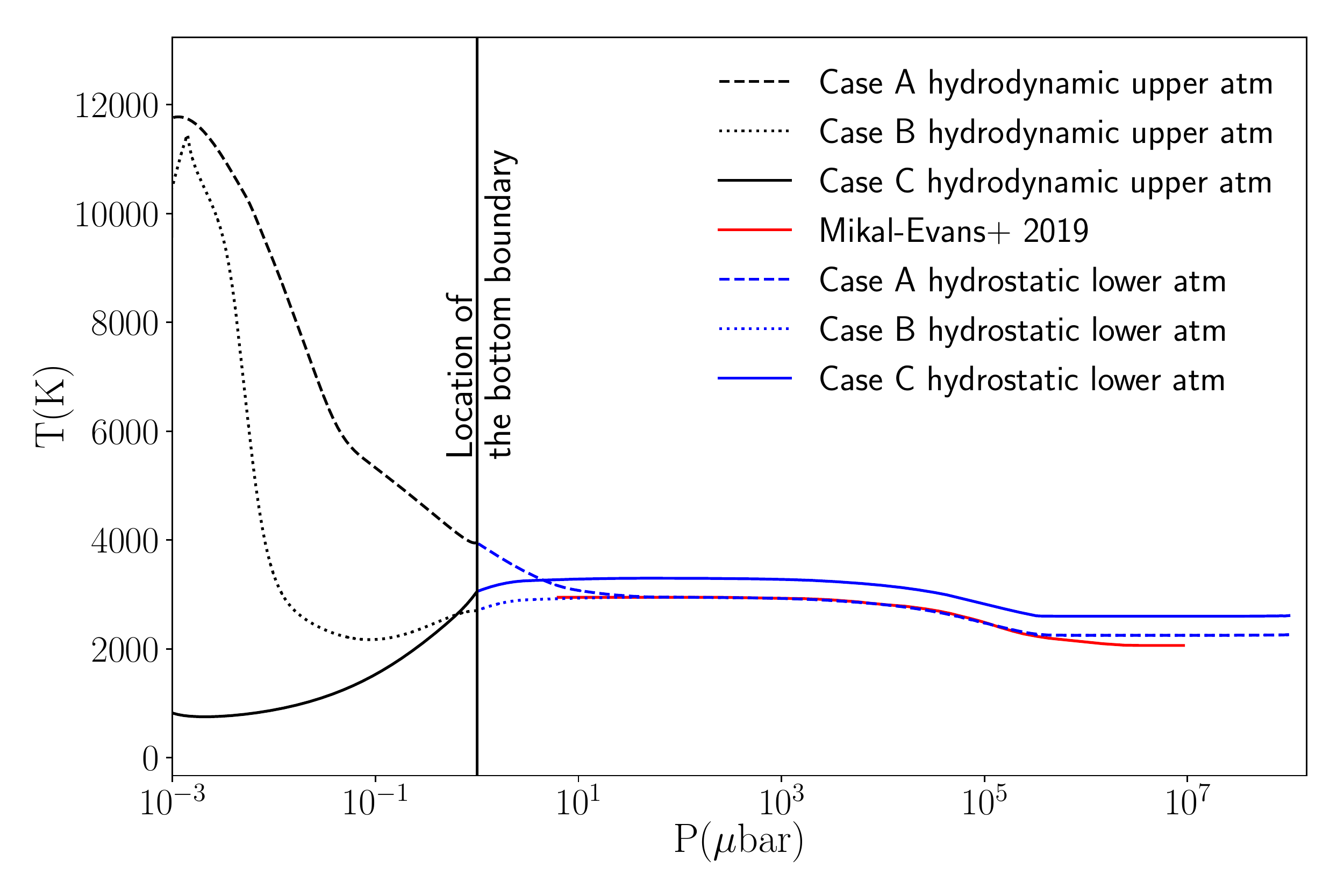}
      \caption{Temperature profiles applied in the lower and middle atmosphere photochemical model.  The blue dashed line shows the temperature profile of case A, the same profile as the one shown in Figure~\ref{fig:Tcomb_notide}, which does not include the tidal potential.  The blue dotted line shows the temperature profile of case B, which is smoothly connected with an upper atmosphere temperature profile that considers the tidal potential.  The blue solid line shows the temperature profile of the best-fit case D model, which uniformly increases the temperature of Case B by 350 K.}
      \label{fig:Tcomb}
    \end{center}
  \end{figure}

The calculation of the transmission spectrum is complicated by the fact that the geometric layout of the atmospheric cross-section is no longer spherically symmetric and we have to try and approximate it by using a 1-D model.
Inspired by the equivalent sphere of the Roche lobe discussed in \citet{Eggleton1983}, we map our simulated 1-D atmosphere profile along the substellar direction to an equivalent spherical obstacle for the transmission spectrum calculation.
As demonstrated in Figure~\ref{fig:demo}, the outflow along the polar direction is suppressed by the tidal force compared to the equatorial direction.  
As a representative direction of the atmosphere in the terminator plane, we choose the atmosphere properties along the $\theta=45^\circ$ direction in the $y-z$ plane to construct the equivalent spherically symmetric atmosphere.
The radius of the RL in this representative direction is listed as $R_{RL}$ in Table~\ref{tab:derived}.

To map the simulated atmosphere profile from the substellar direction to the $y-z$ plane within the RL, we assume the atmospheric temperature and number densities are uniform across the equipotential surface based on the assumption that the energy is efficiently redistributed around the planet.
For regions beyond the L1 point, we conserve the number densities, temperature, and radial velocity, while scaling the radii in the atmosphere by the ratio of $R_{RL}$ to $R_{L1}$ at the RL boundary, which is approximately 2/3. 

After this conversion, the reduced mass-loss rate per solid angle $\rho v_r r^2$ outside of the RL is naturally conserved along the radial direction, as in the original substellar solution. 
The mass-loss rate of our equivalent spherical atmosphere is suppressed by a factor of $\sim (2/3)^2 = 4/9$ compared to the 1D substellar atmosphere model.
The similarity of this ratio to the factor of 2 difference in mass-loss rates based on 1D and 2D models found by \citet{Guo2013} provides some justification for the conversion that we apply.

Within the RL, however, the mass-loss rate is not conserved along the radial direction if the outflow velocity is assumed to be uniform across the equipotential surface.
Therefore, we adjust the outflow velocity below the RL boundary to conserve mass based on the converted number densities and radii.
Since the outflow velocity within the RL makes only a minimal contribution to the spectral line broadening and is smaller than the effect of rotational velocity, this treatment of the radial velocity does not have a noticeable impact on the line profiles in the calculated transmission spectrum.

The use of a 1-D model to simulate the atmospheric structure within strong tidal potential has obvious limitations. The deviation of the escaping cloud of plasma from a spherical shape near the planet, however, may be more limited than first imagined, and the metal line cores in our best-fit model only extend to a radius of about 4.3 $R_p$ ($R_p/R_* =$~0.52) (see Figure~\ref{fig:full_tide}). For example, for a planet like HD209458b, Figure 8 in \citet{Shaikhislamov2020} shows the distribution of \ion{C}{2} ions responsible for \ion{C}{2} $\lambda$1336 \AA\ line absorption in the $y$-$z$ plane, as seen in transit, simulated using their 3D atmospheric hydrodynamic escape model. This distribution is approximately spherical symmetric.
Similarly, using the 3D atmospheric hydrodynamic model, \citet{Wang2021a} showed that the absorption of the He 10830\AA~line by the atmosphere of WASP-69b, including regions outside the RL, is approximately spherically symmetric.

We repeat the steps in Sections~\ref{sec:observation} and \ref{sec:normalization} to compare the simulated transmission spectra with the NUV and optical observations.
Although the atmosphere in the substellar direction becomes more extended with the tidal potential, this effect is offset by the suppression due to the substellar-to-terminator profile mapping described above.
As a result, compared to the results of case A that do not include the tidal effect, the line center transit depths of most features in case B are very similar.
This is reflected in Table~\ref{tab:summary}, where the \ion{Ca}{2} line core transit depths of case A and B are very close.
Because of the higher outflow velocity of case B, the absorption profiles are broader.
Therefore, although the line center transit depths of \ion{Mg}{2} $\lambda$2796 are similar, the broader line width causes the larger case B transit depths in the Mg II lines listed in Table~\ref{tab:summary}.

The Balmer lines constitute an exception.
As a result of the higher neutral hydrogen column density in case B, less stellar \Lya\ radiation reaches the atmosphere near the RL.
Furthermore, the lower atmospheric temperature within the RL leads to less collisional excitation induced \Lya\ radiation.
The combination of these two factors causes lower \Lya\ intensity and thus weaker Balmer line absorption.

\section{A parameter Space Investigation}
\label{sec:compare}

Although the \WA\ system is one of the best-observed exoplanet systems, there are still uncertainties in the measured system parameters.
We adjust several parameters to explore their effects on the atmospheric properties and transmission spectrum, and look for models that can reproduce all the observed spectral features within the range of their uncertainties. 
In the following, we first describe the quantities that we adjusted to obtain a better match to the observations, and then conduct a parameter space investigation around the best-fit model and evaluate the impact of each factor.

\subsection{Matching the observations}
\label{sec:adjust}

We first adjust surface gravity, based on the suggestion of \citet{GM2019} that the mass-loss rates and absorption line profiles of UHJs are sensitive to surface gravity. \WA\ is closer to the Roche limit than KELT-9b, the subject of their study, and due to the prevalence of RLOF, changes to surface gravity should play a greater role on \WA. Therefore, the mass loss from \WA\ is expected to be even more sensitive to surface gravity.

Compared to directly observable quantities such as the rotational period, transit duration, and the planet-to-star radius ratio, the uncertainties in the radius and mass of the planet and the star are much larger. Accordingly, when we vary planetary properties to discuss their effect on the model, we keep the rotational period, transit duration, and radius ratio unchanged by varying the radius and mass of the planet and the star as well as the orbital separation. Our intent is to explore the effect of these parameters on the transit fit, and to explore if the uncertainties in the system parameters for WASP-121b allow for a combination of parameters that lead to larger transit depths.

Because case B underestimates the observed transit depths, we search for input parameters that produce a lower planetary surface gravity.
First, we discuss a case in which we keep the planet's radius unchanged but decrease the planet's mass by 1$\sigma$ (5\%).
The surface gravity of the planet in this case is 1.4$\sigma$ smaller than the canonical value.
The physical parameters used by the model are listed in the lower part of Table~\ref{tab:derived}. Lowering the gravity of the hydrodynamic model of the upper atmosphere further can also be achieved by using higher temperatures for our lower and middle atmospheric model, which is retrieved from secondary eclipse observations and is also subject to large uncertainty.
The 1$\sigma$ uncertainty of the model-dependent MCMC temperature retrieval provided by \citet{Evans2019} is $\sim$200 K.  
Here, we also uniformly increase the lower and middle atmospheric temperature by 350 K, as shown by the blue solid line in Figure~\ref{fig:Tcomb}.
The resulting larger atmospheric scale height effectively increases the altitude of the 1~$\mu$bar level.

Even with these measures in place to reduce the surface gravity and to increase the outflow velocity and mass-loss rate, the model with a solar abundance of iron falls short of the observed NUV transit depths. For a better agreement with NUV absorption, we increase the abundance of iron by a factor of 4 in the model, while retaining solar abundances for the other elements.
Because Fe has almost no impact on the thermal balance of the atmosphere for models with tidal potential (see Section~\ref{sec:atmosph-prop}), this change improves the agreement of the model with the NUV transmission spectrum without significantly changing atmospheric structure.
A supersolar metal abundance would be in agreement with the conclusions of \citet{Evans2018} based on optical-NIR HST transmission observation.  
However, it would be inconsistent with the claim by \citet{Gibson2022} based on the UVES/VLT optical band transmission spectrum that the Fe/Mg abundance ratio is subsolar.

The model with the above adjustments is referred to as the case C model, and the corresponding transit depths for this model are also listed in Table~\ref{tab:summary}.
Except for the Balmer lines whose core depth and width are overestimated, the simulated NUV and optical transmission spectra of case C are agree with the observations.

\begin{figure}
  \begin{center}
    \includegraphics[width=0.5\textwidth]{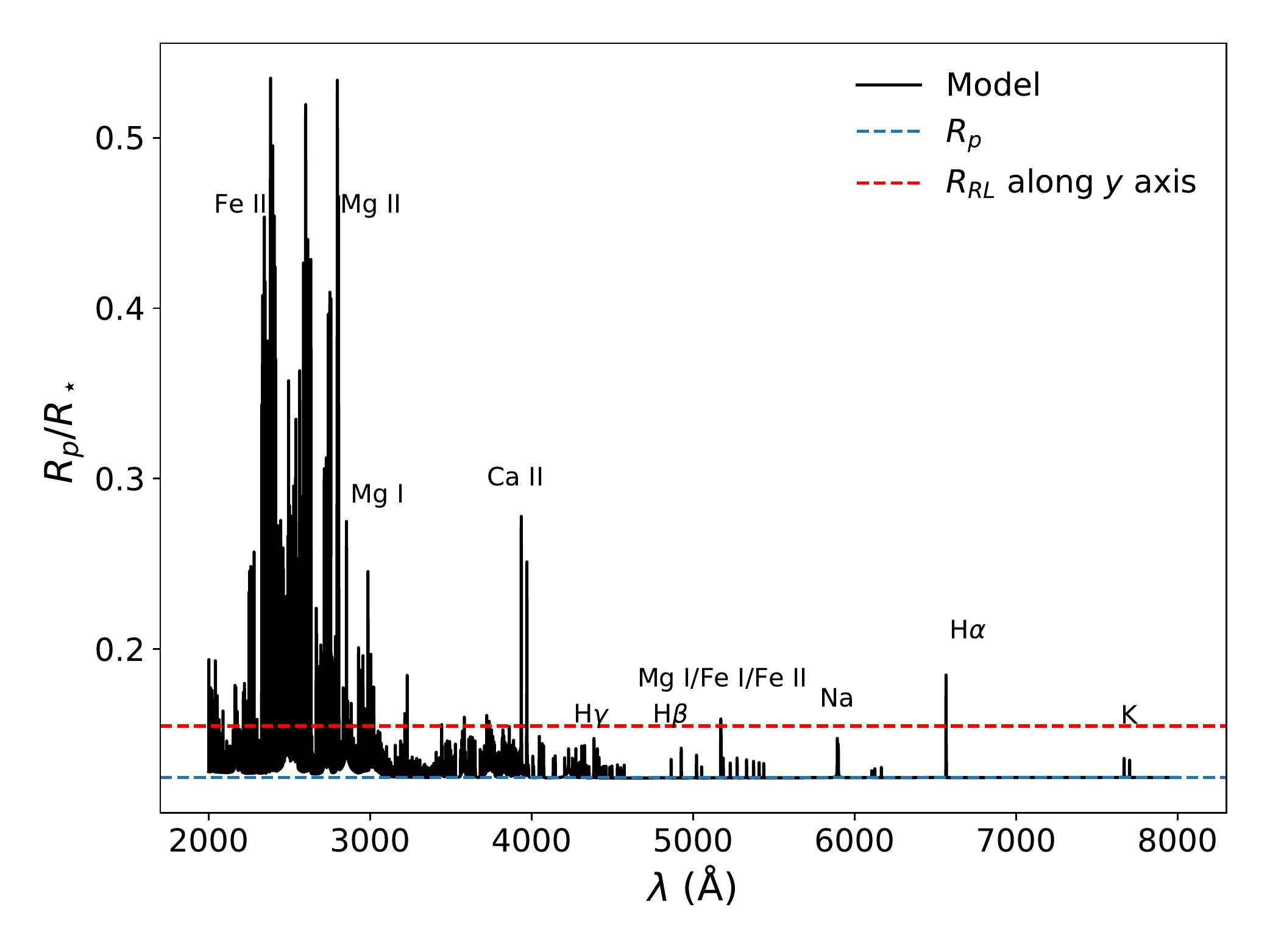}
    \caption{Simulated transmission spectrum of case D. }
    \label{fig:full_tide}
  \end{center}
\end{figure}

\begin{figure}
  \begin{center}
    \includegraphics[width=0.5\textwidth]{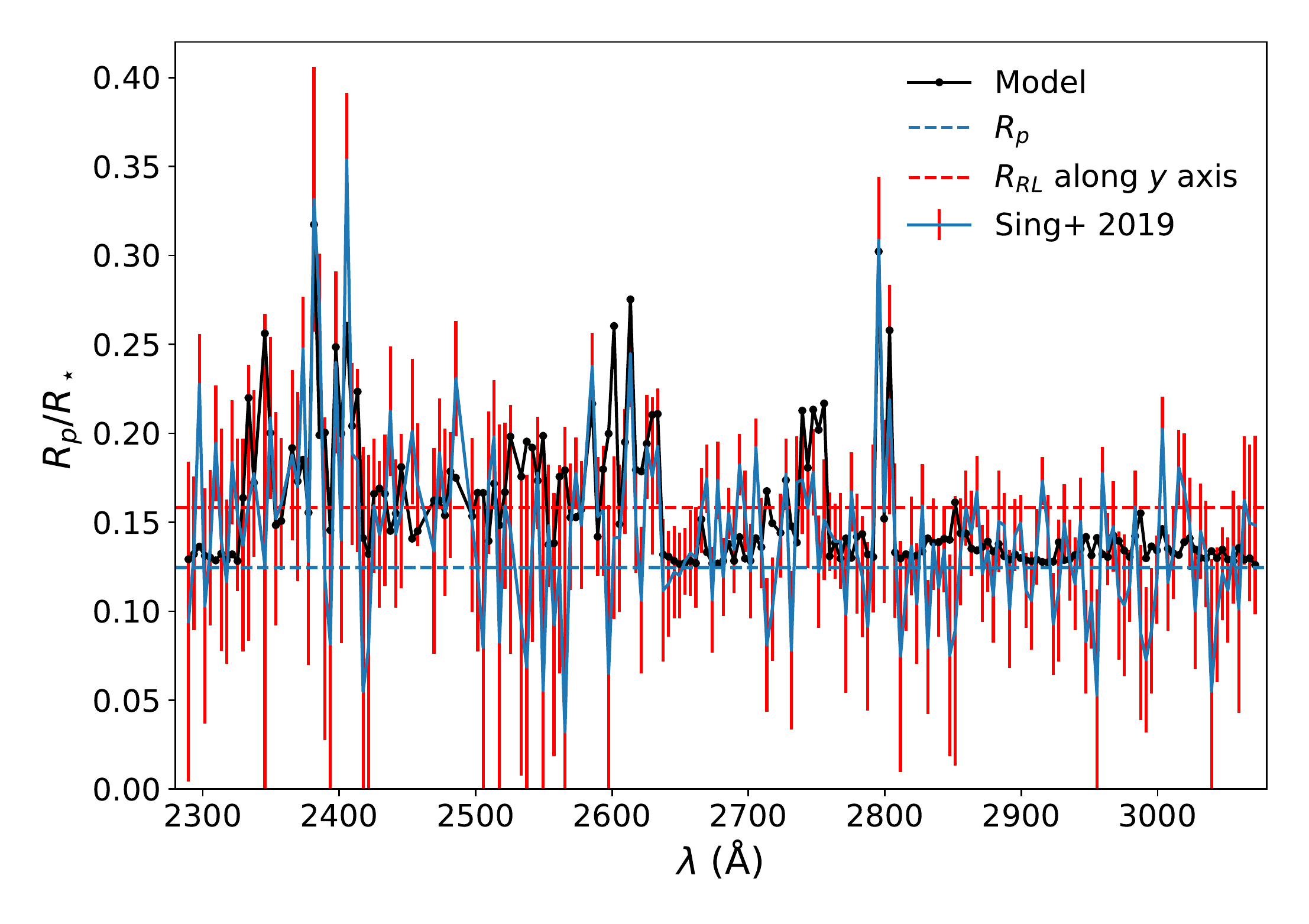}
    \caption{Binned simulated NUV transmission spectrum of case D compared to HST observation \citep{Sing2019}. $\chi^2/N=1.20$.}
    \label{fig:NUV_tide}
  \end{center}
\end{figure}

\begin{figure*}
  \begin{center}
    \includegraphics[width=0.9\textwidth]{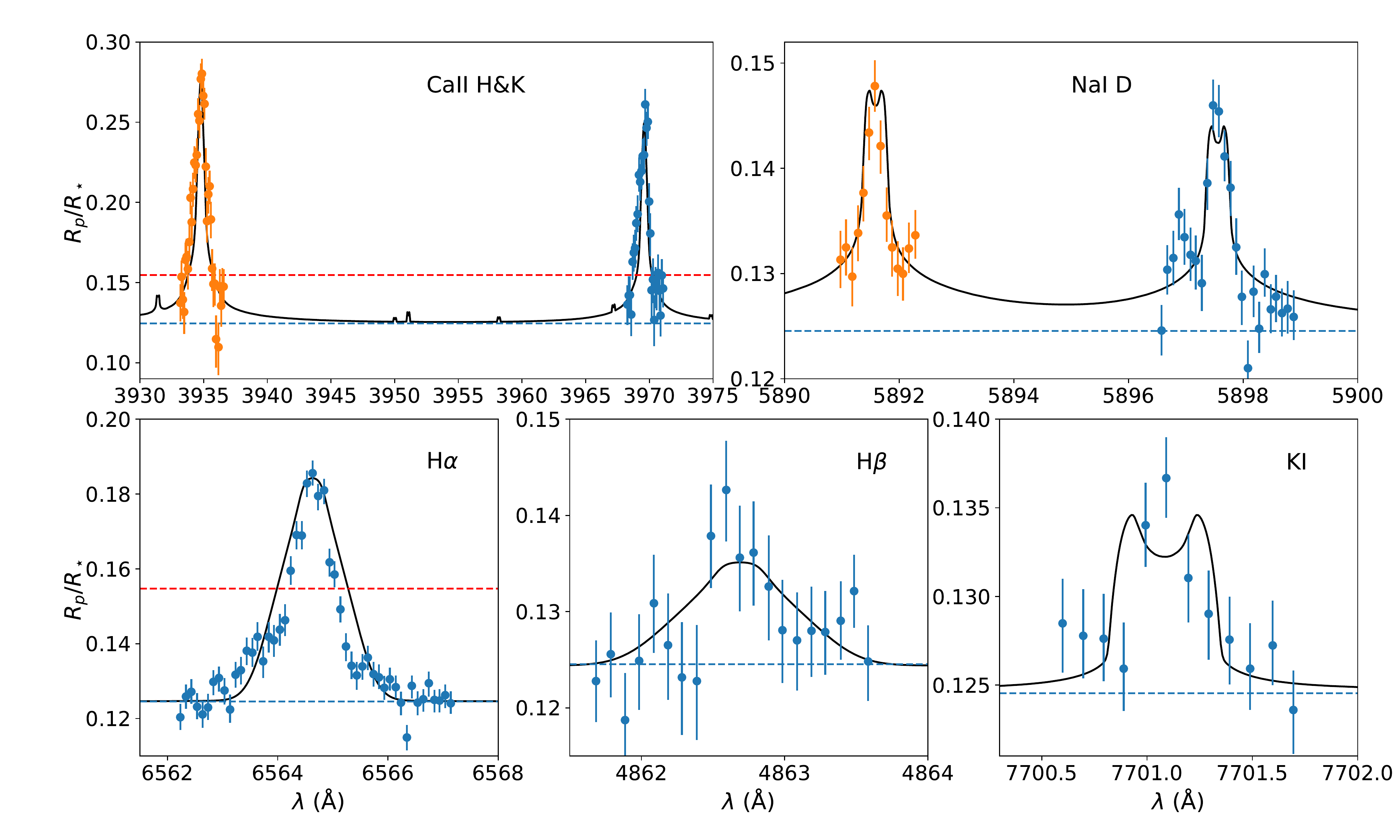}
    \caption{Comparison of optical features observed by \citet{Borsa2021} with simulated profiles of case D.}
    \label{fig:optical_tide}
  \end{center}
\end{figure*}

Other relevant uncertainties include the intensity of the stellar XUV flux used in the hydrodynamic model and the \Lya\ flux used in the \Lya\ radiative transfer calculation.
\citet{Huang2017} showed that the Balmer lines are much more sensitive to the stellar XUV and \Lya\ flux than, say, the Na lines because of their dependence on \Lya\ intensity inside the atmosphere. To reduce the simulated Balmer line absorption in case C, we reduce the stellar flux at wavelengths shorter than 1700 \AA\ to 74\% of the reference spectrum, shown as the magenta line in Figure~\ref{fig:Spec}, and weaken the stellar \Lya\ radiation intensity to 35\% of the canonical value in the case C model. The new model with these adjustments will be referred to as the case D model. The transmission spectrum for case D, shown in Figures~\ref{fig:full_tide}--\ref{fig:optical_tide}, can broadly reproduce all of the observed spectral features that we consider in this study.

\subsection{Parameter variations}
\label{sec:impact}

On the basis of the good fit to the observations, together with the fact that all the input parameters to the model are broadly consistent with their uncertainties, we select the case D model as the best-fit model.
In the following, we present some other simulations that explore the parameter space more fully to illustrate the effect of the uncertainties on the interpretation of the observations. The input parameters and results of these simulations are summarized in Table~\ref{tab:summary}. 

For the purposes of this exercise, we create case E and F models that apply higher and lower stellar XUV fluxes, respectively, than the case D model while the other planetary system parameters remain the same.  In these models, the stellar \Lya\ fluxes are reduced/increased accordingly to maintain a good match to the Balmer lines. The purpose of these models is to illustrate the dependency of the transit depths on the modeled stellar XUV spectrum.
Model G also has the same planet parameters as the best-fit case D, but a spherically symmetric gravitational potential is applied.
Model H also uses spherically symmetric potential but in this model, the planet mass is 3$\sigma$ lower than the canonical value and the radius is larger by 3$\sigma$.  The lower atmospheric temperature in model H is also increased by 1000~K to raise the altitude at which the atmospheric pressure is 1~$\mu bar$. The purpose of models G and H is to investigate whether a hotter model with lower surface gravity could match the observations without RLOF.
Case I is comparable to case D, but the planet mass is only reduced by 0.5$\sigma$.
In this case, the lower atmospheric temperature has to be increased to match the observed features of the metal ions.
Cases J and K assume a larger planet radius by 0.5$\sigma$ with the same planet mass as in case D, along with adjustments to the lower and middle atmosphere temperature to fit observations.

Comparing cases C--F, we learn that either reducing the XUV flux or reducing the \Lya\ flux can effectively decrease the transit depths in the Balmer lines.
Varying the stellar \Lya\ flux affects the proton and electron number densities (see Section~\ref{sec:best-fit}), but has a minimal impact on the distribution of metal species.
Ca and Mg are mostly doubly ionized, and Na and K are mostly singly ionized in the upper atmosphere focused in this work.  Reducing the ionization fraction due to a lower stellar XUV flux can increase the number densities of \ion{Ca}{2}, \ion{Mg}{2}, \ion{Na}{1}, and \ion{K}{1} in the upper atmosphere. However, its effect on the modeled metal lines is offset by the reduction in atmospheric expansion caused by the lower XUV flux.

Comparing the mass-loss rates of case B and C, for a planet close to the Roche limit, a small decrease in planetary surface gravity can significantly increase the mass-loss rate. 
Comparing the mass-loss rates of cases C--F or case J vs. K, we learn that the reduction in the atmosphere mass-loss rate is smaller than the reduction in the stellar XUV flux.
The impact of the surface gravity on the mass-loss rate is much less significant if the tidal potential is not considered, as shown by cases G and H.
It is clear that without considering the tidal potential, the metal species in the atmosphere cannot extend to the altitudes that are probed by the observation even if we significantly reduce the planet's surface gravity.
This is in line with the results of \citet{Koskinen2022} who pointed out that for gas giants very close to the host star, the main mechanism of atmospheric loss is Roche Lobe overflow.
For \WA, heating of the middle atmosphere by the stellar bolometric flux is an important additional source of energy for atmospheric escape. This also highlights the importance of carefully treating lower and middle atmosphere in atmospheric escape models for such systems.

Although the mass-loss rate in case A is over an order of magnitude lower than in case D, the model can still reproduce the observed \Ha\ transit profile if we reduce the stellar XUV flux or \Lya\ flux. The dependency of the H Balmer line transit depths on the stellar Ly$\alpha$ and XUV flux makes these lines an ambiguous probe of mass loss from exoplanetary atmospheres. In general, the results indicate that a single transmission feature cannot reliably constrain the properties of escaping atmospheres because of the number of degenerate free parameters in the atmosphere models. By combining many high-quality spectral features with various absorption depths, wavelengths, and absorbers, models can provide useful constraints on the atmospheric properties of UHJ, but care still needs to be taken to avoid drawing overly strong conclusions due to the neglect of possible parameter degeneracies. At the same time, while models D--F, and I--K can all deliver a reasonable match to all observed features, the mass-loss rates predicted by these models vary only by a factor of about three. This is because sufficient density and velocity at high altitudes around the planet are required to match the transit observations of the metal lines, regardless of the parameter degeneracies, and the mass-loss rate is based on the density and velocity profiles. Therefore, it is not surprising that the best-fit models yield similar values for the mass-loss rate.

\section{Atmospheric Properties and Spectral Features of the Best-Fit Model}
\label{sec:best-fit}

In this section, we discuss the atmospheric properties and dominant physical processes that control the upper atmosphere of WASP-121b based on our results, relying on the case D model. We also discuss the caveat of using a 1-D model to simulate an atmosphere that undergoes Roche lobe overflow, the possible impact of stellar winds, and point to directions for future research.

\subsection{Atmospheric Properties}
\label{sec:atmosph-prop}

\begin{figure}
  \begin{center}
    \includegraphics[width=0.5\textwidth]{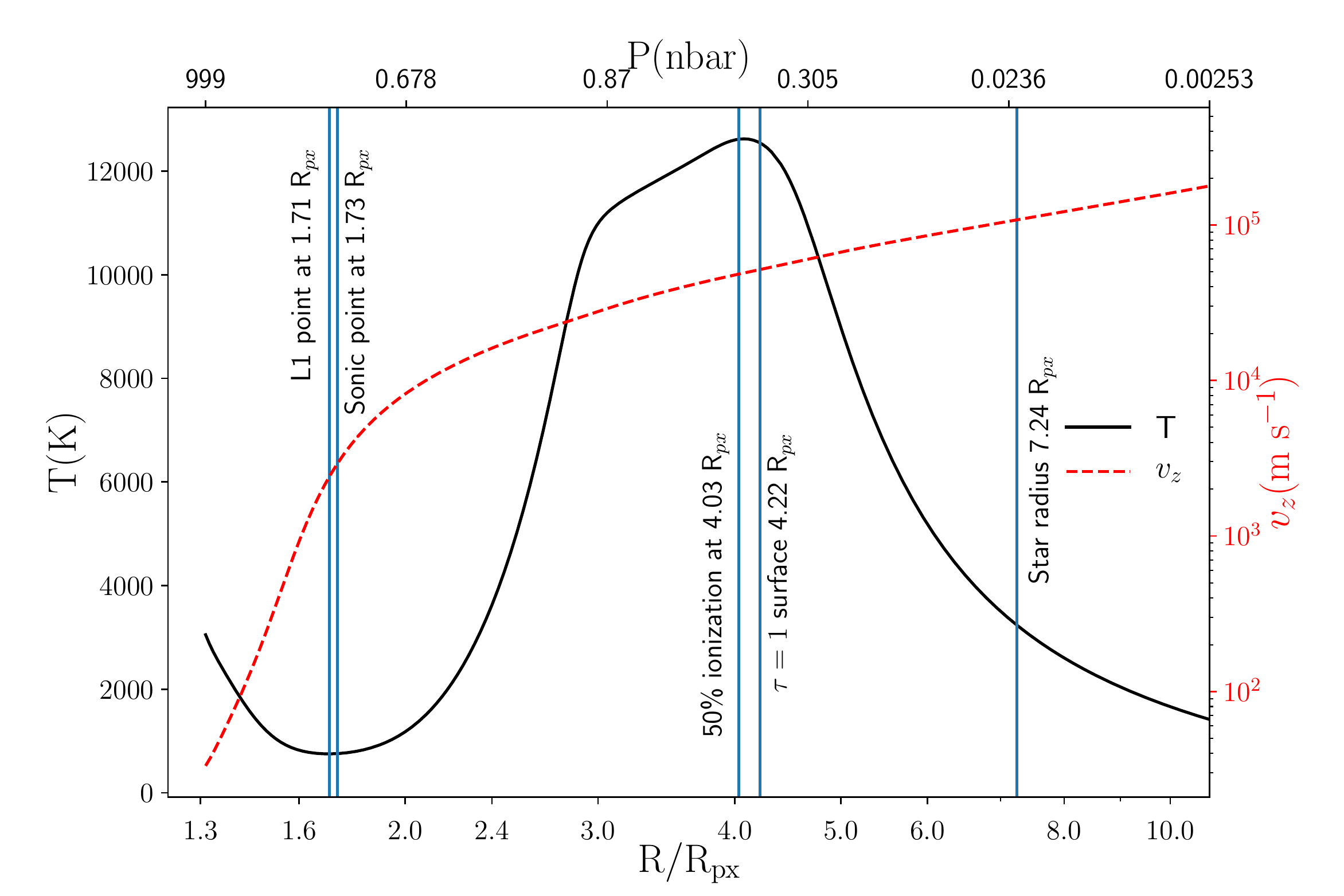}
    \caption{Temperature and radial velocity along the substellar stream line simulated by the hydrodynamic model for case D.}
    \label{fig:Tvr_best}
  \end{center}
\end{figure}

The surface gravity in our case D model is 1.4$\sigma$ lower than the value based on the \citet{Delrez2016} system parameters. \citet{Bourrier2020}, however, reported measurements of planetary parameters for WASP-121b, with $M_p=1.157\pm0.07 ~M_J$ and $R_p=1.773^{+0.041}_{-0.033}~R_J$.
The surface gravity of our best-fit model is only 2.5\% (or 0.3$\sigma$) lower than the surface gravity based on this set of new parameters.
Using the estimated upper limit on the mass-loss rate based on model D, we conclude that the planet can retain its atmosphere for longer than the estimated 1.5$\pm$1 Gyr age of the star \citep{Delrez2016}.

The temperature and bulk outflow velocity along the substellar streamline are shown in Figure~\ref{fig:Tvr_best}.
Within the RL, the temperature decreases significantly with altitude due to the effective adiabatic cooling caused by the strong outflow.
In agreement with the analytical estimate given by \citet{Lubow1975}, the atmospheric escape velocity exceeds the sound speed slightly above the RL.
In this case, the whole thermosphere (i.e., the region primarily heated by the stellar XUV radiation) is located outside of the RL.
In addition, because the effective gravity points towards the star and away from the planet above the RL, when the temperature rises rapidly, pressure may increase with radius. At the L1 point, the pressure of the atmosphere is $\sim 2.0$~nbar.

This atmospheric structure is noticeably different from the \WA\ predictions presented by \citet{Yan2021}.
In addition to the inclusion of a more comprehensive treatment of the chemistry and transport of metals and the inclusion of ionization and heating by excited-state H in our model, an important difference between the two models that drive the difference in atmospheric structure is the treatment of the bottom boundary conditions of the upper atmosphere and escape model.
In both studies, the pressure at the bottom boundary of the escape model is 1 $\mu$bar.
However, in \citet{Yan2021}, the radius corresponding to this pressure is set at the planet transit radius $R_p$.
In contrast, the radius of the 1 $\mu$bar level in our case D model is 1.46~$R_p$, due to the inclusion of lower and middle atmosphere structure as described in Sections~\ref{sec:boundary} and \ref{sec:atm-tide}.   

\begin{figure}
  \begin{center}
    \includegraphics[trim={0 0cm 0 1cm},clip,width=0.5\textwidth]{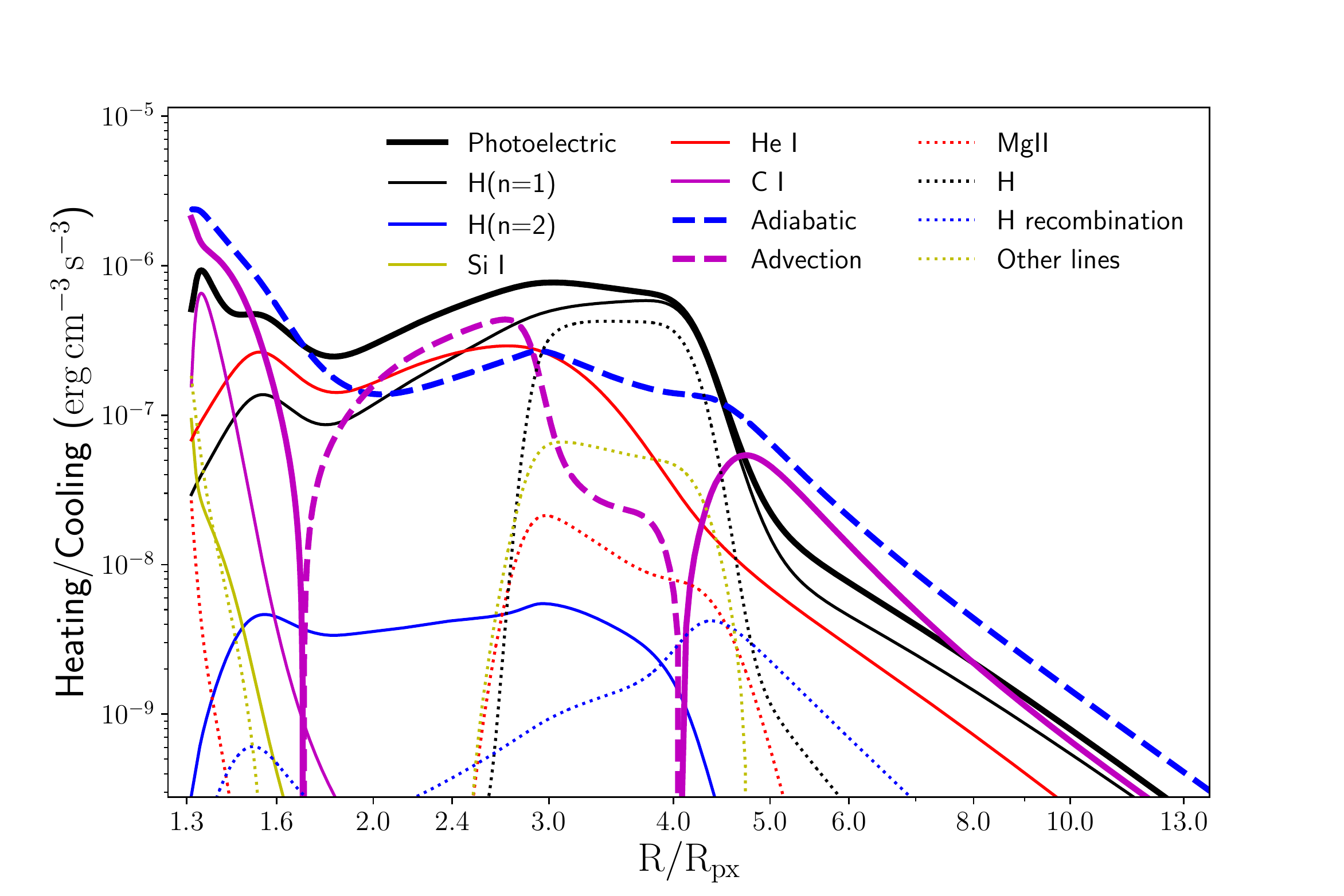}
    \caption{Rate of heating (solid line) and cooling (dashed or dotted line) process for case D.}
    \label{fig:Qr}
  \end{center}
\end{figure}

The rates of the major heating and cooling processes are shown in Figure~\ref{fig:Qr}.  From the top to the bottom of the upper atmosphere, the atomic species that contribute most to heating by photoionization are, in order, H, He, and C.
The higher mass-loss rate predicted by the models with the tidal force can transport more neutral H to the upper atmosphere and partly offset photoionization.
Because of absorption by neutral H atoms, the EUV radiation in such models is absorbed at a higher altitude compared to case A and thus the peak photoionization heating rate is reduced.

In contrast to the KELT-9b model of \citet{GM2019}, photoionization of excited state H is not the main heating source at any altitudes of the atmosphere.
The NUV flux of the A-type star KELT-9, which is an order of magnitude stronger than that of WASP-121, can partially explain this difference.
In addition, the different choices of \Lya\ redistribution function as described in Section~\ref{sec:Hn2} and illustrated by Figure~\ref{fig:Jbar} may also play a role.

Due to the high mass-loss rate, adiabatic cooling dominates the cooling rate of almost the entire simulated atmosphere.
Because of the steep changes in temperature, heating and cooling by advection that originate from the transport of hotter and colder fluid by the flow can also play an important role.
Although radiative cooling by metal lines is negligible in case D, its role in the case A model with a lower atmospheric mass-loss rate, as shown in Figure~\ref{fig:Qr_noT}, means that it can be more important for planets with less significant mass loss, for example, on KELT-9b.

\begin{figure}
  \begin{center}
    \includegraphics[width=0.5\textwidth]{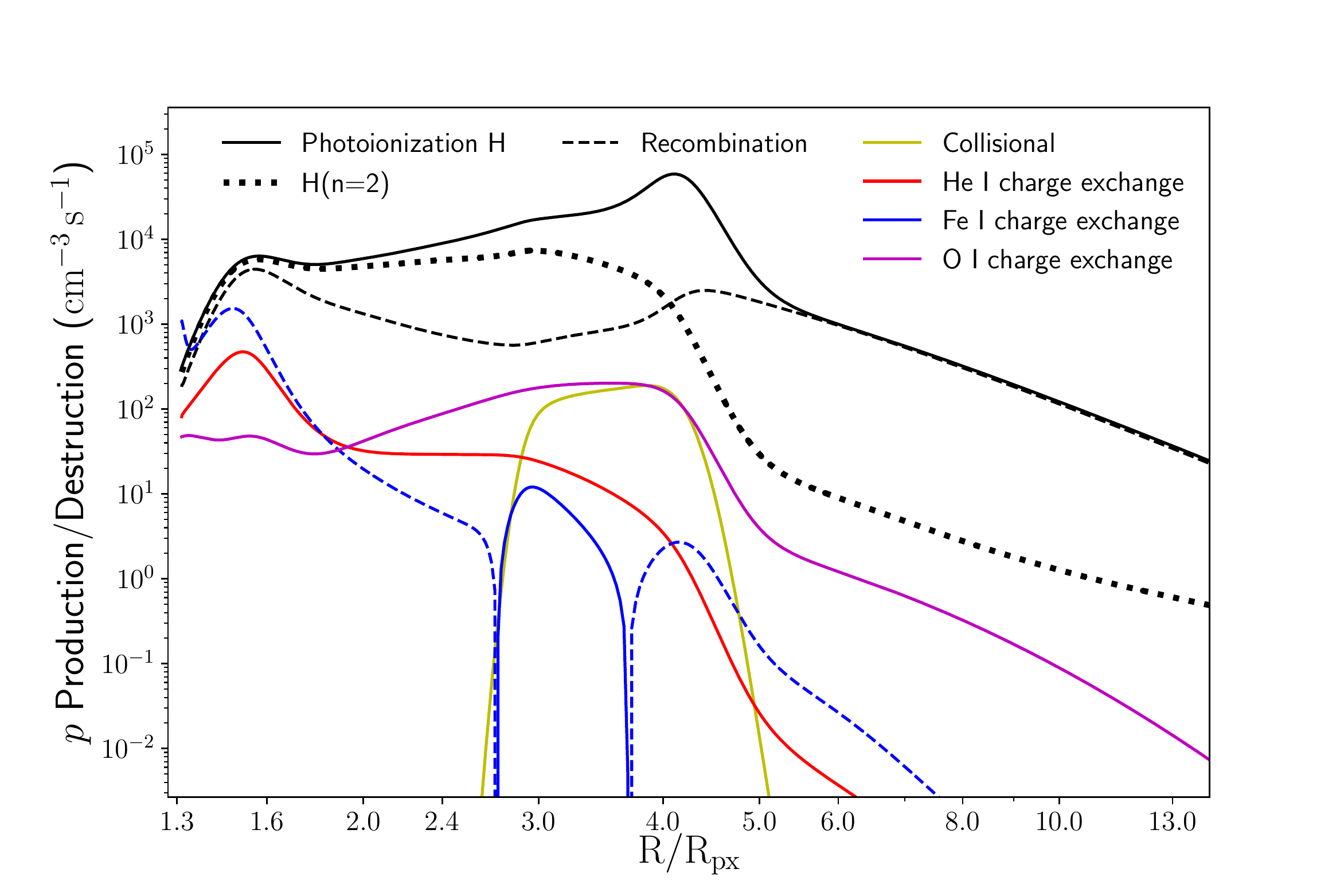}
    \caption{Rates of H ionization and recombination of case D.  Solid lines indicate processes that ionize H and dashed lines indicate processes that recombine H. As a component of the H photoionization, the rate of photoionization from H($n=2$) state is shown in the black dotted line.}
    \label{fig:ionize}
  \end{center}
\end{figure}

\begin{figure}
  \begin{center}
    \includegraphics[width=0.5\textwidth]{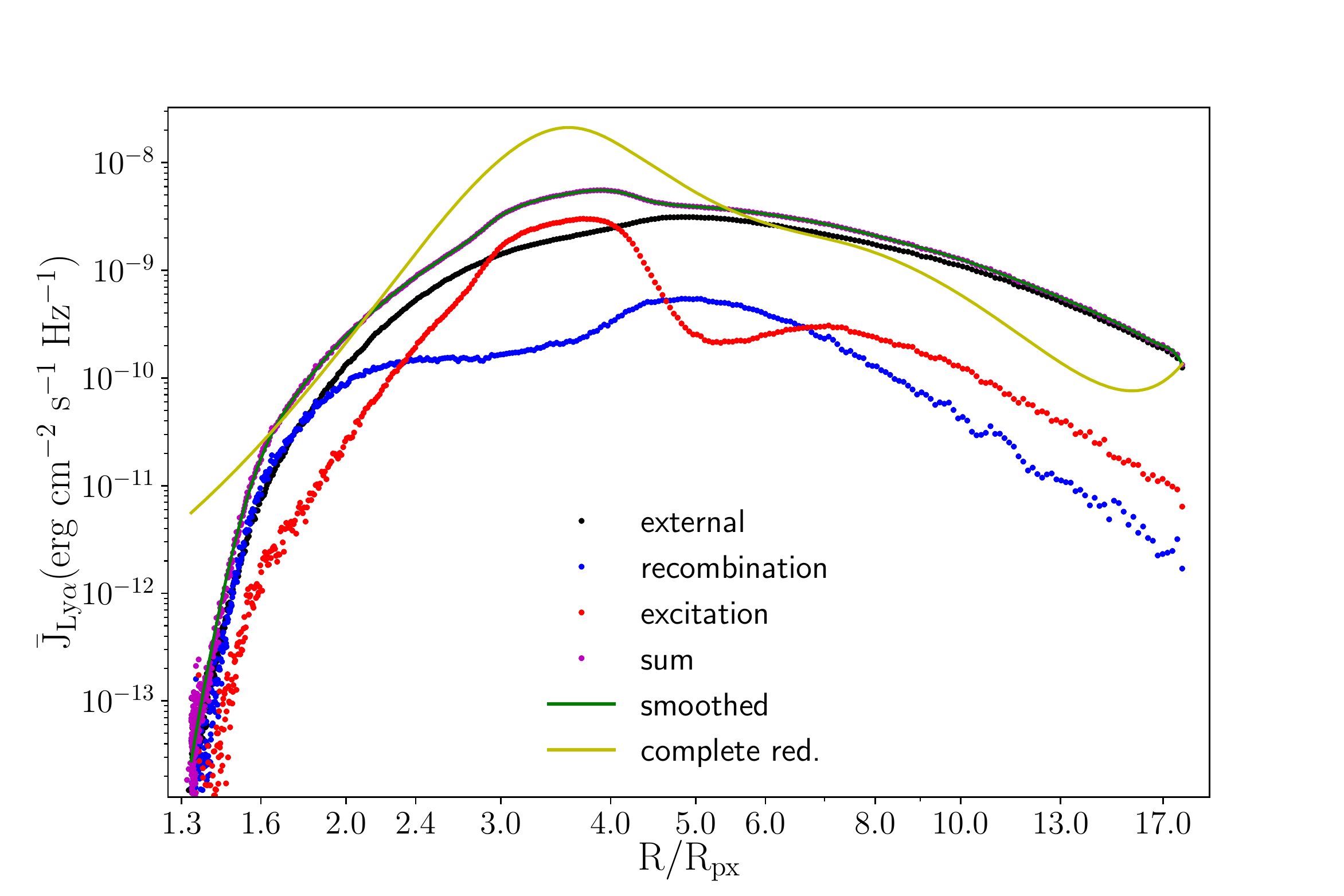}
    \caption{Line profile averaged \Lya\ mean intensity calculated with Monte Carlo radiative transfer for case D. The contribution originated from stellar irradiation, H recombination, and electron collisional excitation of H($1s$) are shown in black, blue, and red dotted lines, respectively. The sum of them is shown in the magenta dotted line.  Being smoothed using B-spline as mentioned in Section~\ref{sec:Hn2}, the profile of this overall \Lya\ mean intensity is shown as the green solid line. The result based on the complete frequency redistribution function, instead of the Hummer-IIB partial redistribution function, is shown by the yellow line.}
    \label{fig:Jbar}
  \end{center}
\end{figure}

The rates of the major proton production and destruction processes are shown in Figure~\ref{fig:ionize}.
Although absorption of starlight in the Balmer continuum is not an important heating mechanism in our model, it dominates proton production below 2$R_{px}$.
As shown in Figure~\ref{fig:Jbar}, stellar \Lya\ radiation is typically the main source of \Lya\ intensity near the base of the atmosphere.
As a result, the proton number density in this region is sensitive to the stellar \Lya\ flux.

\begin{figure}
  \begin{center}
    \includegraphics[width=0.5\textwidth]{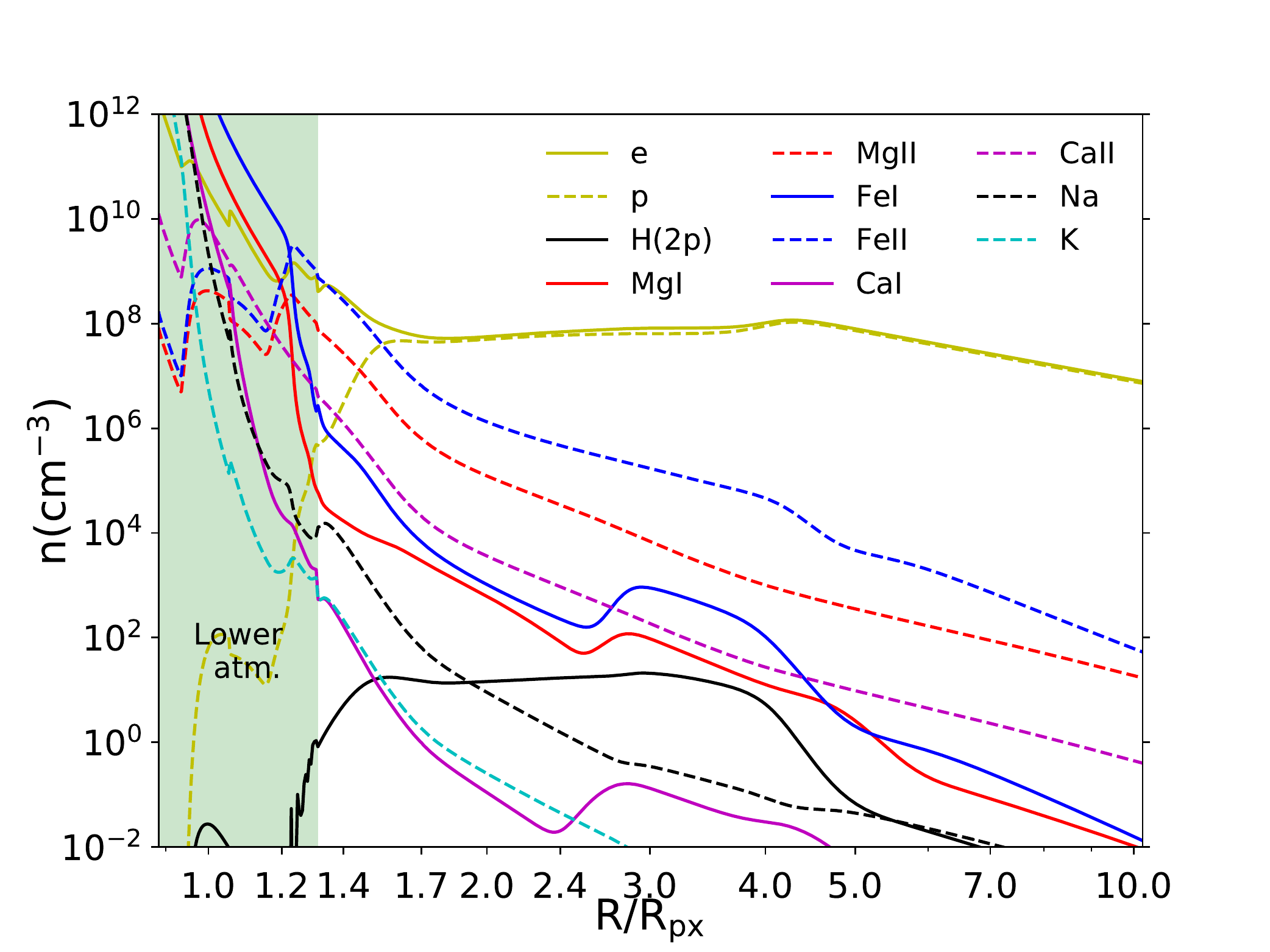}
    \caption{Number density distributions of major species along the substellar direction from both the case D upper atmosphere and the corresponding lower atmosphere model.}
    \label{fig:np_tide}
  \end{center}
\end{figure}

Proton charge exchange with Fe can be more important in the removal of H than recombination near the base of the hydrodynamic model.
\ion{Fe}{1} and \ion{Mg}{1} number densities increase with the sharp increase in temperature between 3 and 4$R_p$, as shown in Figure~\ref{fig:np_tide}.
The higher temperature increases the rates of charge exchange with hydrogen and drives the ionization fractions of Fe and Mg closer to the ionization fraction of H, which is more neutral.
It is worth noting that the charge exchange rate between Fe$^+$ and H used in the calculation originates from a cursory theoretical estimation.
Considering the large impact of this charge exchange rate on the model, experimental measurement or careful theoretical consideration of this rate would be helpful to the understanding of the planetary upper atmosphere.

\subsection{Stellar wind and diurnal temperature variations}

Strong stellar wind may modulate atmospheric outflow and shape the distribution of the escaping plasma away from the planet~\citep{Wang2018}.  Here we discuss some aspects of the potential impact of the stellar wind on \WA\ based on model D. 
We note that the stellar wind properties of WASP-121 are unconstrained by observations. Instead, we use an estimate of the stellar mass-loss rate and the scaling proposed by \citet{Preusse2005} to estimate the thermal, ram, and magnetic pressures of the wind at the orbital distance of WASP-121b.

\citet{Delrez2016} noted that WASP-121 has similar stellar type and RV jitter as $\tau$ Bootis A.  The Mount-Wilson S-index of $\tau$ Boo is 0.185 \citep{Mengel2016}, corresponding to a $\log10(R'_{\rm HK})$ of -4.70.  This larger activity index compared to WASP-121, which has a $\log10(R'_{\rm HK})$ of -4.81 \citep{Borsa2021}, indicates that $\tau$ Boo is more active than WASP-121.  As an upper limit, we use the stellar wind parameters of $\tau$ Boo \citep{Nicholson2016} to estimate its impact on the planetary atmosphere.  We have the stellar wind mass-loss rate of $2.3\times 10^{-12}\rm M_\odot yr^{-1}$ with a constant velocity of $u=220 \rm km~s^{-1}$.  This yields the plasma density of $n_p(0.02545\rm AU) = 2.2\times 10^6 \rm cm^{-3}$.  Assuming magnetic field strength varying as $r^{-2}$ and $B=0.03$ G at 0.049 AU from the host star, as the value at $\tau$ Boo b, we have $B(0.02545\rm AU)=$0.12 G.  Furthermore, we estimate the electron temperature by scaling as $T_p(0.02545\rm AU)=10^6~\rm K (r/0.049 AU)^{-0.7} = 1.6\times 10^6 ~\rm K$ and $T_e=T_p*2=3.2\times 10^6 ~\rm K$.  Combining these, the thermal pressure of the stellar wind at the distance of the planet $P_{th}=1.5\times 10^{-3} ~\rm \mu bar$, the ram pressure $P_{ram}=\rho (v_{wind}^2+v_{orbit}^2) = 3.5\times 10^{-3} ~\rm \mu bar$, and the magnetic pressure $P_B = 6\times 10^{-4} ~\rm \mu bar$.  Therefore, we have the total stellar wind pressure at the planet should be less than $5.6\times 10^{-3} ~\rm \mu bar$.

WASP-121b has a near-polar orbit \citep{Delrez2016} and is therefore subjected to the stellar wind in the polar direction along its orbit, an area that has been studied to a lesser extent.  In the absence of any information on the spatial structure of the stellar wind in systems like WASP-121b, we estimate the dependence of the stellar wind pressure on latitude based on observations of the solar wind by Ulysses \citep{Verscharen2021}.  During solar minimum, the solar wind velocity is higher in the polar direction compared to the equatorial direction, while the number density is lower in the polar direction.  \citet{Verscharen2021} suggested that the momentum flux ($n_p m_p r^2 v^2$) and sonic Mach number ($v^2/T$) exhibit little latitude variation during solar minimum.  Therefore, it is expected that the ram pressure and thermal pressure do not change dramatically with solar latitude.

The total stellar wind pressure balances the pressure of planetary wind in model D at 7.5 $R_{px}$.  In comparison, as we noted before, the line cores of the \ion{Mg}{2} lines, which probe the highest altitudes, reach the effective radius of 4.3$R_p$.  At 4.3$R_{px}$ along the substellar direction, the total atmospheric pressure is $8.5\times 10^{-3} ~\rm \mu bar$. Additionally, the L1 point, which roughly coincides with the sonic point along the substellar streamline, is at about 1.7$R_{px}$.  Therefore, unless the magnetic pressure of the stellar wind of WASP-121 is significantly stronger, it is unlikely to have a significant impact on the dayside mass-loss rate estimated in this model. This is consistent with the recent conclusion by \citet{Mitani2022}. Along the terminator, the stellar wind would push material away from the planet and potentially enhance mass-loss rates, although its relative importance is limited by the relatively low stellar wind pressure, as the thermal escape rates are already high. The stellar wind can also shape the density and velocity distribution of the escaping material. Modeling this is beyond the scope of the current work.

Our globally averaged approach and approximations of the atmosphere under Roche lobe overflow based on a 1-D model ignore day-night variations of density and temperature in the atmosphere. Using a phase curve observed with the HST/WFC3 instrument in the near-IR, \citet{Mikal-Evans2022} inferred diurnal temperature variations in the lower atmosphere of \WA~in the pressure range of 1--100 mbar. The results are supported by more recent observations with the JWST NIRSpec instrument \citep{Mikal-Evans2023}. As shown in Figure~\ref{fig:Tcomb_notide}, the lower atmosphere temperature applied in this study is roughly comparable to their dayside temperature, although it lacks the curious steep increase with altitude and is therefore likely to be more representative of an average temperature profile. 

The much lower night side temperature observed in the near-IR is below the condensation temperature of Mg and Fe, suggesting that mineral clouds can form on the night side. As noted by \citet{Mikal-Evans2022}, however, the clear detection of escaping Mg II and Fe II, with absorption lines that also trace the middle and upper atmosphere, indicates that these species are present at the day-night terminator in the gas phase. This is possible, for example, if efficient mixing acts to prevent day-night cold trapping in the lower atmosphere. We point out that escape on the dayside and along the terminator must also be strong because the NUV spectrum cannot be explained without sufficient densities and velocities at high altitudes around the planet.

In addition, we note that the near-IR phase curve is probing the lower atmosphere whereas our focus is on the upper atmosphere. In the extended upper atmosphere, radiation can penetrate to the night side and leave a relatively small region around the anti-stellar point in the shadow. In addition, WASP-121b is very close to the host star and stellar rays incident on it are not all parallel. The maximum angle of rays incident on the planet's terminator is about 13--17$^{\circ}$ from the parallel direction. The diffuse radiation field, together with the extent of the upper atmosphere, can maintain ionization and heating along the terminator and on the night side. Should the nightside mass-loss rate be reduced, however, the extreme assumption of no mass loss from the night side would lead to a reduction of the global mass-loss rate by a factor of two. Since our radiation field is globally averaged, this geometrical factor is already partly accounted for. Beyond that, the mass-loss rate on the dayside can be further reduced by day-night winds that redirect flows and can extract energy available to power mass loss, although we suspect that the very strong effect of RLOF on the dayside of WASP-121b makes this of limited importance.

\subsection{Spectral Features}
\label{sec:spectral-features}

\begin{figure}
  \begin{center}
    \includegraphics[width=0.5\textwidth]{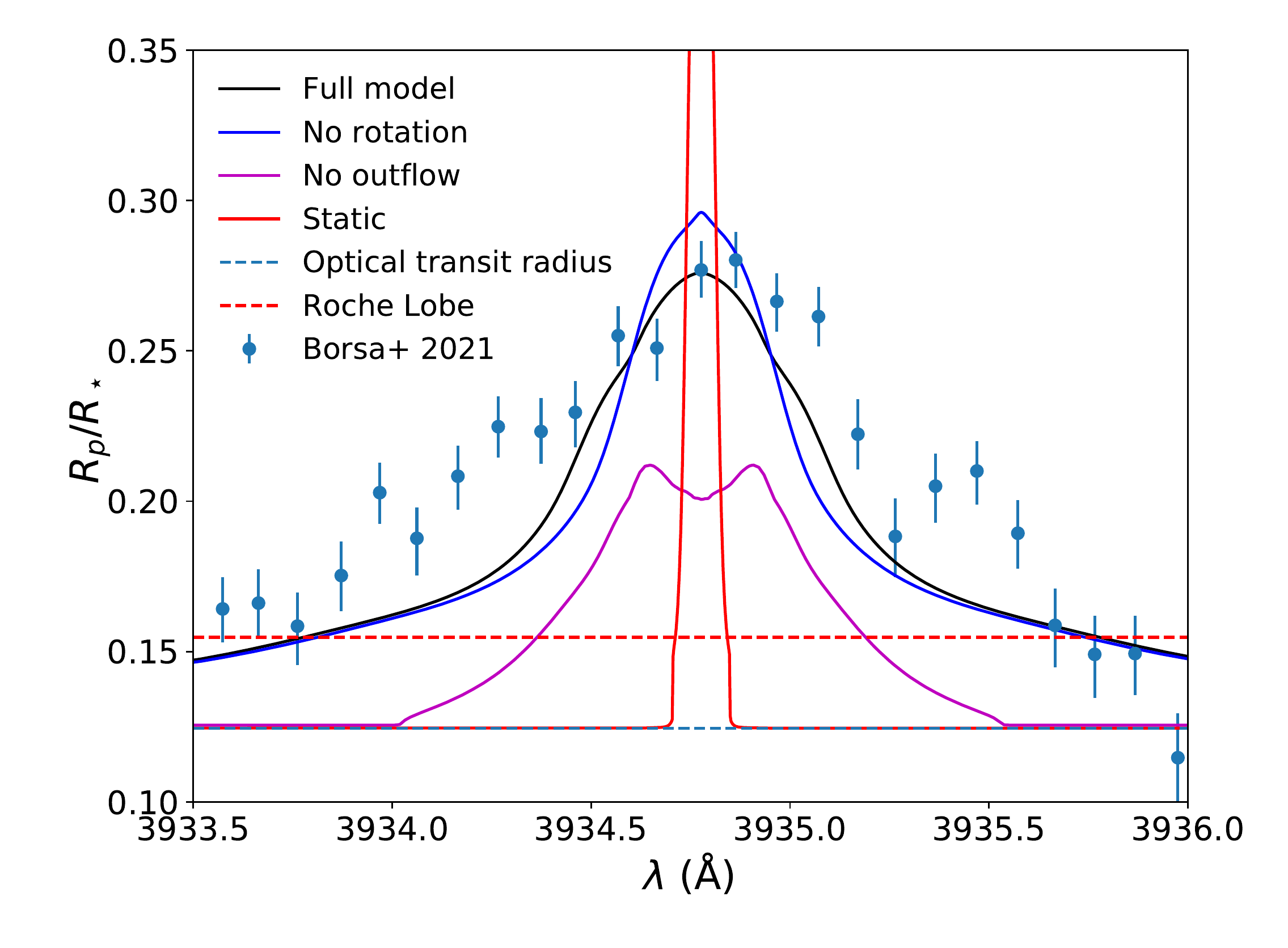}
    \caption{Comparison of \ion{Ca}{2} K$\lambda$3934\AArm\ line profiles given by different treatments for velocity broadening caused by atmospheric outflow and planet rotation.  The black solid line shows the enlarged profile of the same best-fit model as in Figure~\ref{fig:optical_tide}.  The red solid line shows the line profile that only includes thermal broadening.  In addition to thermal broadening, the magenta and blue solid lines show the line profile that also includes the effect of rotation and outflow velocity, respectively.}
    \label{fig:CaIIK}
  \end{center}
\end{figure}

As we have shown, velocity broadening due to planetary rotation and escape makes important contributions to the interpretation of the observations. Figure~\ref{fig:CaIIK} illustrates the impact of line broadening caused due to the atmospheric bulk outflow velocity and the rotation of the planet, which we discussed in Section~\ref{sec:broad}. The red solid line shows the line profile based on the best-fit model with thermal and natural broadening only.
It is only about 1/10 of the width of the observation, which indicates that thermal and natural broadening alone cannot explain the width of the observed line profile.
When rotation is included without radial outflow, the line profile exhibits the double-peaked structure described in Section~\ref{sec:broad}.
In this case, because absorption at a given wavelength originates in narrow strips, all at effectively different line-of-sight velocities, shown in Figure~\ref{fig:sector}, the transit depth is shallower due to the relatively small area of the strips.
Introducing broadening due to the outflow velocity, the line profile is closer to the observed line profile.

When some of the key number density profiles along the substellar direction of the case D model shown in Figure~\ref{fig:np_tide} are compared to those of the case A model shown in Figure~\ref{fig:np_noT}, the excited state H(2$p$) profile in the case D model has a lower peak number density and a shallower decrease with radius.
This means that \WA\ has an extended excited-state hydrogen distribution, which is marginally optically thick at \Ha.
This coincides with the physical image reflected by the observed large line center absorption depths ratio between the Balmer lines, which are close to the ratio of their spontaneous transition rates.
In case D, because the absorption originates from material at higher altitudes and with higher outflow velocity, the velocity broadening leads to a feature that is broader than that of case A and somewhat broader than the observed line profile.
This difference, combined with the fact that the model slightly underestimates the absorption depth at \Hb\ may imply that the slope of the H(2$p$) number density profile should be slightly steeper than it is in the case D model.

\citet{Sing2019} did not detect absorption by \ion{Mg}{1}.
However, as shown in Figure~\ref{fig:full_tide}, the \ion{Mg}{1} $\lambda$2853\AA\ line is clearly visible with a line center transit depth of $R_p/R_* =$~0.275.
Compared to the \ion{Mg}{2} lines, the absorption by \ion{Mg}{1} originates from a region at lower altitudes where the outflow velocity and rotational velocity are lower, and \ion{Mg}{1} $\lambda$2853\AA\ absorption has a much narrower line width.
As a result, after binning to 4 \AA\ wavelength band, Figure~\ref{fig:NUV_tide} shows that the non-detection of \ion{Mg}{1} $\lambda$2853\AA\ line is not statistically significant.

Comparing the simulated NUV spectra with observations, the biggest discrepancy is that the model fails to explain the absorption signal observed at 2438\AA\ and 2485\AA.
\citet{Sing2019} suggested the absorption feature at 2485 \AA\ may be due to the \ion{Fe}{1} $\lambda$2484\AA\ line.
The \ion{Fe}{1} $\lambda$2484\AA\ line, which is weaker than the \ion{Mg}{1} $\lambda$2853\AA~line, however, is not strong or broad enough to match the observations with a bin width of 4 \AA.

\begin{figure}
  \begin{center}
    \includegraphics[width=0.5\textwidth]{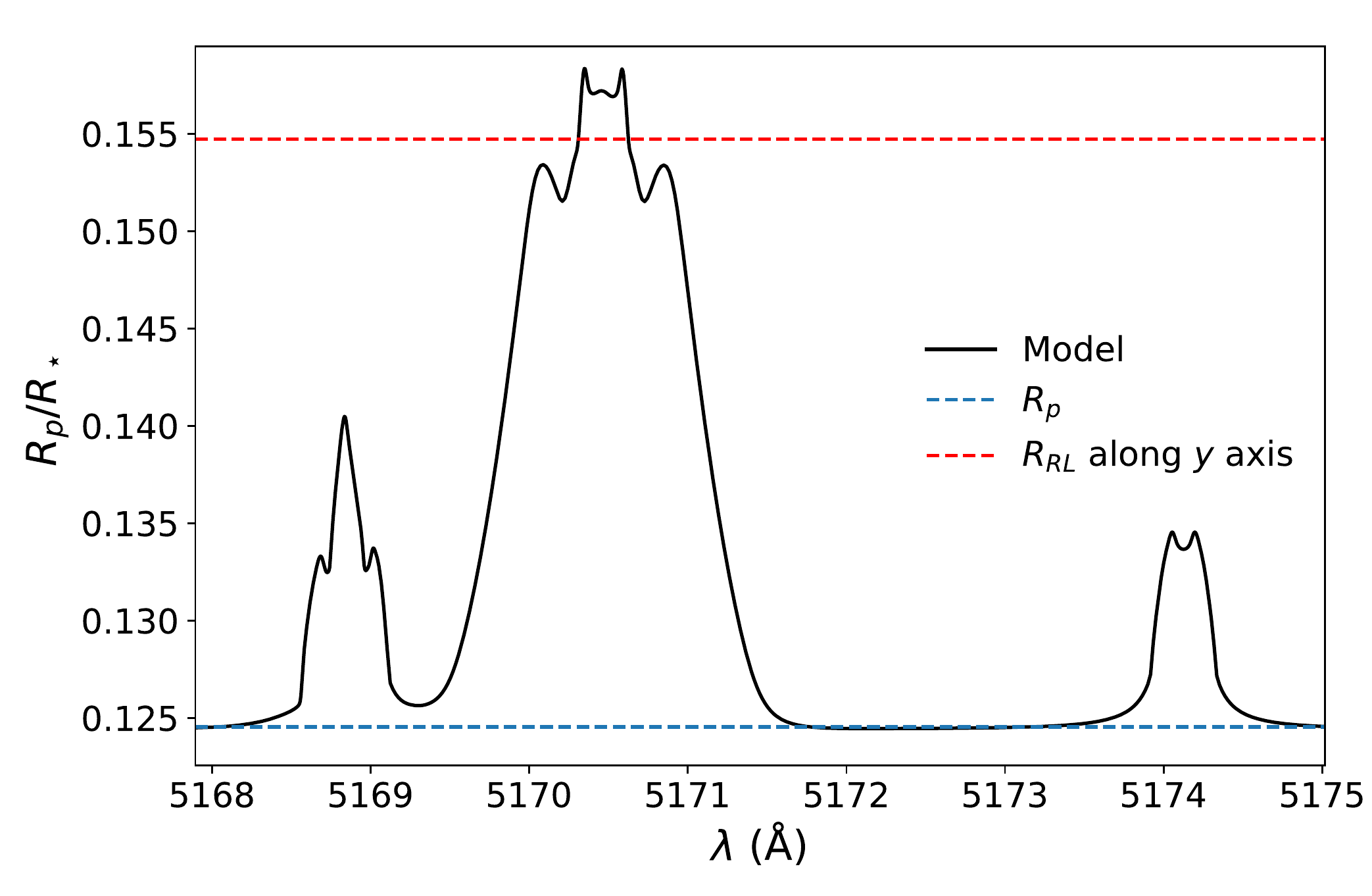}
    \caption{Line profile of \ion{Fe}{2} $\lambda$5170\AA, \ion{Fe}{1} $\lambda$5169\AA, and \ion{Mg}{1} $\lambda$5169\AA, $\lambda$5174\AA\ for case D.}
    \label{fig:FeIIoptical}
  \end{center}
\end{figure}

Other notable features in the simulated spectrum are the \ion{Fe}{2} $\lambda$5170\AA\ ($3d^54s^2$ $^6\mathrm{S} \to 3d^64p$ $^6\mathrm{P}^o$) and the \ion{Fe}{1} $\lambda$5169\AA \ ($3d^74s$ $^3\mathrm{F} \to 3d^64s4p$ $^3\mathrm{D}^o$) lines shown in Figure~\ref{fig:FeIIoptical}, positioned between the \ion{Mg}{1} triplets.
In particular, the line center transit depth of the \ion{Fe}{2} $\lambda$5170\AA\ line reaches the RL in case D.
The lower levels of the two Fe transitions are at the energy of 2.9~eV for \ion{Fe}{2} and 1.5~eV for \ion{Fe}{1}.
The intensities of these two features are sensitive to the temperature profile and radiative intensity within the atmosphere.
For example, case J, where the stellar EUV flux is lower compared to case D, predicts a much weaker \ion{Fe}{2} $\lambda$5170\AA\ line with a stronger \ion{Fe}{1} $\lambda$5169\AA\ line.
We note that \citet{Borsa2021} detected the \ion{Mg}{1} triplet while they did not mention the Fe I lines.
\citet{Asnodkar2022}, on the other hand, resolve individual \ion{Fe}{2} features in the optical transmission spectrum and include the \ion{Fe}{2} $\lambda$5170\AArm\ line displayed in Figure~\ref{fig:FeIIoptical}, but their results are for KELT-9b.
If our model significantly overestimates the absorption by these two Fe I and II lines, the most likely explanation is our assumption that the energy levels of Fe satisfy the Boltzmann distribution. If these lines remain undetected, this assumption should be adjusted in future work.

\section{Conclusions}

In this work, we present a hydrodynamic escape model of the upper atmosphere that includes excited-state H and metals such as Mg and Fe to interpret the NUV and optical transit observations of the ultra-hot Jupiter WASP-121b. The density profiles predicted by the model are combined with density profiles from a photochemical model for the lower and middle atmosphere to provide a consistent, one-dimensional, globally averaged model for the whole atmosphere. Based on the atmospheric structure in this model and adjusting some of the free parameters in the model, we can broadly reproduce the observed NUV spectrum and some of the key optical absorption lines of WASP 121b within uncertainties if we enhance the densities and velocities of the escaping material based on Roche lobe overflow. This allows us to place firmer constraints on the mass-loss rate from one of the most extreme hot Jupiters among the known transiting planets.

As expected, the model demonstrates the importance of the high mass-loss rate caused by Roche lobe (RL) overflow for this extremely close-in UHJ.  In particular, the deep NUV \ion{Fe}{2} transit depths and the broad \ion{Ca}{2} line widths are difficult to explain with thermal broadening alone or without invoking RL overflow. On the other hand, the Na D and K lines probe the atmosphere below the Roche lobe and are not good tracers of escape. At the same time, the transit depths in the H$\alpha$ and H$\beta$ lines are sensitive to the incident stellar radiation field and do not generally provide an unambiguous probe of the properties of the escaping material.

Our model demonstrates that careful consideration of the lower and middle atmosphere that provides the lower boundary conditions for the escape model is important for interpreting the observations and correctly predicting mass-loss rates. For planets such as WASP-121b that lie close to the fundamental stability limit for rapid Roche lobe overflow of the middle atmosphere \citep{Koskinen2022}, this is especially important. Here, we also show that the mass-loss rate from planets like WASP-121b is very sensitive to surface gravity.

According to the best-fit case D model, photoionization of excited H($n$=2) plays an important role in determining the overall ionization fraction of H, especially at radii less than about 2 R$_p$.
In contrast to recent work on KELT-9b by \citet{GM2019}, we do not find it to be an important heating mechanism in the atmosphere of WASP-121b. On average, the energy input per ionization event is much lower than, for example, from ionization of ground-state H. In addition, the upper atmospheric structure on WASP-121b differs significantly from KELT-9b due to strong cooling induced by rapid escape, and the input stellar spectrum is very different from KELT-9b. Rapid escape also means that adiabatic cooling overshadows radiative cooling by metal species on WASP-121b. We note, however, that radiative cooling by metal lines can be important on planets with a lower mass-loss rate, and may be constrained by similar observations of such planets.

In addition, our results highlight the fact that a single transmission feature cannot exclusively determine upper atmosphere structure and, therefore, the mass-loss rate from a planet.
There are a large number of degenerate free parameters in the models, and different atmospheres can produce the same transmission profile for a given feature in spatially unresolved observations.
Combining high-quality spectral features of different species and covering different transit depths and wavelengths simultaneously can effectively break the degeneracy and provide a more reliable constraint on the atmospheric structure.
Even so, for many planets, it can still be hard to resolve parameter degeneracy in the fits, even when several features are available. For example, we found multiple combinations that can fit the observations of WASP-121b with our 1D model.  
Comparing the line shape of high S/N line profile measurements with detailed 3D simulations would help to impose better constraints on the atmosphere from the observations.
More accurate measurement of the system parameters can also significantly benefit the effort to interpret observations of atmospheric escape.

The models of the atmosphere presented in this work have their limitations, and the chief among them is that they are one-dimensional.
One-dimensional models obviously cannot represent multidimensional physics such as the Coriolis effect, acceleration by stellar radiation pressure, the interaction with the stellar wind, and the effects of the stellar or planetary magnetic field.
Although we show the latter three factors are unlikely to significantly change the estimated mass-loss rate, three-dimensional dynamics is likely necessary to explain the observed blueshift of the optical absorption lines and the asymmetry of the Ca II line profiles that may be due to horizontal winds driven by day-night temperature differences in the planet's atmosphere and/or by radiation pressure or stellar wind in the extended exosphere. 
It would still be computationally difficult to incorporate the detailed physical processes discussed in this work into 3D atmospheric models without additional simplifications that would require more analysis of the 1D results to justify. While the observations of WASP-121b clearly provide strong motivation for multidimensional modeling, such modeling is therefore beyond the scope of the present work. Instead, our results provide an overview of the relevant physics and chemistry that future multi-dimensional models can use as guidance while in development.

We note that the mass-loss rate of (4--9)~$\times$~10$^{13}$ g~s$^{-1}$ based on our 1D models is likely to differ from a similar 3D model by a limited factor of a few, or at most an order of magnitude \citep{Koskinen2022}. Furthermore, despite the degeneracy in the parameters when fitting the observations, the mass-loss rates based on the best-fit combinations that we explored are consistent roughly within a factor of three.
In all cases, the models without the RL enhancement on the mass loss, therefore, fail to match the observations. This indicates that WASP-121b is likely to suffer, and to have suffered, significant envelope loss during its main sequence lifetime. 

\section*{Acknowledgements}
We thank Orly Gnat for helping us to set up the Cloudy script to reproduce their results.  C.H. thanks Xiao Fang for the useful information about the transitional rates of \ion{Ca}{1}.  We also thank David Sing for a helpful discussion. T.T.K and C.H. acknowledge support by the NASA Exoplanet Research Program through grant 80NSSC18K0569.

\bibliography{main}{}
\bibliographystyle{aasjournal}

\appendix
\section{List of charge exchange rates}
See Table~\ref{tab:charge_ex}.

\startlongtable
\begin{deluxetable}{c c}
  \tablecaption{Charge exchange rates}

  \tablehead{\colhead{Reactants} & \colhead{Rate\tablenotemark{a} ($\rm cm^{3}~s^{-1}$)}} 
  \startdata
  He + H$^+$ \tablenotemark{b} & $1.75\times 10^{-11}(T/300)^{-0.75}\exp(-12.75/T_4)$ \\
  He$^+$ + H \tablenotemark{b} & $1.25\times 10^{-15}(T/300)^{0.25}$ \\
  Mg + H$^+$ \tablenotemark{c} & $9.76\times 10^{-12}T_4^{3.14}[1+55.54\exp(-1.12T_4)]$ \\
  Mg$^+$ + H \tablenotemark{c} & $2.95\times 10^{-12}T_4^{3.28}[1+55.54\exp(-1.12T_4)]\exp(-6.91/T_4)$ \\
  Mg$^+$ + H$^+$ \tablenotemark{c} & $7.6\times 10^{-14}[1-1.97\exp(-4.32T_4)]\exp(-1.67/T_4)$ \\
  Mg$^{2+}$ + H \tablenotemark{c} & $8.58\times 10^{-14}T_4^{2.49\times 10^{-3}}[1+0.0293\exp(-4.33T_4)]$ \\
  Fe + H$^+$ \tablenotemark{d} & $4.0\times 10^{-9}$ \\
  Fe$^+$ + H \tablenotemark{d} & $1.16\times 10^{-9}(T/300)^{0.072}\exp(-6.61/T_4)$ \\
  Fe$^+$ + H$^+$ \tablenotemark{e} & $2.3\times 10^{-9}\exp(-3.0/T_4)$ \\
  Fe$^{2+}$ + H \tablenotemark{e} & $1.26\times 10^{-9}T_4^{0.0772}[1-0.41\exp(-7.31T_4)]$ \\
  Si + H$^+$ \tablenotemark{b} & $7.41\times 10^{-11}(T/300)^{0.85}$ \\
  Si$^+$ + H \tablenotemark{b} & $4.71\times 10^{-11}(T/300)^{0.95}\exp(-6.32/T_4)$ \\
  Si$^+$ + H$^+$ \tablenotemark{c} & $4.1\times 10^{-10}T_4^{0.24}[1+3.17\exp(-4.18\times10^{-3}T_4)]\exp(-3.18/T_4)$ \\
  Si$^{2+}$ + H \tablenotemark{c} & $1.26\times 10^{-9}T_4^{0.24}[1+3.17\exp(-4.18\times10^{-3}T_4)]$ \\
  Si + He$^+$ \tablenotemark{f} & $3.32\times 10^{-13}\sqrt{T} + 1.2\times 10^{-16}T +4.2\times 10^{-9}/\sqrt{T}-7.9\times 10^{-13}$\\
  Si$^+$ + He \tablenotemark{f} & $1.415\times(3.32\times 10^{-13}\sqrt{T} + 1.2\times 10^{-16}T +4.2\times 10^{-9}/\sqrt{T}-7.9\times 10^{-13})T^{0.103}\exp(-19.1/T_4)$\\
  O + H$^+$ \tablenotemark{g} & $2.08\times 10^{-9}T_4^{0.405}+1.11\times 10^{-11}T_4^{-0.458}$ \\
  O$^+$ + H \tablenotemark{g} & $(1.26\times 10^{-9}T_4^{0.517}+4.25\times 10^{-10}T_4^{6.69\times 10^{-3}})\exp(-227/T)$ \\
  C + H$^+$ \tablenotemark{h} & $1.31\times 10^{-15}(T/300)^{0.213}$ \\
  C$^+$ + H \tablenotemark{h} & $6.3\times 10^{-17}(T/300)^{1.96}\exp(-17/T_4)$ \\
  C + He$^+$ \tablenotemark{b} & $2.5\times 10^{-15}(T/300)^{1.597}$ \\
  C$^+$ + He \tablenotemark{b} & $6.75\times 10^{-15}(T/300)^{1.654}\exp(-15.5/T_4)$ \\
  C + Si$^+$ \tablenotemark{f} & $(0.724T^{0.0463}\exp(-3.61/T_4))/(1.87\times 10^{8}+5.09\times 10^{10}T^{-0.527})$ \\
  C$^+$ + Si \tablenotemark{f} & $1/(1.87\times 10^{8}+5.09\times 10^{10}T^{-0.527})$ \\
  Mg + Si$^+$ \tablenotemark{i} & $2.9\times 10^{-9}$ \\
  Mg$^+$ + Si \tablenotemark{i} & $9.8\times 10^{-10}(T/300)^{-0.0264}\exp(-0.59/T_4)$ \\
  Fe + Si$^+$ \tablenotemark{i} & $1.9\times 10^{-9}$ \\
  Fe$^+$ + Si \tablenotemark{i} & $1.4\times 10^{-9}(T/300)^{-0.236}\exp(-0.299/T_4)$ \\
  Na + Mg$^+$ \tablenotemark{i} & $1.0\times 10^{-11}$ \\
  Na$^+$ + Mg \tablenotemark{i} & $3.76\times 10^{-11}(T/300)^{-0.0302}\exp(-2.909/T_4)$ \\
  Mg + C$^+$ \tablenotemark{i} & $1.1\times 10^{-9}$ \\
  Mg$^+$ + C \tablenotemark{i} & $3.01\times 10^{-10}(T/300)^{-0.0832}\exp(-4.19/T_4)$ \\
  Mg + N$^+$ \tablenotemark{i} & $1.2\times 10^{-9}$ \\
  Mg$^+$ + N \tablenotemark{i} & $9.48\times 10^{-10}(T/300)^{0.139}\exp(-7.99/T_4)$ \\
  Na + Fe$^+$ \tablenotemark{i} & $1.0\times 10^{-11}$ \\
  Na$^+$ + Fe \tablenotemark{i} & $2.4\times 10^{-11}(T/300)^{-0.0987}\exp(-3.21/T_4)$ \\
  Fe + C$^+$ \tablenotemark{i} & $2.6\times 10^{-9}$ \\
  Fe$^+$ + C \tablenotemark{i} & $1.12\times 10^{-9}(T/300)^{0.0147}\exp(-3.90/T_4)$ \\
  Fe + O$^+$ \tablenotemark{d} & $1.71\times 10^{-9}$ \\
  Fe$^+$ + O \tablenotemark{d} & $5.0\times 10^{-10}(T/300)^{-0.0337}\exp(-6.63/T_4)$ \\
  Mg + S$^+$ \tablenotemark{i} & $2.8\times 10^{-10}$ \\
  Mg$^+$ + S \tablenotemark{i} & $5.35\times 10^{-11}(T/300)^{0.121}\exp(-3.15/T_4)$ \\
  Na + S$^+$ \tablenotemark{i} & $2.6\times 10^{-10}$ \\
  Na$^+$ + S \tablenotemark{i} & $1.86\times 10^{-10}(T/300)^{0.151}\exp(-6.06/T_4)$ \\
  Fe + S$^+$ \tablenotemark{i} & $1.8\times 10^{-10}$ \\
  Fe$^+$ + S \tablenotemark{i} & $5.38\times 10^{-11}(T/300)^{0.052}\exp(-2.85/T_4)$ \\
  S + H$^+$ \tablenotemark{j} & $\exp[-50+13.3\ln T-2.77(\ln T)^2+0.243(\ln T)^3-7.24\times 10^{-3}(\ln T)^4]$ \\
  S$^+$ + H \tablenotemark{j} & $\exp[-50.14+13.3\ln T-2.77(\ln T)^2+0.243(\ln T)^3-7.24\times 10^{-3}(\ln T)^4-3.76/T_4]$ \\      S + C$^+$ \tablenotemark{k} & $3.26\times 10^{-11}(T/300)^{0.289}\exp(-338/T)$ \\
  S$^+$ + C \tablenotemark{k} & $6.97\times 10^{-11}(T/300)^{0.356}\exp(-1.06/T_4)$ \\
  Fe + N$^+$ \tablenotemark{d} & $9.46\times 10^{-10}$ \\
  Fe$^+$ + N \tablenotemark{d} & $1.17\times 10^{-9}(T/300)^{0.07}\exp(-7.696/T_4)$ \\
  O + He$^+$ \tablenotemark{l} & $4.99\times 10^{-15}T_4^{0.379}+2.78\times 10^{-15}T_4^{-0.216}\exp(T_4/81.97)$ \\
  O$^+$ + He \tablenotemark{l} & $3.2\times(5.0\times 10^{-15}T_4^{0.38}+2.78\times 10^{-15}T_4^{-0.22}\exp(T_4/81.97))T^{0.0377}\exp(-12.7/T_4)$ \\
  Na + Ca$^+$ \tablenotemark{m} & $3.0\times 10^{-9}$ \\
  Na$^+$ + Ca \tablenotemark{m} & $1.7\times 10^{-8}(T/300)^{-0.168}\exp(-1.13/T_4)$ \\
  N + H$^+$ \tablenotemark{n} & $\exp[-35.4+1.94\ln T-0.154(\ln T)^2-6.3\times 10^{-3}(\ln T)^3-1.16\times 10^{-3}(\ln T)^4]$ \\
  N$^+$ + H \tablenotemark{n} & $\exp[-40.1+6.4\ln T-1.75(\ln T)^2+0.18(\ln T)^3-5.96\times 10^{-3}(\ln T)^4]$ \\
  Na + H$^+$ \tablenotemark{o,p} & $\exp[48.2-43.0\ln T+8.74(\ln T)^2-0.77(\ln T)^3-0.0256(\ln T)^4]$ \\
  Na$^+$ + H \tablenotemark{o} & $\exp[46.6-41.6\ln T+8.02(\ln T)^2-0.65(\ln T)^3+0.0197\times 10^{-3}(\ln T)^4-9.817/T_4]$ \\
  K + Na$^+$ \tablenotemark{q}  & $1.0\times 10^{-11}$ \\
  K$^+$ + Na \tablenotemark{q} & $6.58\times 10^{-12}(T/300)^{-0.203}\exp(-0.927/T_4)$ \\
  K + H$^+$ \tablenotemark{p} & $\exp[-27.8+0.125\ln T+0.0663(\ln T)^2-0.0237(\ln T)^3-1.36\times 10^{-3}(\ln T)^4]$ \\
  \enddata  
  \tablecomments{{\normalfont\textsuperscript{\normalsize\it a}}$T_4=T/10^4K$, {\normalfont\textsuperscript{\normalsize\it b}}\citet{Glover2007}, {\normalfont\textsuperscript{\normalsize\it c}}\citet{Kingdon1996}, {\normalfont\textsuperscript{\normalsize\it d}}\citet{Rutherford1972}, {\normalfont\textsuperscript{\normalsize\it e}}\citet{Neufeld1987}, {\normalfont\textsuperscript{\normalsize\it f}}\citet{Satta2013}, {\normalfont\textsuperscript{\normalsize\it g}}\citet{Stancil1999}, {\normalfont\textsuperscript{\normalsize\it h}}\citet{Stancil1998}, {\normalfont\textsuperscript{\normalsize\it i}}\citet{Prasad1980}, {\normalfont\textsuperscript{\normalsize\it j}}\citet{Zhao2005}, {\normalfont\textsuperscript{\normalsize\it k}}\citet{Chenel2010}, {\normalfont\textsuperscript{\normalsize\it l}}\citet{Zhao2004}, {\normalfont\textsuperscript{\normalsize\it m}}\citet{Kwolek2019}, {\normalfont\textsuperscript{\normalsize\it n}}\citet{Lin2005}, {\normalfont\textsuperscript{\normalsize\it o}}\citet{Dutta2001}, {\normalfont\textsuperscript{\normalsize\it p}}\citet{Watanabe2002}, {\normalfont\textsuperscript{\normalsize\it q}}\citet{Lavvas2014}.}
  \label{tab:charge_ex}
\end{deluxetable}

\(\)

\end{CJK*}
\end{document}